%% file: edipack_sp.tex
\documentclass[dvipsnames]{SciPost}
\usepackage[bitstream-charter]{mathdesign}
\usepackage[english]{babel}
\usepackage[T1]{fontenc}
\usepackage{lmodern}
\usepackage{amsthm}
\usepackage{microtype}
\usepackage{ragged2e}
\usepackage{makecell}
\usepackage{array}
\usepackage{stackengine}
\newlength\llength
\usepackage{makecell}
\usepackage{subfiles}
\usepackage{comment}
\usepackage{xspace}
\usepackage{array}
\usepackage{tabularx}
\usepackage{ltablex}
\usepackage{xkcdcolors}
\usepackage{dialogue}
\usepackage{lips}
  \hypersetup{
    breaklinks=true,
    colorlinks,
    pdfborder={0 0 0},
    linkcolor={red!50!black},
    citecolor={blue!50!black},
    urlcolor={blue!80!black}
  }

\DeclareSymbolFont{usualmathcal}{OMS}{cmsy}{m}{n}
\DeclareSymbolFontAlphabet{\mathcal}{usualmathcal}
\DeclareMathAlphabet\mathbfcal{OMS}{cmsy}{b}{n}

\renewcommand{\texttt}[1]{%
  \begingroup
  \ttfamily
  \begingroup\lccode`~=`_\lowercase{\endgroup\def~}{/\discretionary{}{}{}}%
  \begingroup\lccode`~=`/\lowercase{\endgroup\def~}{/\discretionary{}{}{}}%
  \begingroup\lccode`~=`[\lowercase{\endgroup\def~}{[\discretionary{}{}{}}%
  \begingroup\lccode`~=`.\lowercase{\endgroup\def~}{.\discretionary{}{}{}}%
  \catcode`/=\active\catcode`[=\active\catcode`.=\active
  \scantokens{#1\noexpand}%
  \endgroup
}

\definecolor{comment-color}{rgb}{0.8,0.1,0.1}
\definecolor{keyword-color}{rgb}{0.3,0.3,1}
\definecolor{string-color}{rgb}{0.58, 0, 0.82}
\definecolor{bg-gray}{gray}{0.96}

\usepackage{listings}

\lstdefinestyle{fstyle}{
  language={[03]Fortran},
  basicstyle=\ttfamily\linespread{1.15}\footnotesize,
  stringstyle=\color{string-color},
  keywordstyle=\color{keyword-color}\footnotesize,
  commentstyle=\color{comment-color}\itshape\footnotesize,
  morecomment=[l]{!\ },
   fontadjust=true,
  mathescape,
   breakatwhitespace=false,
   keepspaces=true,
  showstringspaces=false,
  columns=fullflexible,
  frame=none,
  xleftmargin=1pt,
  xrightmargin=1pt,
  backgroundcolor=\color{xkcdPaleLavender!20!white},
  numbers=left, numberstyle=\tiny, stepnumber=1, numbersep=3pt
}

\lstdefinestyle{cstyle}{
  language=C,
  basicstyle=\ttfamily\footnotesize,
  stringstyle=\color{string-color},
  keywordstyle=\color{keyword-color}\footnotesize,
  commentstyle=\color{comment-color}\itshape\footnotesize,
  captionpos=b,
  mathescape,
  fontadjust=true,
  breakatwhitespace=false,
  keepspaces=true,
  showstringspaces=false,
  columns=fullflexible,
  xleftmargin=3.4pt,
  xrightmargin=3.4pt,
  backgroundcolor=\color{xkcdSky!15!white},
  frame=none
}

\lstdefinestyle{mypython}{
  language=python,
  basicstyle=\ttfamily\footnotesize,
  stringstyle=\color{string-color},
  keywordstyle=\color{keyword-color},
  commentstyle=\color{comment-color}\itshape\footnotesize,
  fontadjust=true,
  morekeywords={np,plt,oplot,solve,EDIpackSolver,c,c_dag,n,ed,MPI,comm},
  morecomment=[l]{\#\ },
  mathescape,
  fontadjust=true,
  mathescape,
  breakatwhitespace=false,
  keepspaces=true,
  showstringspaces=false,
  columns=fullflexible,
  backgroundcolor=\color{xkcdPale!50!white},
  numbers=none, numberstyle=\tiny, stepnumber=1, numbersep=3pt
}

\lstdefinestyle{mybash}{
  language=bash,
  basicstyle=\ttfamily\footnotesize,
  stringstyle=\ttfamily,
  keywordstyle=\color{keyword-color},
  commentstyle=\color{comment-color}\itshape\footnotesize,
  morekeywords={*,git,mkdir,cmake,make,pip,ninja,conda},
  alsoletter=-
  deletekeywords={conda-forge},
  captionpos=b,
  mathescape,
  fontadjust=true,
  breakatwhitespace=false,
  keepspaces=true,
  showstringspaces=false,
  columns=flexible,
  xleftmargin=3.4pt,
  xrightmargin=3.4pt,
  backgroundcolor=\color{bg-gray}, 
  frame=none
}

\lstdefinelanguage{Julia}%
  {morekeywords={abstract,break,case,catch,const,continue,do,else,elseif,%
      end,export,false,for,function,immutable,import,importall,if,in,%
      macro,module,otherwise,quote,return,switch,true,try,type,typealias,%
      using,while},%
   sensitive=true,%
   alsoother={\$},%
   morecomment=[l]\#,%
   morecomment=[n]{\#=}{=\#},%
   morestring=[s]{"}{"},%
   morestring=[m]{'}{'},%
}[keywords,comments,strings]%

\lstdefinestyle{myjulia}{
  frame=lines,
  language=julia,
  basicstyle=\ttfamily\footnotesize,
  stringstyle=\ttfamily,
  commentstyle=\itshape,
  fontadjust=true,
  keywordstyle=\color{keyword-color},
  commentstyle=\color{comment-color}\footnotesize,
  morecomment=[l]{\#\ },
  mathescape,
  fontadjust=true,
  mathescape,
  keepspaces=true,
  showstringspaces=false,
  columns=fullflexible,
  xleftmargin=3.4pt,
  xrightmargin=3.4pt,
  backgroundcolor=\color{xkcdSoftPurple!20!white},
  frame=none
}

\lstdefinelanguage{Monicelli}%
  {
  morekeywords={vaffanzum,Lei,ha, clacsonato,più,meno,per,diviso,maggiore,minore,scappellamento,come fosse,Necchi,Mascetti,Perozzi,Melandri,Sassaroli,voglio,posterdati,mi,porga,stuzzica,brematura, anche,che, cos'è, o, magari,tarapia, genio,blinda,prematurata, la, supercazzola, avvertite,con,il,scherziamo,?,bituma},%
   sensitive=true,%
   alsoother={\$},%
   morecomment=[l]\#,%
   morecomment=[n]{\#=}{=\#},%
   morestring=[s]{"}{"},%
   morestring=[m]{'}{'},%
}[keywords,comments,strings]%

\lstdefinestyle{MonicelliStyle}{
  language=Monicelli, 
  basicstyle=\ttfamily\footnotesize,
  stringstyle=\ttfamily,
  keywordstyle=\color{keyword-color},
  commentstyle=\color{comment-color}\itshape\footnotesize,
  morecomment=[l]{\#\ },
  mathescape,
  fontadjust=true,
  mathescape,
  breakatwhitespace=false,
  keepspaces=true,
  showstringspaces=false,
  columns=fullflexible,
  xleftmargin=3.4pt,
  xrightmargin=3.4pt,
  backgroundcolor=\color{xkcdLightSage!20!white}, 
  frame=none
}

\newcolumntype{T}[1]{>{\tt\footnotesize\raggedright\arraybackslash}p{#1}}
\newcolumntype{D}[1]{>{\it\footnotesize\raggedright\arraybackslash}p{#1}}
\newcolumntype{M}[1]{>{\scriptsize\raggedright\arraybackslash}p{#1}}

\newcommand{\onlinecite}[1]{\cite{#1}}

\newcommand{\equ}[1]
{Eq.\,(\ref{#1})}

\newcommand{\figu}[1]
{Fig.\,\ref{#1}}

\newcommand{\secu}[1]
{Sec.\,\ref{#1}}

\newcommand{\ket}[1]
{|#1\rangle}

\newcommand{\bra}[1]
{\langle #1|}

\def\a{\alpha}       \def\b{\beta}      
\def\e{\varepsilon}  \def\z{\zeta}        
              
               \def\r{\rho}     \def\s{\sigma}

\def\PP{{\cal P}}\def\MM{{\cal M}} 
\def\FF{{\cal F}}\def\HH{{\cal H}}
\def\TT{{\cal T}}\def\NN{{\cal N}}\def\BB{{\cal B}} 
\def\RR{{\cal R}} \def\OO{{\cal O}}
\def\DD{{\cal D}}
\def\AA{{\cal A}}
\def\GG{{\cal G}} \def\SS{{\cal S}}

\def\RRR{\mathbb{R}} \def\CCC{\mathbb{C}}

\def\=={\equiv}

\def\qed{\raise1pt\hbox{\vrule height5pt width5pt depth0pt}}

\def\cG0{{\cal G}_0}
\def\cG{{\cal G}}    
  
\def\up{\uparrow} \def\down{\downarrow} \def\dw{\downarrow}

\def\=={\equiv}
 
\def\Im{{\rm Im}} \def\Re{{\rm Re}} \def\Tr{{\rm Tr}}

\def\ibra{\langle}
\def\iket{\rangle}

\def\ka{{\bf k}}

\def\11{\mathbb{1}}
\def\00{\mathbf{0}}
\def\NAME{{\rm EDIpack}\xspace}

\fancypagestyle{SPstyle}{
\fancyhf{}
\lhead{\colorbox{scipostblue}{\bf \color{white} ~SciPost Physics Codebases }}
\rhead{{\bf \color{scipostdeepblue} ~Accepted }}

\fancyfoot[C]{\textbf{\thepage}}
}

\begin{document}

\pagestyle{SPstyle}

\begin{center}{
\Large \textbf{
\color{scipostdeepblue}{
Next-generation EDIpack: A Lanczos-based package for quantum impurity models featuring general broken-symmetry phases, flexible bath topologies and multi-platform interoperability\\
}
}
}
\end{center}

\begin{center}\textbf{\small
    Lorenzo Crippa\textsuperscript{1,2,3},
    Igor Krivenko\textsuperscript{1},
    Samuele Giuli\textsuperscript{4},
    Gabriele Bellomia\textsuperscript{4},
    Alexander Kowalski\textsuperscript{3},    
    Francesco Petocchi\textsuperscript{5},
    Alberto Scazzola\textsuperscript{6},
    Markus Wallerberger\textsuperscript{7},
    Giacomo Mazza\textsuperscript{8},
    Luca de Medici\textsuperscript{9},
    Giorgio Sangiovanni\textsuperscript{2,3},
    Massimo Capone\textsuperscript{4,10} and 
    Adriano Amaricci\textsuperscript{10}    
}\end{center}

\begin{center}
%
%
  \newcommand{\CNRIOM}{CNR-IOM, Istituto Officina dei Materiali,
  Consiglio Nazionale delle Ricerche, Trieste, Italy}
\newcommand{\SISSA}{SISSA, Scuola Internazionale Superiore di Studi Avanzati, Trieste, Italy}
\newcommand{\ITPHamburg}{I. Institut f\"ur Theoretische Physik,
  University of Hamburg, Hamburg, Germany}
\newcommand{\WZBURG}{Institut f\"ur Theoretische Physik und
  Astrophysik, Universit\"at W\"urzburg, W\"urzburg, Germany}
\newcommand{\CTQMAT}{W\"urzburg-Dresden Cluster of Excellence ct.qmat, Dresden, Germany}
\newcommand{\Geneve}{Department of Quantum Matter Physics, University of
  Geneva, Geneva, Switzerland}
\newcommand{\UPISA}{Department of Physics ``E. Fermi'', University of
  Pisa, Pisa, Italy}
\newcommand{\ESPCI}{LPEM, ESPCI Paris, PSL Research University, CNRS, Sorbonne Universit\'e, Paris, France}
\newcommand{\TUW}{Institute of Solid State Physics, TU Wien, Vienna, Austria}
\newcommand{\ToPoli}{Department of Electronics and Telecommunications, Politecnico di Torino, Torino, Italy}
{\small
{\bf 1} \ITPHamburg\\
{\bf 2} \CTQMAT\\
{\bf 3} \WZBURG\\
{\bf 4} \SISSA\\
{\bf 5} \Geneve\\
{\bf 6} \ToPoli\\
{\bf 7} \TUW\\   
{\bf 8} \UPISA\\
{\bf 9} \ESPCI\\
{\bf 10} \CNRIOM\\
}
\end{center}

\begin{center}
EDIpack group: \href{mailto:edipack@cnr.iom.it}{\small edipack@cnr.iom.it}
\end{center}

\section*{\color{scipostdeepblue}{Abstract}}
\textbf{\boldmath{%
We present a next-generation version of EDIpack, a flexible, high-performance numerical library using Lanczos-based exact diagonalization to solve generic quantum impurity problems, such as those introduced in Dynamical Mean-Field Theory to describe extended strongly correlated materials. 
This new release efficiently solves impurity problems allowing for different broken-symmetry solutions, including superconductivity, featuring local spin-orbit coupling and/or electron-phonon coupling. 
It provides quick access to dynamical correlation functions on the entire complex frequency plane at zero and low-temperatures.    
The modular architecture of the software not only provides Fortran APIs but also includes bindings to C/C++, interfaces with Python and Julia or with TRIQS and w2dynamics research platforms, thus ensuring unprecedented level of inter-operability.   
The outlook includes further extensions to study quantum materials and cold atoms quantum simulators, as well as quantum information applications.
}}

\vspace{\baselineskip}



\vspace{10pt}
\noindent\rule{\textwidth}{1pt}
\tableofcontents
\noindent\rule{\textwidth}{1pt}
\vspace{10pt}

%
%

\subfile{01_intro.tex}


\subfile{02_install.tex}


\subfile{03_edipack2.tex}


\subfile{04_cbinding.tex}


\subfile{05_examples.tex}


\subfile{06_conclusions_acknowledgement_appendix.tex}


\bibliography{references}

\end{document}

%% file: 01_intro.tex
\section{Introduction and Motivation}\label{SecIntro}
Quantum impurity models play a central role in the study of strongly correlated electron systems. They describe the coupling of a small number of localized degrees of freedom to an extended non-interacting environment \cite{Nozieres1980JP,Hewson1993}. 
Originally introduced to describe diluted magnetic impurities in metallic hosts, single-impurity models have played a central role in the field of quantum many-body systems through the decades, starting from the full understanding of the Kondo problem \cite{Anderson1961PR,Kondo1964POTP,Schrieffer1966PR} and subsequently proving crucial to the understanding of a wide range of quantum many-body phenomena\cite{Wilson1975RMP,Georges1996RMP,Kotliar2004PT,Kotliar2006RMP}. More recently, their scope has expanded beyond the realm of condensed matter, influencing the understanding of other areas such as quantum information \cite{Su2013MPLB,Walsh2019PRL,Walsh2020PQ,Walsh2021PNAS,Stocker2022,Bellomia2024PRB} or ultra-cold atomic systems \cite{Dao2007PRL,Amaricci2014PRA,Del-Re2018PRA,Walsh2019PRB,Tusi2022NP}.

At the heart of quantum impurity models is the interaction between a localized ``impurity''  and a surrounding bath of itinerant particles, usually represented as a conduction band of electrons \cite{Hewson1993}. In this context, the impurity  can be defined more generally, so as to include orbital and spin local degrees of freedom as well as a finite number of lattice sites, i.e. quantum clusters, immersed in a suitable environment. 
This setup lies at the core of several quantum embedding methods, including Density-Matrix Embedding Theory (DMET) \cite{Knizia2012PRL,Sun2020PRB} and further extensions \cite{Scott2021PRB,Nusspickel2022PRX}, the Gutzwiller Approximation (GA) \cite{Lanata2015PRX,Mejuto-Zaera2023PRB},
Quantum Cluster methods \cite{Potthoff2003TEPJBCMACS,Potthoff2011ACP,Dionne2023SPC,Dionne2023SPCa}
or Dynamical Mean-Field Theory (DMFT) \cite{Georges1996RMP, Kotliar2004PT,Kotliar2006RMP}, capturing many essential aspects of strong correlation, including screening \cite{Roekeghem2014PRL,Roekeghem2014EL,Werner2016JOPCM,Tomczak2017TEPJST}, decoherence and entanglement in a computationally tractable way \cite{Walsh2021PNAS,Bellomia2024PRB}. 


In several of these embedding schemes, most notably DMFT, the complex physics of the lattice systems gets reduced to an effective self-consistent quantum impurity problem.
By accounting for local quantum fluctuations, DMFT successfully describes the most important physical features of correlated materials~\cite{Georges1996RMP,Kotliar2006RMP}, including Mott insulators~\cite{Georges1992PRL,Zhang1993PRLb,Rozenberg1994PRB,Kotliar1996PRB,Rozenberg1999PRL,Kotliar1999EPJB,Kotliar2000PRL,Bulla2001PRB,Capone2001PRL,Kotliar2002PRL,Limelette2003S8}, heavy fermion compounds~\cite{Shim2007N,De-Leo2008PRL,Haule2009NP,Crippa2024NC,Gleis2024PRX,Crispino2025PRL,Gleis2025PRL} and unconventional superconductors~\cite{Emery1987PRL,Caffarel1994PRL,Capone2001PRL,Capone2002Science,Toschi2005NJP,Toschi2005PRB,Haule2007PRB,Medici2014PRL,Giannetti2016AIP,Mazza2021PRB,Walsh2021PNAS}.

Accurately and efficiently solving a quantum impurity problem and computing the Green's functions, which are central in the DMFT framework, nonetheless remains a critical challenge, particularly for multi-orbital systems, those 
at low temperatures or endowed with lower symmetry.
To this end, 
many advanced numerical techniques have been developed over the years, incorporating cutting-edge strategies to reduce computational loads and leverage state-of-the-art algorithms \cite{Bauer2011JOSMTAE,Parcollet2015CPC}. 
These range from continuous-time quantum Monte Carlo 
methods \cite{Gull2011RMP,Rubtsov2005PRB,Haule2007PRB,Seth2016CPC,Wallerberger2019CPC},
to numerical renormalization group approaches \cite{Zitko2009PRB,Bulla2001PRB,Bulla2008RMP,Debertolis2021PRB} 
and density-matrix renormalization group \cite{Zitko2009PRB,Bulla2001PRB,Bulla2008RMP,Nunez-Fernandez2025A} or Configuration Interaction \cite{Zgid2012PRB,Lu2014PRB,Go2017PRB,Bi2019CPC,Mejuto-Zaera2019PRB} solvers. 
In this context, the Exact Diagonalization (ED) approach plays a significant 
and distinctive role \cite{Caffarel1994PRL,Dolfen2006,Perroni2007PRB,Capone2007PRB,Weber2012PRB,Lu2017TEPJST,Amaricci2022CPC}.

Several open-source ED software packages have recently been made available within the condensed matter community to address quantum many-body problems with an emphasis on efficiency and flexibility. Examples include \texttt{EDLib} \cite{Iskakov2017} and the newer 
\texttt{XDiag} \cite{Wietek2025}, which focus on generic fermionic systems; 
\texttt{Pomerol} \cite{Antipov2015} designed for fermion-boson models and interfaced with TRIQS \cite{Parcollet2015CPC}; \texttt{H$\Phi$} \cite{Kawamura2017CPC,Ido2024CPC} 
for fermions coupled to spins; and general-purpose spin-model solvers such as 
\texttt{QuSpin} \cite{Weinberg2017SP,Weinberg2019SP}, all contributing to the 
rich landscape of ED-based techniques.

In this work we present the improved and renewed \NAME{}: a flexible, high-performance numerical library designed to address these challenges, providing an ED solver for generic quantum impurity problems. 
The library builds on the foundations of
its predecessor \cite{Amaricci2022CPC}, featuring massively parallel Lanczos-based algorithms \cite{Lanczos1950JRNBSB,Lehoucq1998,Maschhoff1996} while including several significant extensions and improvements. The library now supports, within a unified framework, both zero and finite temperature calculations \cite{Amaricci2022CPC,Capone2007PRB} for a wide set of impurity models, making it suitable for addressing a broad class of problems, from spin-orbit coupling in quantum materials to multi-orbital superconductivity. It also provides direct
access to numerically exact dynamical correlation functions across the
entire frequency complex plane, thus enabling calculations of spectral functions, local susceptibilities, and other response properties.

Additionally, \NAME offers access to the Fock space of
the impurity system and thus, for instance, the possibility to evaluate the impurity reduced density matrix, enabling quantum information-inspired analyses providing insights into the entanglement properties, subsystem purity, and quantum correlations of interacting electron systems \cite{Walsh2021PNAS,Bellomia2024PRB}.
In the quantum embedding context, these features are increasingly recognized as essential to understand the emergent behavior of correlated matter, including quantum criticality and superconductivity.

This new \NAME version also includes support for systems with inequivalent impurities. This is a critical feature in DMFT or, more generally, in quantum embedding methods, to study hetero-structures, super-lattices, and disordered systems, where the local environment varies significantly across different sites.

With the growing importance of quantum impurity models in the study of correlated materials, standardization and interoperability across software suites have become critical necessities.
\NAME has been expressly designed following this guiding principle. 
While the library itself is implemented using modern Fortran constructs, it also provides an extensive set of C-bindings that enable integration with other programming languages,
be they compiled (C, C++), JIT-compiled (Julia) or interpreted (Python).  
Especially notable in this sense is the Python API, EDIpack2py, which beyond providing a 
high-level interface for defining and solving impurity problems, it also allows integration with established scientific libraries, such as NumPy and SciPy, facilitating data analysis and
rapid prototyping. 

A key development feature in \NAME, unlocked by the introduction of the Python API, 
is the possibility to directly interface to both TRIQS \cite{Parcollet2015CPC} (Toolbox for Research on Interacting Quantum Systems) and
w2dynamics \cite{Wallerberger2019CPC} software suites, which are two of the most important and widely used computational frameworks in the field.
\NAME slots in as an alternative solver to the native Quantum Monte Carlo 
implementations \cite{Gull2011RMP,Seth2016CPC}, extending the parameter (and especially temperature) range reachable by 
the simulation while minimizing the coding requirements for the end user.
At the same time, this integration provides \NAME users with access to the
extensive data handling and analysis tools in the TRIQS ecosystem. 
Overall, interoperability enhances the collective reach and reliability of these platforms, 
improving the reproducibility and cross-validation of numerical results, with the ultimate goal of achieving standardization for quantum impurity solvers along the guiding principles of initiatives such as FAIRmat (see \href{https://www.fairmat-nfdi.eu/fairmat/}{fairmat-nfdi.eu/fairmat}).

In this paper, we present the mathematical framework, details of the implementation, and key features of \NAME, along with several illustrative applications. We aim to demonstrate how this flexible,
high-performance solver can be used to tackle a wide range of
challenging problems. 
The rest of this work is organized as follows. In \secu{SecInstall} we 
briefly illustrate the overall structure of \NAME together with its dependencies, configuration, and installation procedures. In \secu{SecEDIpack} we present the quantum impurity
problem and review the software implementation in detail. In \secu{SecInterop} we thoroughly discuss the C-binding interface that is at the heart of the 
interoperability capabilities of \NAME. In this section we also discuss 4 interface layers: i) 
the Python API, ii) the TRIQS interface, iii) the w2dynamics interface and iv) an experimental, yet operative, Julia API.
In \secu{SecExamples} we discuss the functionalities and the capabilities of \NAME and  all the discussed extensions with a series of illustrative examples. 
Finally, in \secu{SecConclusions} we present some concluding remarks and considerations. 

\ifSubfilesClassLoaded{
  \bibliography{references}
}{}

\end{document}

%% file: 02_install.tex
\section{Installation}\label{SecInstall}
The configuration and installation of \NAME is handled by CMake, which ensures
cross-platform compatibility and dependency resolution.  
The software builds into two distinct libraries.
The main one is a static Fortran library named {\tt libedipack.a}, which alongside the compiled
modules, wraps the \NAME software.
A second dynamic library, {\tt libedipack\_cbindings.so}, together with the associated header
file {\tt edipack\_cbindings.h}, enables interoperability with other programming languages.

\subsection{Structure}\label{sSecInstallStructure}
\NAME is a modular library organized into two main components. 
At its core lies the ED solver for the quantum
single-impurity Hamiltonian, constituting the primary computational engine.
Building on this, the library includes \NAME{2ineq}, an extension for handling multiple inequivalent impurity problems at once. Finally, a Fortran-C/C++ interface is provided for seamless integration with external software and the development of additional APIs.

\begin{itemize}
\item{\bf EDIpack.}
  This module forms the foundation of the library, implementing a 
  Lanczos-based solver for general quantum impurity problems. It 
  supports systems with a wide range of symmetries leveraging conservation of different quantum numbers, multi-orbital models, and even coupling to local 
  phonons. The \NAME solver is structured hierarchically and modularly, 
  with different sections of the library communicating through a shared 
  memory layer. The top-level module, {\tt EDIPACK}, provides access 
  to the core Fortran API, exposing key procedures for initialization, 
  execution, and finalization, while abstracting the underlying data 
  structures. A detailed overview of this part of the library is   provided in \secu{SecEDIpack}.
  \begin{itemize}
    \item[$\hookrightarrow$]
    {\it EDIpack2ineq.}
  This extension, developed using suitable Fortran interfaces,
  enables the treatment of multiple, independent quantum 
impurities. It is particularly useful in DMFT applications involving 
unit cells with inequivalent atomic sites or systems with broken 
translational symmetry, such as heterostructures, large supercells, 
or disordered materials. This module provides flexible memory 
management and supports the simultaneous solution of multiple impurity 
problems, as discussed in \secu{sSecIneq}.
\end{itemize}

\item{\bf EDIpack C-bindings.}
  For enhanced interoperability, \NAME includes a dedicated module 
implementing Fortran-C/C++ bindings for key library procedures. This 
module relies on the {\tt ISO\_C\_BINDING} capabilities available in 
modern Fortran, allowing direct translation of Fortran functions to C. 
To ensure a straightforward user experience, only the functions and 
variables directly exposed to the user are included in this binding, 
shielding developers from the complexity of the library's internal 
architecture. This interface is intended to facilitate integration 
with third-party software and support the development of custom APIs.
\end{itemize}

\subsection{Dependencies}\label{sSecInstallDependencies}
\NAME directly depends on two external libraries.
\begin{enumerate}
\item {\bf SciFortran}: an open-source Fortran library that provides support
for mathematical and scientific software development, available at \href{https://github.com/SciFortran/SciFortran}{github.com/SciFortran}.
\item {\bf MPI}: a distributed memory parallel communication layer with support for modern Fortran compilers.
\end{enumerate}
 
{\bf SciFortran} provides a solid development platform enabling access to
many algorithms and functions, including standard linear algebra
operations and high-performance Lanczos-based algorithms. This
greatly reduces code clutter and development time.
The use of distributed memory parallel environment 
is required to access scalable parallel diagonalization algorithms,
which speed up calculations for large systems. Nonetheless, thanks to careful design, \NAME can operate serially even without initializing the MPI framework. 

\subsection{Build and Install}\label{sSecInstallBuildInstall}
\subsubsection{Compilation from source}
The software can be installed from source as follows. The code package can
be retrieved directly from its GitHub repository, for instance using
\begin{lstlisting}[style=mybash,numbers=none]
git clone https://github.com/edipack/EDIpack 
\end{lstlisting}
Then, assuming to be in the software root directory ({\tt \$cd EDIpack}), a conventional out-of-source building can be performed as follows:

\begin{lstlisting}[style=mybash,numbers=none]
mkdir build
cd build
cmake ..
make -j
make install
\end{lstlisting}



\noindent
The CMake configuration can be further tuned using the following
variables:
\begin{table}
\centering
\small
\begin{tabular}{ l|l|l } 
 \hline
  {\bf Option}               & {\bf Scope} & {\bf Value (default in {\color{xkcdEmerald}green})}\\
  \hline
  -D{\bf CMAKE\_INSTALL\_PREFIX}          & Install prefix override  & {\color{xkcdEmerald} None}\textbf{/}User-defined path \\
  
  -D{\bf LONG\_PREFIX}          & Install directory (if & {\color{xkcdEmerald} $\sim$/opt/edipack/\textbackslash}\\
  &install prefix unset)&{\color{xkcdEmerald}TAG/\textbackslash} (\textit{e.g. 5.3.3})\\
  &&{\color{xkcdEmerald}PLAT/\textbackslash} (\textit{gnu/intel/...})\\
  &&{\color{xkcdEmerald}BRANCH} (\textit{branch name})\\
  &&$\sim$/opt/edipack/custom \\
  -D{\bf USE\_MPI}       & MPI support  &  \textbf{/}{\color{xkcdEmerald}True}\textbf{/}{False}\\
  -D{\bf WITH\_INEQ}   & Multi-impurities support & {\color{xkcdEmerald}True}\textbf{/}{False}\\
  -D{\bf VERBOSE}      & Verbose CMake output & {\color{xkcdEmerald}True}\textbf{/}{False}\\ 
  -D{\bf BUILD\_TYPE} & Compilation flags & {\color{xkcdEmerald}RELEASE}\textbf{/}TESTING\textbf{/}DEBUG \\
 \hline
\end{tabular}
\caption{{\bf CMake optional flags}. The table lists the CMake
  optional flags available for the configuration step of the \NAME
  installation. Default variables are indicated in green.}
\label{Table1}
\end{table}
The default target builds and installs both the main library and the C-bindings.
Separate build targets for each component are available. A recap message is printed at the end of the CMake configuration step to guide the user through the compilation, installation and OS loading process. 

\subsubsection{Anaconda}
Installation is also available through Anaconda packages for Linux and macOS systems, within a virtual environment containing Python
($\geq$ 3.10).
The Conda package installation procedure for a virtual environment called {\tt myenv} reads
\begin{lstlisting}[style=mybash,numbers=none]
conda create -n myenv
conda activate myenv
conda install -c conda-forge -c edipack edipack
\end{lstlisting}
and installs a bundle of SciFortran and \NAME libraries together with
specific {\tt pkg-config} configuration files, which can be used to
retrieve compilation and linking flags. In order to compile Fortran/C++ 
programs, the {\tt compilers} conda package will need to be installed.

\subsection{Environment Variables}\label{sSecInstallOSloading}
In order to avoid possible conflicts or the requirement for administrative
privileges, the results of the building step are installed by default
in the user's {\tt HOME}
directory.
As a consequence, the environment variables holding the library and include paths will need to be updated by the user.
We offer different ways to perform this action:
\begin{enumerate}
\item  A CMake-generated configuration file for an environment module
  which allows users to load and unload the library at any time. This
  is the preferred solution for HPC systems. 
\item A CMake-generated bash script to be sourced (once or
  permanently) in any shell session to add \NAME library to the
  default environment.
\item A CMake-generated {\tt pkg-config} file to be added in
  the {\tt PKG\_CONFIG\_PATH} itself.  
\end{enumerate}
A recap message with all instructions is automatically generated at the end of the installation procedure.

\ifSubfilesClassLoaded{
  \bibliography{references}
}{}

\end{document}

%% file: 03_edipack2.tex
\section{Implementation}\label{SecEDIpack}
In this section we introduce the generic quantum impurity problem and
present a detailed overview of the \NAME library implementation.

\subsection{The quantum impurity problem}\label{sSecQIM}
We consider a quantum impurity problem described by the Hamiltonian:
$$
\hat{H} = \hat{H}_\mathrm{imp} + \hat{H}_\mathrm{bath} + \hat{H}_\mathrm{hyb} + \hat{H}_\mathrm{ph} + \hat{H}_\mathrm{e-ph}
$$
which characterizes a multi-orbital quantum impurity
coupled to an electronic bath and local phonons (e.g. Holstein
modes). For now, we assume no specific symmetries.

\subsubsection{Impurity Hamiltonian}
The impurity Hamiltonian is split into a quadratic and an 
interacting part,
\begin{equation}\label{Himp}
  \hat{H}_\mathrm{imp}  = \hat{H}^0_\mathrm{imp} + \hat{H}^\mathrm{int}_\mathrm{imp}.
\end{equation}
The first term has the form
\begin{equation}\label{H0imp}
  \hat{H}^0_\mathrm{imp}  =
  \sum_{\a\b\sigma\sigma'}h^{0}_{\a\b\sigma\sigma'}d^\dagger_{\a\sigma}d_{\b\sigma'}
\end{equation}
and represents the non-interacting part, with $d_{\a\sigma}$ ($d^\dagger_{\a\sigma}$) being the annihilation (creation)
operators for impurity electrons in 
orbital $\alpha=1,\dots,N_\a$, where $N_\a$ is the number of orbitals,
and spin $\sigma=\up,\dw$.
The internal structure of this quadratic part is captured by the 
matrix $h^{0}_{\a\b\sigma\sigma'}$, which can include 
orbital-dependent on-site energies or amplitudes, spin-orbit coupling or other single-particle terms.
In \NAME the matrix $\hat{h}_0$ must be set by the user before the solver is initialized using the function {\tt ed\_set\_Hloc}. The latter  accepts complex arrays of rank-2 or 4 as only argument. In a quantum embedding framework, e.g. DMFT, the matrix $\hat{h}_0$ corresponds to the local part of the non-interacting Hamiltonian. Some examples of use can be found in  \secu{SecExamples}.

The interaction part of the impurity Hamiltonian, 
 $\hat{H}^{\mathrm{int}}$ can, in principle, contain any 
set of two-body operators: 
\begin{equation}
\hat{H}^\mathrm{int}_\mathrm{imp} = \frac{1}{2} \sum_{ijkl} d^{\dagger}_{i}d^{\dagger}_{j}U_{ijkl}d_{l}d_{k}.\label{HintUmat}
\end{equation}
where the symbols $\{ijkl\}$ collect both orbital and spin indices, e.g.  $i=\alpha\sigma$.   
However, we typically 
adopt a generic formulation of the local multi-orbital Hubbard-Kanamori interaction \cite{Georges2013ARCMP}:
\begin{equation}\label{Hint}
  \begin{split}
    \hat{H}^\mathrm{int}_\mathrm{imp} &=U\sum_{\a}n_{\a\uparrow}n_{\a\downarrow}+U'\sum_{\a\neq \b}n_{\a\uparrow}n_{\b\downarrow}+(U'-J)\sum_{\a<\b,\sigma}n_{\a\sigma}n_{\b\sigma}\\
    &{\phantom =}- J_X\sum_{\a\neq
      \b}d^\dagger_{\a\uparrow}d_{\a\downarrow}d^\dagger_{\b\downarrow}d_{\b\uparrow}+J_P\sum_{\a
      \neq
      \b}d^\dagger_{\a\uparrow}d^\dagger_{\a\downarrow}d_{\b\downarrow}d_{\b\uparrow},\\
\end{split}
\end{equation}
where we introduced the occupation number operator $n_{\alpha\sigma}=d^\dagger_{\alpha\sigma}d_{\alpha\sigma}$.
The first three terms represent the density-density part of the
interaction, where $U$ is the local intra-orbital Coulomb repulsion,
$U'$ the inter-orbital one and $J$ the Hund's coupling \cite{Werner2008PRL,Werner2009PRB,Mravlje2011PRL,Medici2011PRL,Medici2011PRB,Georges2013ARCMP}.  
The last two terms are, respectively, the spin-exchange ($J_X$) and the pair-hopping ($J_P$).
In the three-orbital case $N_\a=3$ a fully symmetric $SU(3)_\mathrm{orbital}\otimes SU(2)_\mathrm{spin}\otimes
U(1)_\mathrm{charge}$ form of the interaction is obtained by setting $U'=U-2J$ and
$J_X=J_P=J$ \cite{Georges2013ARCMP}. Alternative choices that preserve part
of the combined symmetry group can be made for other numbers of orbitals \cite{Georges2013ARCMP}.

\subsubsection{Bath and Hybridization}
The coupling between the impurity and the bath is described by
\begin{equation}\label{Hbath}
  \begin{split}
    \hat{H}_\mathrm{bath} &=
    \sum_p\sum_{\a\b\sigma\sigma'}h^p_{\a\b\sigma\sigma'}a^\dagger_{p\a\sigma}a_{p\b\sigma'},\\
    \hat{H}_\mathrm{hyb} &= \sum_p\sum_{\a\b\sigma\sigma'}V^p_{\a\b\sigma\sigma'}d^\dagger_{\a\sigma}a_{p\b\sigma'}+H.c., \\
\end{split}
\end{equation}
where $p=1,\dots,N_\mathrm{bath}$ indexes the
bath elements. The operators $a_{p\alpha\sigma}$
($a^\dagger_{p\alpha\sigma}$) correspond to the destruction (creation) of
bath electrons with index $p$, orbital $\alpha$ and spin $\sigma$.
Any bath element can be composed of several electronic levels
according to the chosen bath topology (see \secu{sSecBath}). The  properties of each bath element are described by the
matrix $h^p_{\a\b\sigma\sigma'}$ while the amplitude
$V^p_{\a\b\s\sigma'}$ describes the coupling with the impurity.   
The bath parametrization is handled in \NAME using reverse communication strategy as will be discussed later in \secu{sSecBath}.

\subsubsection{Electron-Phonon coupling}
Finally, we include a local electron-phonon coupling on the impurity site
described by the Hamiltonian terms: 
\begin{equation}\label{Hph}
  \begin{split}
    \hat{H}_\mathrm{ph}&=\sum_m \omega_{0m} b_m^\dagger b_m,   \\
    \hat{H}_\mathrm{e-ph} &= \sum_m [ \hat{O}_m - A_m](b_m+b_m^\dagger), \\
\end{split}
\end{equation}
where $m=1,\dots,M$ indexes the number of local phonon modes, and $b_m$
($b_m^\dagger$) are the destruction (creation) operators for a phonon with
frequency $\omega_{0m}$. Each phonon mode is coupled to a bilinear fermionic operator $\hat{O}_m= \sum_{\alpha \beta, \sigma }g^{m}_{\alpha \beta} d^\dagger_{\alpha, \sigma} d_{\beta, \sigma}$ on the impurity, while $A_m$ is a constant acting as a displacement field.  
For single-orbital impurities, the relevant case is $\hat{O}_1 = g_1 n_{1d} = g_1 \sum_{\sigma} d^{\dagger}_{1\sigma}d_{1\sigma}$, which defines the Holstein model \cite{Holstein1959APNY}. For two-orbital models we included also the two Jahn-Teller modes \cite{Capone2010AICMP}, for which $\hat{O}_2 = g_2(n_{1d} - n_{2d})$ \cite{Scazzola2023PRB} and $\hat{O}_3 = g_3 \sum_{\sigma} (d^{\dagger}_{1\sigma}d_{2\sigma}+ H.c.)$. 

For each phonon mode, one has to introduce a cut-off in the phonon number to avoid unbounded growth of the phonons Fock space. The cutoff has to be chosen in a problem dependent way, also considering the value of the electron-phonon coupling \cite{Capone2003PRL,Capone2006PRB} and its interplay with the electron-electron interaction \cite{Capone2004PRL,Capone2004PRLsc,Sangiovanni2005PRL,Paci2005PRL,Sangiovanni2006PRL,Sangiovanni2006PRBa,Paci2006PRB}. 
The larger the number of excited phonons in the ground state, the larger the cutoff. 
For practical calculations a cutoff of about 20 phonons is usually sufficient to achieve converged results, except for the deep polaronic regime \cite{Capone2003PRL}.

In many cases, one can reduce the number of excited phonons in the ground state, and
consequently the required cutoff, by appropriately choosing the shift $A_m$.
In practice one can use $A_m = \langle \hat{O}_m \rangle $ and redefine a shifted bosonic fields $\tilde{b}_m^{(\dagger)} = b_m^{(\dagger)} + \frac{\langle \hat{O}_m \rangle }{\omega_0}$. This choice would ensure that $\langle \tilde{b}_m^\dagger + \tilde{b}_m \rangle =0$ effectively reducing the number of phonons required for the calculation. Observables on the physical bosons are linked to the new one via the previous transformation.
Although feasible, dealing with more than one phonon mode quickly becomes computationally very demanding, thus we will consider only one mode and we drop the $m$ index in the rest of this work.

The phonon frequency is passed to the \NAME program via the variable { \tt W0\_PH } to be read from the input file. The displacement field acting on the phonon is passed through the variable {\tt A\_PH} from the input file. For every phonon mode, the $g$ coupling can be written as a matrix $g_{\alpha \beta}$ which can be passed in two different ways. If the matrix is diagonal ($g_{\alpha \beta} = g_\alpha \delta_{\alpha \beta}$), from the input file one can simply write in order the values $g_\alpha$ in the variable {\tt G\_PH}. For any general shape one can pass to the input file a {\tt GPHfile } variable containing the name of the file storing the $g_{\alpha \beta}$ array.
If {\tt GPHfile=NONE}, the matrix $g_{\alpha\beta}$ is read from the variable {\tt G\_PH}, otherwise it is read from the {\tt GPHfile}.

\subsubsection{System setup}
We consider a quantum impurity system comprised of a single multi-orbital impurity, a bath discretized into a number $N_\mathrm{bath}$ of {\it elements} possibly endowed with an internal structure, and a number of available phonons. 
The total size of the system is determined by three contributions: i) the number of impurity orbitals $N_\a$, ii) the total count of electronic bath levels $N_\mathrm{b}$, and iii) by the maximum  number of allowed phonons $N_\mathrm{ph}$, introducing a cutoff to the unbounded dimension of the local phonon Hilbert space.
The number of bath degrees of freedom $N_\mathrm{b}$ is a
function of the bath topology and of $N_\mathrm{bath}$, which is determined internally by \NAME.
For instance, for a simple case of a purely electronic problem, each bath
element corresponds to an independent electronic level coupled to the
impurity, thus $N_\mathrm{b}\equiv N_\mathrm{bath}$. 
The total number of electronic levels in quantum impurity systems is indicated with $N_\mathrm{s}$.

The setup of the quantum impurity problem is implemented in {\tt
  ED\_SETUP} through the input variables {\tt Nspin}$=N_\sigma$,
{\tt Norb}$=N_\a$ and {\tt Nbath}$=N_\mathrm{bath}$ globally defined in {\tt
  ED\_INPUT\_VARS}. The variable {\tt Ns}$=N_\mathrm{s}$, evaluated in
{\tt ed\_setup\_dimensions}, corresponds to the total number of
electronic levels. 
The local non-interacting Hamiltonian
$h^0_{\a\b\s\s'}$ is specified using the function {\tt ed\_set\_hloc} from {\tt ED\_AUX\_FUNX}.
The setup of the bath matrices $h^p_{\a\b\s\s'}$
requires a more involved procedure, which will be illustrated in
\secu{sSecBath}.

\subsection{The Fock basis states}\label{sSecBasis}
The Fock space of the quantum impurity problem is defined as
$\FF=\FF_\mathrm{e}\otimes \FF_\mathrm{ph}$, where: $$
\FF_\mathrm{e}=\bigoplus_{n=0}^{N_\mathrm{s}}
S_-\HH_\mathrm{e}^{\otimes n}
$$ 
is the electronic Fock space built 
from the local electronic Hilbert space $\HH_\mathrm{e}$, and 
$$
\FF_\mathrm{ph}=\bigoplus_{n=0}^{N_q}S_+\HH_\mathrm{ph}^{\otimes n}
$$
is the
phonon Fock space, with local phonon Hilbert space
$\HH_\mathrm{ph}=\{\ket{0},\ket{1},\dots,\ket{N_\mathrm{ph}}\}$. 
Here ($S_-$) $S_+$ is the (anti-)symmetrization operator.  
The total dimension of the Fock space is
$$
D=4^{N_\mathrm{s}} (N_\mathrm{ph}+1)=D_\mathrm{e}D_\mathrm{ph},
$$ 
highlighting the exponential
growth with the number of electronic levels. 

The quantum states in the space $\FF$ are naturally represented in
the occupation number formalism of second quantization,
i.e.~the Fock basis.
For a system of $N_\mathrm{s}$ electrons, each Fock state
is given as $\ket{p}\ket{\vec{n}}$, with
\begin{equation} \label{eq:FockState_|nupndw>}
    \ket{\vec{n}}=\ket{n_{1\up},\dots,n_{N_\mathrm{s}\up},n_{1\dw},\dots,n_{N_\mathrm{s}\dw}},
\end{equation}
where $p=1,\dots,N_\mathrm{ph}$ is the number of local phonons and
$n_{a\sigma}=0,1$ indicates the absence or 
presence of an electron with spin $\sigma$ at level $a$.

The electronic part of the Fock state $\ket{\vec{n}}$ is represented
as a binary string of length $2N_\mathrm{s}$. Thus, any such state can be encoded
in a computer using a sequence of $2N_\mathrm{s}$ bits, or equivalently, as an integer $I=0,\dots 2^{2N_\mathrm{s}}-1$ such that $\ket{\vec{n}}=\ket{I}$.  
The electronic destruction and creation operators, $c_{a\sigma}$ and $c^\dagger_{a\sigma}$ 
respectively, act on the Fock space as:
\begin{align*}
  c_{a\sigma}\ket{\vec{n}} &=
    \begin{cases}
      (-1)^{\#_{a\sigma}}\ket{\dots,n_{a\sigma}\!-\!1,\dots}
      &\text{if $n_{a\sigma}\!=\! 1$}\\
      0 &\text{otherwise}
    \end{cases};\\
    c^\dagger_{a\sigma}\ket{\vec{n}} &=
     \begin{cases}
      (-1)^{\#_{a\sigma}}\ket{\dots,n_{a\sigma}\!+\!1,\dots}
      & \text{if $n_{a\sigma}\!=\! 0$}\\
      0 & \text{otherwise}
    \end{cases}    
\end{align*}
where $\#_{a\sigma}=\sum_{i\sigma'<a\sigma} n_{i\sigma'}$ accounts for the fermionic sign imposed by the Pauli principle.
The bosonic operators $b$ and $b^\dagger$ act on the phonon part $\ket{p}$ of the Fock state as
\begin{align*}
  b\ket{p} &= \sqrt{p} \ket{p-1}, \\
  b^\dagger\ket{p} &= \sqrt{p+1}\ket{p+1}.
\end{align*}
The implementation of Fock states and operators can be found in 
{\tt ED\_AUX\_FUNX}. In particular, in that module we define the bitwise action 
of the fermionic creation and annihilation operators as functions 
{\tt CDG} and {\tt C}, used throughout the code. 
Additionally, 
we provide the function {\tt bdecomp} for reconstructing the Fock  state bit sequence from the integer representation.

\subsection{Conserved quantum numbers}\label{sSecQNs}
To mitigate the exponential scaling of the Fock space dimension, it 
is essential to exploit suitable symmetries. Specifically, if the 
Hamiltonian $H$ commutes with a set of operators ${\cal \vec{Q}}=[{\cal Q}_1,{\cal Q}_2,\dots,{\cal Q}_M]$ such that 
$[H, {\cal Q}_i] = 0$, the Fock space can be decomposed into smaller, 
disjoint symmetry sectors labeled by quantum numbers $\vec{Q}$.

In the context of quantum impurity problems, two commonly used 
symmetries are: i) conservation of the total charge $N$ and ii) 
conservation of the total magnetization $S_z$. Although the total 
spin operator $S^2$ may also be conserved, its implementation is 
computationally challenging and typically provides only marginal 
gains, making it less practical in many cases.
As the symmetries apply to the electronic terms, here we discuss the organization of the electronic Fock states only. The presence of phonons is accounted for by exploiting the tensor structure of the global Fock space.

In \NAME, three different symmetry configurations 
controlled by the input variable {\tt ed\_mode} = {\bf normal}, 
{\bf superc}, {\bf nonsu2} are available. A summary of the sector properties is provided in
Table~(\ref{Table2}). 

\begin{itemize}
\item{} The {\bf normal} case conserves both the electrons' total occupation $N$
and total magnetization $S_z$, or equivalently, the total number
of electrons with spin up $N_\up$ and down $N_\dw$.
Optionally, the symmetry
can be extended to act independently on each orbital and spin
component, i.e.~$\vec{N_\s}=[N^1_\s,\dots,N^{N_\a}_\s]$. This
specific case is discussed extensively
Ref. \cite{Amaricci2022CPC}, so we will cover it no further here recalling that any consideration concerning the {\tt normal} case extends directly to this case as well.

\item{} The {\bf superc} case maintains only the $S_z$ conservation,
allowing the total charge to not be conserved. This setting captures
systems with $s$-wave superconductivity, including 
intra- and inter-orbital pairing.

\item{} The {\bf nonsu2} case conserves the total
charge $N$ while allowing the spin symmetry group to be broken.
This scenario captures effects such as local spin-orbit coupling
$\vec{L}\cdot\vec{S}$, in-plane spin ordering \cite{BellomiaKMH} or
in-plane spin-triplet exciton condensation \cite{Amaricci2023PRB,Blason2020PRB}.  
\end{itemize}

\begin{table}
  \centering
  \begin{tabular}{ |c|c|c| }
 \hline
  {\tt ed\_mode} & {\it Quantum Numbers} & {\it Sector Dimension} \\
  \hline
  {\bf normal} & $[N,S_z]\Leftrightarrow[N_\up,N_\dw]$ &
                                                $\binom{N_\mathrm{s}}{N_\up}\binom{N_\mathrm{s}}{N_\dw}$
  \\
  \hline
  {\bf superc} & $S_z\equiv N_\up-N_\dw$ &  $\sum_m 2^{N_\mathrm{s}-S_z-2m}\binom{N_\mathrm{s}}{N_\mathrm{s}-S_z-2m}\binom{S_z+2m}{m}$
  \\
  \hline
  {\bf nonsu2} & $N \equiv N_\up+N_\dw$ & $\binom{2N_\mathrm{s}}{N}$ \\ 
 \hline  
\end{tabular}
\caption{
 {\bf Quantum Numbers}.
A table summarizing the possible values of the input variable {\tt
  ed\_mode} selecting the symmetries of the problem (first column). To each value
correspond different quantum numbers listed in the second column.
The third column reports the dimension of the electronic symmetry sector $\SS_{\vec{Q}}$ for a
given value of the quantum numbers $\vec{Q}$.}
\label{Table2}
\end{table}

From a computational perspective, constructing a symmetry
sector amounts to defining an injective map
$\MM:\SS_{\vec{Q}}\rightarrow \FF_e$ that relates the consecutive indexing of the electronic states  $\ket{i}$
within a given symmetry sector $\SS_{\vec{Q}}$ to the Fock states $\ket{I_e}$ in the electronic Fock space in occupation number representation~\cite{Wallerberger2022PRR}. 
This map is typically implemented as a rank-1 integer array, whose size corresponds to the dimension of the sector $D_{\SS_{\vec{Q}}}$. 
The global Fock state $\ket{I}=\ket{I_e}\ket{p}$, including the phonon part, is univocally determined by the relation: $I\to I_e + p\,2^{2 N_s}$.

The {\bf normal} case deserves a special note. Since $N_\up$ and
$N_\dw$ are conserved independently, the local Hilbert
space and the electronic Fock space can be factorized as
$\HH_e=\HH_{e\up}\otimes\HH_{e\dw}$ and $\FF_\mathrm{e} = \FF_{e\up}\otimes \FF_{e\dw}$, respectively.  
Consequently, each electronic Fock state can be written as a product $\ket{\vec{n}_\up}\otimes \ket{\vec{n}_\dw}$. This factorization splits the
symmetry sector as $\SS_{\vec{Q}} = \SS_{N_\up}\otimes\SS_{N_\dw}$,
and finally the sector map can be expressed as the product $\MM = \MM_\up
\otimes \MM_\dw$.
The electronic part of each sector state $\ket{i}=\ket{i_\up} \otimes \ket{i_\dw}$ in this factorized basis
is labeled by two integers $[i_\up,i_\dw]$, with 
$i_\sigma=1,\dots,D_{\SS_\sigma}$ such that $i=i_\up + i_\dw
D_{\SS_\dw}$.
The maps $\MM_\sigma$ then connect these basis states to Fock states
$\ket{I_e}=\ket{I_\up}\ket{I_\dw}$, labeled by two integers
$[I_\up,I_\dw]$ as $I_e=I_\up +   I_\dw 2^{N_\mathrm{s}}$.
For a more detailed discussion on the structure of the Fock basis in this case, see Ref. \cite{Amaricci2022CPC}. 

The presence of a symmetry reduces the electronic Hamiltonian matrix to
a block-diagonal form, where each block labeled by $\vec{Q}$ has dimension
$D_{\SS_{\vec{Q}}}$. The sector Hamiltonian matrix $H_\SS$ is represented in the
basis $\ket{i}\in\SS_{\vec{Q}}$ as a sparse matrix.
In the {\bf normal} case this block structure is particularly
symmetric due to the factorized nature of the sectors as discussed
above, see also Ref. \cite{Amaricci2022CPC}.
The analysis of the spectrum is then reduced to inspecting the
Hamiltonian within each symmetry sector, or to a subset thereof if
additional constraints are present.


In \NAME, the implementation of symmetry sectors is managed through the
{\tt sector} object, which is defined in {\tt
  ED\_VARS\_GLOBAL}. This object contains all the relevant
information for defining the symmetry, including the sector
dimensions, quantum numbers and the map $\MM$. The constructor
(destructor) for this object is defined in the {\tt
  ED\_SECTORS} module via the function {\tt build\_sector} ({\tt
  delete\_sector}).
These functions use different algorithms depending on the nature 
of the quantum numbers $\vec{Q}$, i.e. the value of {\tt ed\_mode}. The key idea is to loop over the Fock states and enforce the quantum number constraint.
The following code snippets summarize the various available implementations and serve as a basis to understand other parts of the software, see \secu{CodeSparseMap}.  
\begin{center}
\begin{minipage}[t]{0.32\linewidth}
\textbf{Normal}
\begin{lstlisting}[style=fstyle,frame=leftline,numbers=none,basicstyle={\ttfamily\scriptsize}]
i=0
do Iup=0,2**Nbit-1     
  nup_ = popcnt(Iup) 
  if(nup_ /= Nups(1))cycle
  i = i+1
  H(iud)%map(i) = lup
enddo
i=0
do Idw=0,2**Nbit-1
  ndw_= popcnt(Idw)
  if(ndw_ /= Ndws(1))cycle
  i = i+1
  H(iud+Ns)%map(i)=Idw
enddo
\end{lstlisting}
\end{minipage}
\begin{minipage}[t]{0.32\linewidth}
\textbf{Superc}
\begin{lstlisting}[style=fstyle,frame=leftline,numbers=none,basicstyle={\ttfamily\scriptsize}]
i=0
do Idw=0,2**Ns-1
  ndw_= popcnt(idw)
  do Iup=0,2**Ns-1
    nup_ = popcnt(iup)
    sz_  = nup_ - ndw_
    if(sz_ /=self%Sz)cycle
    i=i+1
    self%H(1)%map(i)= &
        Iup+Idw*2**N
  enddo
enddo
$\phantom{.}$
$\phantom{.}$
\end{lstlisting}
\end{minipage}
\begin{minipage}[t]{0.32\linewidth}
\textbf{Nonsu2}
\begin{lstlisting}[style=fstyle,frame=leftline,numbers=none,basicstyle={\ttfamily\scriptsize}]
i=0
do Idw=0,2**Ns-1
  ndw_= popcnt(Idw)
  do Iup=0,2**Ns-1
    nup_ = popcnt(Iup)
    nt_  = nup_ + ndw_
    if(nt_/=self%Ntot)cycle
    i=i+1
    self%H(1)%map(i)= &
        Iup+Idw*2**Ns
  enddo
enddo
$\phantom{.}$
$\phantom{.}$
\end{lstlisting}
\end{minipage}
\end{center}

In addition to the basic construction and destruction routines, the
{\tt ED\_SECTORS} module also includes functions to
retrieve sector indices and quantum number
information. Furthermore, it provides a set of
essential functions for applying creation ({\tt apply\_op\_CDG}), destruction ({\tt apply\_op\_C}) or arbitrary linear combinations
({\tt apply\_Cops}) of these to a given Fock state $\ket{v}\in\SS$, i.e.:
$$
\OO\ket{v} = \sum_i \left( a_i C_{\a_i,\s_i}+ b_i C^{\dagger}_{\b_i,\r_i}\right)\ket{v}
$$ 
with $a_i, b_i \in\CCC$, $\a_i,\b_i=1,\dots,N_\a$ and $\s_i,\r_i=\up,\dw$. These functions are widely used throughout the code.

\subsection{Interaction setup}\label{sSecIntSetup}
\NAME offers two distinct methods to setup the interaction terms $\hat{H}^\mathrm{int}_\mathrm{imp}$ in \equ{Himp}, which are controlled by the input variables {\tt
  ED\_USE\_KANAMORI} and {\tt ED\_READ\_UMATRIX}. 
  
The generalized Hubbard-Kanamori interaction, as defined in \equ{Hint}, is natively
supported and controlled by the input parameters {\tt ULOC}$=U$, {\tt
  UST}$=U'$, {\tt JH}$=J$, {\tt JX}$=J_{X}$ and {\tt JP}$=J_{P}$.
These quantities can be specified either in the input file or directly
via the command line, provided the logical input parameter {\tt ED\_USE\_KANAMORI=T} and setting {\tt ED\_READ\_UMATRIX=F}. In this mode, the maximum number of impurity orbitals {\tt NORB}$=N_\a$ is limited to $5$.

Alternatively, setting {\tt ED\_USE\_KANAMORI=F} and {\tt ED\_READ\_UMATRIX=T}, a list of two-body interaction operators can be supplied by the user in a properly
formatted text file, whose name is specified by the input variable {\tt UMATRIX\_FILE}.  
The file must have the suffix {\tt <UMATRIX\_FILE>.re\-star\-t} and should be formatted as follows: 
\begin{lstlisting}[style=mybash,numbers=none]
<NORB> BANDS
i1 j1 k1 l1 U_i1j1k1l1
i2 j2 k2 l2 U_i2j2k2l2
...
\end{lstlisting}
where {\tt NORB} is the number of orbitals $N_\a$ and $\{ijkl\}$ refer to the combined spin ({\tt u/d}) and internal (e.g. orbital index) degrees of freedom. For example, if {\tt i1} refers to orbital 1, spin up, it is written explicitly as {\tt 1 u}.
Each line defines a two-body second-quantized operator of the form 
$$
\hat{H}^{\mathrm{int}}=\frac{1}{2}d^{\dagger}_{i}d^{\dagger}_{j}U_{ijkl}d_{l}d_{k}.
$$

Note that the coefficient indices $l$ and $k$ are swapped in the input file with respect to the two-body operator they implement. The $\tfrac{1}{2}$ prefactor is applied internally by the parsing routine and need not be accounted for by the user.
Empty lines and lines starting with {\tt \#, \%, !} are ignored.

While {\tt ED\_READ\_UMATRIX} and {\tt ED\_USE\_KANAMORI}
are mutually exclusive, it is allowed to set both to {\tt F}. 
In this case, the user can specify additional interaction terms in the calling program or script by means of the procedure {\tt add\_twobody\_operator}, before the impurity problem is solved. The input of this function has the same form of each line of the {\tt UMATRIX} file. 
A list of user-provided interaction terms will be added to those read from file or from the Hubbard-Kanamori coefficients in case the relative flags are {\tt T}. To clear the list, the procedure {\tt reset\_umatrix} has to be called.

For reference and future use, when the impurity problem is solved, a list of the interaction operators used in each call to the solver is saved in the output file 
{\tt <UMATRIX\_FILE>.used}.
The reading, writing, parsing and initialization of the two-body interaction terms are implemented in the module {\tt ED\_PARSE\_UMATRIX}.

\subsection{Classes}
The use of objects and classes greatly simplifies the implementation of critical mathematical concepts required for solving quantum 
impurity problems. This section provides an overview of the main 
classes used in \NAME, focusing on their structure and functionality.

\subsubsection{Sparse matrix}\label{CodeSparseMatrix}
Sparse matrix storage is handled through a dedicated class in the 
{\tt SPARSE\_MATRIX} module. This module defines the 
{\tt sparse\_matrix\_csr} object, which stores sparse matrices as 
hash tables. Each key corresponds to a row index, while the associated 
value contains a pair of dynamic arrays: one for the non-zero matrix 
elements and one for their respective column indices. 
In an MPI framework, each matrix instance on any given thread stores only a given set $Q=D_\SS/N_{cpu}$ of matrix rows. In addition, the elements local in the memory are stored in a separate set of rows, i.e. {\tt loc}, and used to speed up the execution of matrix-vector product. 
We keep track, for each thread, of the relative position of the matrix rows $q=1,\dots,Q$ with respect to the global indexing $i=1,\dots,D_\SS$, through the definition of additional indices {\tt istart, iend, ishift}. These correspond, respectively, to the initial global index $i_1$ for $q=1$, the final global index $i_Q$ for $q=Q$ and the integer shift mapping the index $q$ into $i$.    
Each instance of {\tt sparse\_matrix\_csr} can be stored serially (one copy per process) or distributed across multiple 
processes with rows dynamically assigned to each process. 

\begin{lstlisting}[style=fstyle,numbers=none]
! sparse row: contains the NNZ elements on any row of the sparse matrix.
type sparse_row_csr
   integer                                   :: size  !size of the row = NNZ
   real(8),dimension(:),allocatable          :: dvals !rank-1 array for double precision
   complex(8),dimension(:),allocatable       :: cvals !rank-1 array for double complex
   integer,dimension(:),allocatable          :: cols  !rank-1 array for column indices
end type sparse_row_csr
! Sparse Matrix: allocatable array of sparse rows
type sparse_matrix_csr
   type(sparse_row_csr),dimension(:),pointer :: row  !array of {\tt sparse\_row\_csr} 
   integer                                   :: Nrow !total number of rows
   integer                                   :: Ncol !total number of columns
   logical                                   :: status=.false. !Allocation status
#ifdef _MPI
   type(sparse_row_csr),dimension(:),pointer :: loc      !array for the diagonal blocks
   integer                                   :: istart=0 !start index MPI range
   integer                                   :: iend=0   !end index MPI range
   integer                                   :: ishift=0 !shift index MPI range
   logical                                   :: mpi=.false.
#endif
end type sparse_matrix_csr
\end{lstlisting}

Matrix elements are inserted using the 
{\tt sp\_insert\_element} procedure, which leverages the Fortran 
intrinsic {\tt move\_alloc} for faster execution compared to 
implicit reallocation, i.e.~{\tt vec=[vec,new\_element]}. 
This approach offers several key advantages, including efficient 
memory management for matrices with unknown numbers of non-zero 
elements per row, and $O(1)$ element access time, both of which 
are critical for the efficient implementation of Krylov subspace 
methods.

\subsubsection{Sparse map}\label{CodeSparseMap}
As will be discussed in \secu{sSecRDM}, the construction of a symmetry 
sector often requires associating each sector state $\ket{i}$ with 
information about the corresponding electronic Fock state $\ket{I_e}$. 
In this context, the electronic Fock state can be decomposed into bit chunks, 
e.g. $\ket{\vec{n}} = \ket{\vec{i}_\up\,\vec{b}_\up\,
\vec{i}_\dw\,\vec{b}_\dw}$, reflecting a natural grouping into impurity 
and bath components and where we highlight the overall ordering in
spin-$\up$ and spin-$\dw$ parts (see \equ{eq:FockState_|nupndw>}).

The module {\tt ED\_SPARSE\_MAP} provides an efficient hash table 
implementation, {\tt sparse\_map} which is part of the {\tt
  sector} object and stores the relation between these bit chunks. 
Specifically, for each impurity configuration 
$\vec{i}_\s$ ({\it key}), it maintains a list of compatible bath 
configurations $\vec{b}_\s$ ({\it values}) consistent with the conserved 
quantum numbers of the given sector (see \secu{sSecQNs}).
The {\tt sparse\_map} objects ${\cal P}$ are constructed upon call of the {\tt
  build\_sector} (i.e. the {\tt sector} constructor) using different algorithms 
for each value of {\tt ed\_mode}.

\begin{lstlisting}[style=fstyle,numbers=none]
!Sparse row: contains bath state values for each impurity key 
type sparse_row
   integer                               :: size          !current size
   integer                               :: bath_state_min!smallest bath state
   integer                               :: bath_state_max!largest bath state
   integer,dimension(:),allocatable      :: bath_state    !Values:rank-1 bath states
   integer,dimension(:),allocatable      :: sector_indx   !bath state sector indices
end type sparse_row
!Sparse map: array of sparse rows 
type sparse_map
   type(sparse_row),dimension(:),pointer :: imp_state     !key: impurity states
   integer                               :: Nimp_state    !key upper limit 
   logical                               :: status=.false.!allocation status
end type sparse_map
\end{lstlisting}

In the {\bf normal} mode, the {\tt sector} object contains two separate {\tt
  sparse\_maps} objects ${\cal P}_{\s=\up,\dw}$, reflecting the
factorization of the Fock and sector states into independent spin components: 
$\ket{J_e}=\ket{J_\up}\otimes\ket{J_\dw} \xleftarrow{\MM}
\ket{j_\up}\otimes\ket{j_\dw}=\ket{j}$.
The sparse maps are built as follows. 
For any spin Fock state appearing in the sector $\ket{j}\xrightarrow{\MM}\ket{J_\s}=\ket{{\vec{I}_{\s}\vec{B}_{\s}}}$,
the {\it key} is determined by the integer $I_\s$ corresponding to the
impurity bitset $\vec{I}_\s$. The {\it values} are given by all 
integers $B_\s$ corresponding to any bath bit set
$\vec{B}_\s$ associated with $\vec{I}_\s$ by the sector symmetry constraint: $\#\vec{I}_\s + \#\vec{B}_\s = \mathtt{popcnt}(I_\s) + \mathtt{popcnt}(B_\s)= N_\s$.
Thus, any given {\it key-value} combination 
reconstructs an integer $J_\s$ representing a Fock state in the
given spin sector according to the rule $J_s = I_\s +
2^{N_\mathrm{imp}}B_\s$, where $N_\mathrm{imp}$ is the number of impurity bits. 

In both {\bf superc} and {\bf nonsu2} modes a single sparse map is
used.
Here, each electronic Fock state $\ket{J_e}$ is associated with four integers $I_\up, B_\up, I_\dw,
B_\dw$ representing the bit decomposition of the state. 
These integers satisfy the relationship
\begin{equation}\label{spMapEq1}
J_e =I_\up +  B_\up2^{N_\mathrm{imp}} + (I_\dw +  B_\dw2^{N_\mathrm{imp}})2^{N_\mathrm{s}}
\end{equation}
where $N_\mathrm{imp}$ is the number of impurity bits and 
$N_\mathrm{s}$ is the total number of spin orbitals.
We define the \emph{key} for the sparse
map as $I=I_\up + I_\dw2^{N_\mathrm{imp}}$ and the corresponding
\emph{values} as the integers $B=B_\up + B_\dw2^{N_\mathrm{b}}$.
Note that, for the purpose of storing the key-value pairs, the impurity and bath indices are determined using an algorithm different from \equ{spMapEq1}.   
These relations ensure a better scaling and can be easily inverted to obtain $I_{\sigma=\up,\dw}$ from
$I$ and $B_{\sigma=\up,\dw}$ from $B$ in order to reconstruct the Fock
state $\ket{J}$.
The four integers satisfy the symmetry constraints:   
$\mathtt{popcnt}(I_\up + B_\up2^{N_\mathrm{imp}}) + \mathtt{popcnt}(I_\dw + B_\dw2^{N_\mathrm{imp}}) = N$
for the \textbf{nonsu2} case and
$\mathtt{popcnt}(I_\up + B_\up2^{N_\mathrm{imp}}) - \mathtt{popcnt}(I_\dw + B_\dw2^{N_\mathrm{imp}}) = S_z$
for the \textbf{superc} case.
This structure ensures an efficient and compact representation 
of the many-body Hilbert space, enabling a rapid state lookup 
during the diagonalization process.

An overview of the implementation leveraging the Fortran intrinsic {\tt ibits(I,pos,len)} is reported in the following listing:
\begin{center}
\begin{minipage}[t]{0.49\linewidth}
\textbf{Normal}
\begin{lstlisting}[style=fstyle,frame=leftline,numbers=none,basicstyle={\ttfamily\scriptsize}]
i=0
do Iup=0,2**Nbit-1     
  nup_ = popcnt(Iup) 
  if(nup_ /= Nups(1))cycle
  i=i+1 
  !...
  iImp  = ibits(iup,0,Norb)
  iBath = ibits(iup,Norb,Norb*Nbath)
  call sp_insert_state(self%H(1)%sp,
                       iImp,&
                       iBath,&
                       i)
enddo
!
i=0
do Idw=0,2**Nbit-1
  ndw_= popcnt(Idw)
  if(ndw_ /= Ndws(1))cycle
  i=i+1 
  !...
  iIMP  = ibits(idw,0,Norb)
  iBATH = ibits(idw,Norb,Norb*Nbath)
  call sp_insert_state(self%H(2)%sp,&
                       iImp,&
                       iBath,&
                       i) 
enddo
\end{lstlisting}
\end{minipage}
\begin{minipage}[t]{0.49\linewidth}
\textbf{Superc/Nonsu2}
\begin{lstlisting}[style=fstyle,frame=leftline,numbers=none,basicstyle={\ttfamily\scriptsize}]
i=0
do Idw=0,2**Ns-1
  ndw_= popcnt(idw)
  do Iup=0,2**Ns-1
    nup_ = popcnt(iup)
    select case (ed_mode)
      case('superc') 
        sz_  = nup_ - ndw_
        if(sz_ /= self%Sz)cycle
      case('nonsu2') 
        nt_  = nup_ + ndw_
        if(nt_ /= self%Ntot)cycle 
    end select
    i=i+1 
    !...
    iImpUp = ibits(iup,0,Norb)
    iImpDw = ibits(idw,0,Norb)
    iBathUp= ibits(iup,Norb,Norb*Nbath)
    iBathDw= ibits(idw,Norb,Norb*Nbath)
    iImp   = iImpUp  + iImpDw*(2**Norb)
    iBath  = iBathUp + iBathDw*(2**Ns)
    call sp_insert_state(self%H(1)%sp,&
                         iImp,&
                         iBath,&
                         dim)
  enddo
enddo
\end{lstlisting}
\end{minipage}
\end{center}

\subsubsection{Eigenspace}\label{CodeEigenspace}
The module {\tt ED\_EIGENSPACE} contains the implementation of the 
class {\tt sparse\_espace}, which maintains a linked list 
of selected eigenvalues, eigenvectors, and quantum numbers for the 
low-lying spectrum of the quantum impurity Hamiltonian. 
In MPI mode, the memory footprint of each eigenvector 
is minimized by automatically distributing it among all processors in suitable shares
according to the value of {\tt ed\_mode}. 
While for zero-temperature calculations only the ground state (with its
degeneracy) is stored, for finite temperature the list includes all the necessary excited states.
If the input variable {\tt ED\_TWIN=T}, the {\it twin sectors}, i.e. sectors with symmetric values of the quantum numbers ($S_z\to -S_z$ and $N\to N_s-N$), are excluded from the list and their contribution is reconstructed by symmetry where needed. 
A truncation mechanism is put in place to avoid the unbounded growth of this list. On the first call, a fixed number of
states per sector  
\texttt{lanc\_nstates\_sector} is collected, up to a
maximum total \texttt{lanc\_nstates\_total}, both set via input.  
The list is then truncated to retain only the states satisfying the condition 
$$
e^{-\beta(E_i-E_0)} < \mathtt{cutoff}, 
$$
where
$E_i$ is the energy of the $i^{\rm th}$ state in the list, $E_0$ is the
ground state energy, $\beta=1/T$ is the inverse temperature ($k_B=1$) and \texttt{cutoff} is an input parameter fixing an {\it a priori} energy threshold.
In the following we indicate with $N_{\tt state\_list}$ the number of eigen-solutions stored in the list. 
By successive call to the \NAME solver, the list dynamically adjusts to optimize the distribution of states 
across sectors, balancing memory use and computation time. The
variable \texttt{lanc\_nstates\_step} controls the increase or
decrease in the number of states per sector. 
A histogram  of the number of states per sector is produced after each diagonalization to monitor this distribution.

\subsubsection{gfmatrix}
One of the main goals of the \NAME is to evaluate dynamical correlations
functions (DCF), e.g. $C(t) = \ibra \TT[ A(t) A^\dagger ]\iket$. Using Krylov methods,
these DCFs can be expressed as truncated K\"all\'en-Lehmann spectral sums of the form
$$
C(z) = \tfrac{1}{Z}\sum_{n=1}^{N_{\tt state\_list}} e^{-\beta E_n} \sum_{m=1}^{N} \tfrac{|w_{mn}(\AA)|^2}{ z
  - dE_{mn}}
$$
where $w_{mn}(\AA)$ is the weight determined by the amplitude of operator $\AA$ between the
$m^{\rm th}$ eigenstate and the $n^{\rm th}$ Krylov basis vector, 
$dE_{mn}=E_m-E_n$ is the corresponding excitation energy.

The module {\tt ED\_GFMATRIX} implements the class {\tt gfmatrix}, which provides an efficient and flexible representation of DCFs. It is structured as a hierarchical, or {\it multi-layered}, data container that organizes spectral weights and poles across three levels: (i) the initial eigenstate $\ket{n}$ of the Hamiltonian spectrum, (ii) the amplitude channel associated with this state, corresponding to the applied operator, and (iii) the set of excitations from $\ket{n}$, each characterized by a pole and its corresponding weight.
This structure allows the {\tt gfmatrix} class to simultaneously handle multiple initial states and operators, enabling on-the-fly evaluation of DCFs at arbitrary complex frequencies $z \in \mathbb{C}$. By avoiding precomputed grids and redundant storage this approach minimizes memory footprint and computational overhead. Through the use of ${\tt gfmatrix}$ EDIpack achieves efficient memory management, consistent storage of spectral data and high-performance post-processing.

\subsection{Bath parametrization}\label{sSecBath}
The quantum impurity problem is characterized by the coupling between
a local impurity and a surrounding bath. Following the structure of
\equ{Hbath}, the bath is parametrized by two components: the
Hamiltonian matrices, $h_{\a\b\s\s'}^p$, and the hybridization amplitudes,
$V_{\a\b\s\s'}^p$, for $p = 1, \dots, N_\mathrm{bath}$.
Internally, the bath is represented by a dedicated object, {\tt
  effective\_bath}, defined in the {\tt ED\_VARS\_GLOBAL} module. From
the user's perspective, all parameters are consolidated into a rank-1
array of doubles, managed using a reverse communication strategy. 

The bath topology, which defines the connectivity between the
electronic levels assigned to the bath, is specified
through the input variable {\tt bath\_type}. 
In \NAME this variable can take one of four possible values: 
{\bf normal}, {\bf hybrid}, {\bf replica}, and {\bf general}.

\paragraph{{\tt bath\_type}={\bf normal}.} In this configuration, the
bath consists of $N_\mathrm{bath}$
electronic sites coupled to each impurity orbital, resulting
in $N_\mathrm{b} = N_\a N_\mathrm{bath}$ bath levels and a total size of $N_\mathrm{s}=(N_\mathrm{bath}+1)N_\a$.
For {\tt ed\_mode}={\bf normal} the bath Hamiltonian is parametrized as diagonal matrices in both the orbital and spin spaces:
$$
h^p_{\a\b\s\s'}=\varepsilon_{\a\s}\delta_{\a\b}\delta_{\s\s'},
$$
where $\varepsilon_{\a\s}$ represents  the on-site energies for each orbital and spin. 
If {\tt ed\_mode}={\bf superc},  anomalous
amplitudes connecting bath levels with opposite spins must also be
considered. In this case, the bath Hamiltonian gains 
an additional set of parameters $\Delta^p_{\a}\delta_{\a\b}\delta_{\s \bar{\s}}$, where $\Delta^p_\a$ represents the orbital diagonal pairing amplitudes. 
This choice corresponds to a Nambu representation for each component, with the Hamiltonian matrices taking the form:

$$
h_{\a\b\s\s'}^p = \e_\a^p\delta_{\a\b}\tau^z_{\s\s'} + \Delta^p_\a\delta_{\a\b}\tau^x_{\s\s'} .
$$
where $\tau^{i=0,x,y,z}$ are Pauli matrices.
The hybridization amplitudes between the impurity and the bath levels
are similarly structured. For all values of {\tt ed\_mode} these
parameters are diagonal in both
spin and orbital space: $V^p_{\a\b\s\s'}=V^p_{\a\s}\delta_{\a\b}\delta_{\s\s'}$.
If {\tt ed\_mode}={\bf nonsu2}  an additional set including terms describing spin-flip processes (as total magnetization is not conserved) should be included, $V^p_{\a\b\s\s'} =
W^p_{\a}\delta_{\a\b}\delta_{\s \bar{\s}}$. 

The resulting total number of parameters used to determine the bath is $N_\mathrm{param}=2N_\a N_\mathrm{bath}$ for {\tt ed\_mode}={\bf normal} or $N_\mathrm{param}=3N_\a N_\mathrm{bath}$ for the other two values of {\tt ed\_mode}.

\paragraph{{\tt bath\_type}={\bf hybrid}.}
The hybrid topology is the simplest generalization of the bath pa\-ra\-me\-tri\-za\-tion which captures the effects of locally
coupled impurity orbitals. However, this setup typically requires a more challenging optimization process, especially
with few available bath levels (see \secu{sSecFit}). 
The bath is formed by $N_\mathrm{bath}$ sites coupled to all the
impurity orbitals, so that there are $N_\mathrm{b}\equiv  N_\mathrm{bath}$ bath levels and   $N_\mathrm{s}=N_\mathrm{bath}+N_\a$. 
The parametrization of the bath Hamiltonian is 
diagonal in both orbital and spin spaces, i.e.
$$
h^p_{\a\b\s\s'}=\varepsilon_{\a\s}\delta_{\a\b}\delta_{\s\s'}.
$$
For {\tt ed\_mode}={\bf superc} additional anomalous components 
$\Delta_\a^p\delta_{\a\b}\delta_{\s\bar{\s}}$ diagonal in the orbital basis are included.

The key difference compared to the previous case is that the hybridization matrix elements are generally off-diagonal in the orbital indices:
$$
V^p_{\a\b\s\s'}=V^p_{\a\b\s}\delta_{\s\s'}.
$$
If full spin symmetry is not preserved, i.e. for {\tt ed\_mode} = {\bf nonsu2}, a further 
set of parameters must be specified to account for spin-flip processes:
$V^p_{\a\b\s\s'} = W^p_{\a\b}\delta_{\s \bar{\s}}$. 

The resulting total numbers of available bath parameters are 
 $N_\mathrm{param}=N_\mathrm{bath}+ N_\a N_\mathrm{bath}$ for {\tt ed\_mode}={\bf normal},  
 $N_\mathrm{param}=2N_\mathrm{bath}+ N_\a N_\mathrm{bath}$ if  {\tt ed\_mode}={\bf superc} or 
$N_\mathrm{param}=N_\mathrm{bath}+ 2N_\a N_\mathrm{bath}$ for  {\tt ed\_mode}={\bf nonsu2}.

\paragraph{{\tt bath\_type}={\bf replica}, {\bf general}.}
This configuration provides a more flexible parametrization of the 
bath. The original core idea \cite{Capone2004PRB,Civelli2006,Koch2008PRB}
is to assign each bath element a structure 
that {\it replicates} the internal structure of the impurity, while 
maintaining diagonal coupling between bath elements and the impurity. 
This offloads the complexity of representing structured quantum 
impurities to the bath Hamiltonian rather than the hybridizations.

From a broader perspective, this can be formalized by considering  
a user-defined matrix basis of dimension $N_\mathrm{sym}$, 
$\vec{\Gamma}=\{ \Gamma^\nu_{\a\b\s\s'} \}_{\nu=1,\dots,N_\mathrm{sym}}$ in the (Nambu-)spin-orbital
space, and expressing the bath Hamiltonian as
$$
h^p = \sum_{\nu=1}^{N_\mathrm{sym}} \lambda^p_\nu
\Gamma^\nu\equiv \vec{\lambda}^p\cdot \vec{\Gamma},
\label{eq:replica_bath_sym}
$$
where $\vec{\lambda}^p\in\RRR^{N_\mathrm{sym}}$ is a vector of variational parameters. 
The choice of the matrix basis can be inspired by symmetry group considerations \cite{Koch2008PRB,Senechal2010PRB}, by the internal structure
of the quantum impurity or determined
case-by-case depending on the desired properties of the solution.
In this scheme, the total number of bath levels is  $N_\mathrm{b}
= N_\a N_\mathrm{bath}$.

For the {\bf replica} topology, the coupling between the impurity 
and each bath element is independent on spin, orbital, and internal 
bath structure: $V^p_{\a\b\s\s'} = V^p$.
In the {\bf general} case, the coupling includes explicit 
diagonal
dependence on the internal spin and orbital indices: $V^p_{\a\b\s\s'}
= V^p_{\a\s}\delta_{\a\b}\delta_{\s\s'}$. 
The number of independent bath parameters to optimize is either 
$N_\mathrm{param} = (N_\mathrm{sym}+1)N_\mathrm{bath}$ for {\tt bath\_type}={\bf replica} or 
$N_\mathrm{param} = (N_\mathrm{sym}+N_\a N_\s)N_\mathrm{bath}$ for {\tt bath\_type}={\bf general}.

We summarize the total number $N_s$ of electronic levels of the
quantum impurity system according to the value of $N_\a$,
$N_\mathrm{bath}$ and the bath topology in Table~\ref{Table3}:

\begin{table}
\centering
\begin{tabular}{ |c||c|c|c|c| } 
 \hline     
{\tt bath\_type} & {\bf normal} & {\bf hybrid} & {\bf replica} & {\bf general}\\
 \hline 
  \hline    
  $N_s$ & 
 $(N_\mathrm{bath}+1)N_\a$ & 
 $N_\mathrm{bath}+N_\a$ & 
 $N_\a(N_\mathrm{bath}+1)$ & 
 $N_\a(N_\mathrm{bath}+1)$  \\ 
 \hline
\end{tabular}
\caption{
  {\bf Total number of electronic levels}.
  The table summarizes the possible number of the electronic levels $N_s$ (second row) according to the value of the {\tt bath\_type}
(first row), the number of impurity orbitals $N_\alpha$ and the number
of the bath sites $N_{\rm bath}$.}
\label{Table3}
\end{table}

All the procedures concerning the bath
are grouped into a set of modules wrapped by {\tt ED\_BATH}. 
We divided the set of modules into three categories, according to their scope.

\paragraph{Bath Auxiliary Tools.}
This group of modules provides user-oriented utilities, including 
standard symmetry operations on the bath array, e.g. orbital 
symmetry and particle-hole symmetry, available in 
{\tt ED\_BATH\_USER}. 

The critical functionality here is the 
determination of the total bath array dimension, implemented in 
{\tt ED\_BATH\_DIM}. The function {\tt ed\_get\_bath\_dimension} 
calculates the required dimension $N_\mathrm{param}$ for user-allocated bath 
arrays, ensuring they contain exactly the necessary number of 
parameters. 

Table~\ref{Table4} summarizes the total number of parameters $N_\mathrm{param}$, as returned by the function {\tt ed\_get\_bath\_dimension} and required to describe the bath according to the values of {\tt ed\_mode} and {\tt bath\_type}:

\begin{table}
\centering
\begin{tabular}{ |c||c|c|c|c| } 
 \hline     
 $N_\mathrm{param}$ & {\bf normal} & {\bf hybrid} & {\bf replica} & {\bf general}\\
 \hline 
 \hline    
 {\bf normal} & 
 $2N_\a N_\mathrm{bath}$ & 
 $(N_\a+1)N_\mathrm{bath}$ & 
 $(N_\mathrm{sym}+1)N_\mathrm{bath}$ & 
 $(N_\mathrm{sym}+N_\a N_\sigma)N_\mathrm{bath}$\\
 \hline
   {\bf superc} & $3N_\a N_\mathrm{bath}$ & 
 $(N_\a+2)N_\mathrm{bath}$ & 
 $(N_\mathrm{sym}+1)N_\mathrm{bath}$ & 
 $(N_\mathrm{sym}+N_\a N_\sigma)N_\mathrm{bath}$\\
 \hline
   {\bf nonsu2} & $3N_\a N_\mathrm{bath}$ & 
 $(2N_\a+1)N_\mathrm{bath}$ & 
 $(N_\mathrm{sym}+1)N_\mathrm{bath}$ & 
 $(N_\mathrm{sym}+N_\a N_\sigma)N_\mathrm{bath}$\\
 \hline
\end{tabular}
\caption{
  {\bf Total number of bath parameters}. The table lists the number of
  bath parameters $N_{\rm param}$ according to the conserved
  symmetry {\tt ed\_mode} (first column), and the
  bath topology {\tt bath\_type} (first
  row). The number of parameters is obtained as a function of the
  impurity orbitals $N_\alpha$, the number of spin channels
  $N_\sigma$, the number of bath sites $N_{\rm bath}$ and, where
  applicable, the size of the matrix basis used to describe the {\tt
    replica/general} bath structure $N_{\rm sym}$.}
\label{Table4}
\end{table}

\paragraph{Bath replica/general.}
The module {\tt ED\_BATH\_REPLICA} defines the class {\tt Hreplica}/{\tt Hge\-ne\-ral} for the {\bf replica}/{\bf general} bath parametrization. In  addition to standard object construction, destruction, reading and saving, this module provides the setup for either the matrix basis  
$\{ \Gamma^\nu \}_{\nu=1,\dots,N_\mathrm{sym}}$ and the initial variational parameters $\vec{\lambda}$ through the function  {\tt set\_Hreplica}/{\tt set\_Hge\-ne\-ral}. 
The bath Hamiltonian is constructed with the function 
{\tt build\_Hreplica}/{\tt build\_Hge\-ne\-ral} via the relation
$
h^p = \vec{\lambda}^p \cdot \vec{\Gamma}
$.

\paragraph{Effective Bath.}
The module {\tt ED\_BATH\_DMFT} implements the class 
{\tt effective\_bath} (defined in {\tt ED\_VARS\_GLOBAL}), which 
manages all the actual bath parameters for any choice of 
{\tt ed\_mode} and {\tt bath\_type}. For the 
{\bf replica/general} topologies, it directly references the 
{\tt Hreplica} class. A global instance of this class, 
{\tt dmft\_bath}, is shared throughout the code, providing a 
centralized structure for bath parameter management.

The constructor for this class includes the function 
{\tt init\_dmft\_bath}, which either initializes bath parameters 
from scratch or reads them from a file specified by the input 
variable {\tt Hfile} as {\tt <Hfile>.restart}. Additionally, the class offers methods to 
translate bath parameters between the internal {\tt dmft\_bath} 
representation and the user-defined bath arrays via 
{\tt get/set\_dmft\_bath}.

\subsection{Lanczos based Diagonalization}\label{sSecHam}

In presence of symmetries (see \secu{sSecQNs}) the matrix representing
the Hamiltonian operator $\hat{H}$ can be reduced to a block-diagonal form. Each block corresponds to a symmetry sector
with a fixed set of quantum numbers $\vec{Q}$ and is represented by
a Hamiltonian matrix $H_{\SS_{\vec{Q}}}$. 
Consequently, the analysis of the energy spectrum of the quantum impurity reduces to solving
the secular equation within each individual symmetry sector:
$$
H_{\SS_{\vec{Q}}}\ket{v}=E_{\SS_{\vec{Q}}}\ket{v}.
$$

Even though the dimension of a given sector $\SS_{\vec{Q}}$ is much smaller than the
total Fock space dimension $D_\SS \ll D_\FF$, solving the
corresponding eigenvalue problem completely remains a formidable challenge
already for $N_\mathrm{s}\simeq 8$.
To outwit this, various
algorithms exploiting the sparse nature of
$H_\SS$ have been developed, including Krylov subspace methods and related approaches
\cite{Lanczos1950JRNBSB,Lin1993CIP,Lehoucq1998,Maschhoff1996,Siro2012CPC,Siro2016CPC}.

In \NAME, the preferred algorithm is P-ARPACK, a state of the art
Lanczos-based eigensolver with distributed memory support
\cite{Lehoucq1998} ensuring the accurate convergence of the desired eigenpairs. 
However, it is also possible to select a simpler,
though less efficient, parallel Lanczos algorithm using the input variable {\tt lanc\_method}.

The core computational task in any Krylov-based diagonalization
algorithm is the Matrix-Vector Product (MVP), i.e.~the application
of the Hamiltonian to a given input vector:
\begin{equation}
H_\SS \ket{v} \to \ket{w}\qquad \ket{v},\ket{w}\in \SS_{\vec{Q}}.
\end{equation}
This operation amounts to more than 80\% of the overall
computational effort, making its optimization a critical aspect of
high-performance diagonalization algorithms. In Ref. \cite{Amaricci2022CPC} we discussed in greater details the massively parallel strategies  employed in \NAME to optimize the MVP operations and presented various benchmarks of the parallel scaling. 
Here, we provide a brief outline of the main ideas following the notation introduced
in Ref.~\onlinecite{Amaricci2022CPC}, and focusing on the impact of the choice of {\tt ed\_mode}. 
For the sake of clarity, here we focus on the purely electronic case, i.e. neglecting the electron-phonon part. The product structure of the electron and phonon Fock spaces ensure that the following approach applies with few algorithmic modifications in presence of electron-phonon coupling.

In the {\bf normal} case, the tensor structure of the Fock space and the sector
symmetries are fully leveraged. Within a given sector, the electronic
part of the Hamiltonian can be expressed as:
\begin{equation}
H^e_{\SS} = H_\mathrm{d} + H_\up\otimes \11_\dw + \11_\up\otimes
H_\dw + H_\mathrm{nd},
\label{HssNormal}
\end{equation}
where $H_\mathrm{d}$ represents the diagonal (local) part of the
Hamiltonian, including density-density interactions, while
$H_\sigma$ captures hopping processes for electrons with spin
$\sigma=\up,\dw$. The term $H_\mathrm{nd}$ encompasses all remaining
non-diagonal contributions, such as spin-exchange and pair-hopping
terms.
In this formulation, a sector state $\ket{v}$ can be
represented in a matrix basis $\hat{v}$ with rows (columns)
corresponding to $\up$ ($\dw$) configurations. This structure allows
for efficient MVP operations using {\tt MPI\_All2AllV}, which aims to
optimize memory locality. However, terms in $H_\mathrm{nd}$ that
break this structure require {\tt MPI\_AllGatherV}, introducing a
slight communication overhead which partially reduces the parallel efficiency.

For the {\bf superc/nonsu2} modes, symmetries enforce specific constraints between
$\up$ and $\dw$ configurations, preventing the direct use of the
factorized Fock space representation. The electronic
part of the Hamiltonian is typically structured as:
\begin{equation}
H^e_\SS = H_\mathrm{imp} + H_\mathrm{int} + H_\mathrm{bath} + H_\mathrm{imp-bath},
\end{equation}
where $H_\mathrm{imp}$ encodes the local impurity Hamiltonian
determined by $h^0_{\a\b\s\s'}$,
$H_\mathrm{int}$ contains the interaction terms, $H_\mathrm{bath}$
describes the effective bath contributions corresponding to
$h^p_{\a\b\s\s'}$, and $H_\mathrm{imp-bath}$ accounts for the
impurity-bath couplings $V^p_{\a\b\s\s'}$.
In an MPI setup, the first three terms contains also elements local to each node and
stored separately within each {\tt sparse\_matrix} instance representing $H_\SS$ (see \secu{CodeSparseMatrix}).
The MVP is carried out using the {\tt MPI\_AllGatherV} algorithm which, as noted  
Ref.~\onlinecite{Amaricci2022CPC}, requires reconstruction of distributed
arrays and incurs significant communication overhead.
\vspace{2mm}

\noindent
The diagonalization process itself is divided into two primary phases:
\paragraph{\bf Global Setup.} This phase allocates the required memory, initializes
the MPI environment, and configures the appropriate MVP procedure
based on the symmetries of the problem. This is implemented in
a set of independent modules, {\tt ED\_HAMILTONIAN\_<ed\_mode>}. Any module includes two functions: {\tt build\_Hv\_sector\_<ed\_mode>} and  {\tt vecDim\-\_\-Hv\-\_\-sector\-\_\-<ed\_mode>}.

The {\tt build\_Hv\_sector\_<ed\_mode>} function builds the symmetry sector and
allocates the MVP function by setting the shared abstract function pointer {\tt
  spHtimesV\_p} to the correct MVP function. 
  This latter step is controlled by the value of
{\tt ed\_sparse\_H=T/F}. If {\tt T}, the Hamiltonian $H_\SS$
gets stored in a {\tt sparse\_matrix} instance and used to perform
the MVPs. The corresponding algorithms are implemented in the modules {\tt
  ED\_HAMILTONIAN\-\_\-<ed\_mode>\-\_\-STORED\-\_HxV}.
If {\tt ed\_sparse\_H} is {\tt F}, the MVP is operated on-the-fly,
i.e. each element of $H_\SS$ is directly applied to the input
vector $\ket{v}$, either in serial or parallel mode as implemented in {\tt
  ED\_HAMILTONIAN\_<ed\_mode>\_DIRECT\_HxV}.  

The function {\tt vecDim\_Hv\_sector\_<ed\_mode>} returns the 
dimension of the vector used in the
MVP. In an MPI-parallel execution the returned value is the
dimension $d_i$ of the vector chunk per each node such that
$\sum_{i=1}^{N_\mathrm{nodes}}d_i = D_\SS$. Yet, the specific value of $d_i$
depends on the MPI algorithm used for the MVP function.

\paragraph{\bf Diagonalization.}
This step is managed by the {\tt ED\_DIAG\_<ed\_mode>}
modules, which contain the main functions {\tt diagonalize\_impurity\_<ed\_mode>}.
This phase includes: (i) selecting sectors for diagonalization, (ii)
performing diagonalization within each sector, and (iii) analyzing
the resulting {\tt state\_list} of conserved eigenstates.

\subsection{Dynamical correlation functions}\label{sSecGF}
The determination of the low-energy part of the Hamiltonian's spectrum
enables evaluation of the Dynamical Correlation Functions (DCFs) using Krylov subspace algorithm.
This capability is central to the library when using \NAME as an
impurity solver within the DMFT framework.

Before delving into the implementation specifics, we outline the
generic approach. Consider the generic DCF:
\begin{equation}
  \label{eqGaa}
  C_\AA = \ibra \TT_\pm[ \AA(t) \AA^\dagger ]\iket,
\end{equation}
where $\AA(t)=e^{iHt}\AA e^{-iHt}$, $\TT_\pm$ is the time-ordering
operator for fermions ($+$) or
bosons ($-$), and $\ibra \cdot \iket=\frac{1}{Z}\Tr{ \left[ e^{-\beta
      H}\cdot \right]}$, with $Z=\sum_n e^{-\beta E_n}$, the thermodynamic
average.
Using spectral decomposition, the expression \equ{eqGaa} reduces to a
K\"all\`en-Lehmann form:
\begin{equation}\label{KLgf}
  \begin{split}
    C_\AA(z) 
    &=  \ibra \AA \frac{1}{z-H} \AA^\dagger\iket
    \mp \ibra \AA^\dagger \frac{1}{z+H} \AA\iket \cr
    & \simeq \frac{1}{\tilde{Z}}\sum_{n=1}^{N_{\tt state\_list}} e^{-\beta E_n}\sum_m
  \frac{ |\bra{\psi_m}\AA^\dagger \ket{\psi_n}|^2 }  {z-(E_m-E_n)}
  \mp
  \frac{ |\bra{\psi_m}\AA \ket{\psi_n}|^2 }  {z+(E_m-E_n)},\cr
\end{split}
\end{equation}
where $z\in\CCC$, $\tilde{Z}=\sum_{n=1}^{N_{\tt state\_list}} e^{-\beta E_n}$ and $\{\ket{\psi_n}, E_n\}$ are the eigenpairs of the
Hamiltonian $H$. 
This form is appealing but computationally prohibitive, as it requires
the full spectrum of the Hamiltonian. However, the first line in \equ{KLgf}
highlights that the $C_\AA$ essentially corresponds to a specific
matrix element of the resolvent operator $(z-H)^{-1}$, which can be
efficiently approximated using the Krylov subspace method.

To illustrate this approach, consider the normalized initial state
$$
\ket{\phi_n}=\AA^\dagger\ket{\psi_n}/\NN_n,
$$
where  $\ket{\psi_n}\in\SS_{\vec{Q}}$ is an eigenstate of the sector Hamiltonian and $\NN_n=\sqrt{\bra{\psi_n}\AA
  \AA^\dagger\ket{\psi_n}}$.
The Krylov basis is constructed by repeated applications of the sector Hamiltonian via MVP: 
$$
\mathcal{K}_{N} (\ket{\phi_n})=\{\ket{\phi_n}, H\ket{\phi_n}, \dots,
H^N\ket{\phi_n}\}\equiv \{\ket{v^n_0}, \ket{v^n_1},\dots, \ket{v^n_N}
\}
$$ 
with $1 \ll N \ll \DD_\SS$.
Given that any eigenstate $\ket{\psi_n}$ can be expressed within the Krylov basis as
$$
\ket{\psi_n} = \sum_i  \ibra v^n_i|\psi_n \iket  \ket{ v^n_i} =
\sum_i a^n_i \ket{v^n_i}, 
$$
we can approximate the expression \equ{KLgf} as:  
\begin{equation}
  \label{eqGKrylov}
  \begin{split}
    C_\AA(z)  &\simeq \frac{1}{\tilde{Z}}\sum_{n=1}^{N_{\tt state\_list}} e^{-\beta E_n}
    \sum_{m=1}^{N} \frac{\bra{\psi_n}\AA \AA^\dagger\ket{\psi_n} |a^n_m|^2}{
      z - (E_m-E_n)} \mp \frac{\bra{\psi_n}\AA^\dagger \AA\ket{\psi_n}
      |a^n_m|^2}{ z + (E_m-E_n)}\cr
    &= \frac{1}{\tilde{Z}}\sum_{n=1}^{N_{\tt state\_list}}\sum_{m=1}^{N} \sum_{\nu=\pm}\frac{w^{\nu}_{mn}(\AA)}{z - dE^\nu_{mn}(\AA)} 
    \equiv \frac{1}{\tilde{Z}}\sum_n
    \sum_{m=1}^{N} \sum_{\nu=\pm} g_\AA(z;\nu, w^{\nu}_{mn},  dE^\nu_{mn}), \cr
  \end{split}
\end{equation}
where the terms are grouped as a sum over spectral weights and
poles. We introduced the notation $g(z;\nu, w^{\nu}_{mn},  dE^\nu_{mn})$
for the {\tt gfmatrix} object containing the weights $w^{\nu}_{mn}(\AA)$ and
poles $dE^{\nu}_{mn}(\AA)$ for the operator $\AA$, for every initial state $\ket{\psi_n}$
contributing to the low-energy spectrum, for every order $m$ of the
Krylov subspace algorithm and for channel $\nu$.

This method is  limited to diagonal DCF like
\equ{eqGaa}. In practice, many applications also require building off-diagonal
functions of the form
$$
C_{\AA\BB}(t) = \ibra T_{\pm}[\AA(t) \BB^\dagger]\iket\, .
$$
This can be addressed by introducing auxiliary operators, e.g.  $\OO=\AA +\BB$ and
$\PP=\AA-i\BB$, enabling extraction of the desired function via
simple algebraic combinations:
$$
C_{\AA\BB} = \frac{1}{2}\left[C_\OO + C_\PP - (1-i)C_\AA -(1-i)C_\BB\right]
$$

In \NAME, the computation of the impurity Green's functions 
$$
G_{\a\b\s\s'}(t)=\ibra \TT_\pm[ c_{\a\s}(t) c_{\b\s'}^\dagger ]\iket
$$
is handled by the module \texttt{ED\_GREENS\_FUNCTIONS}.
It integrates more specialized methods based on the symmetry
classification defined by the \texttt{ed\_mode} parameter.
Moreover, in the {\bf normal} mode, \NAME also includes spin, charge, pair, and excitonic susceptibility functions, providing a comprehensive framework for dynamical response calculations.
From a computational perspective, the construction of the Krylov basis
$\mathcal{K}_N(\OO\ket{\psi_n})$ for each eigenstate within the low-energy
spectrum is typically the most resource-intensive step.
As with the
diagonalization process, a significant performance gain is achieved through
the parallel execution of the MVP at the core of
the Hamiltonian tri-diagonalization algorithm.
The input variable
\texttt{lanc\_gfniter} regulates the maximum order of the Krylov basis,
determining the upper limit on the number of excitations considered in
\equ{eqGKrylov}.
Operationally, each symmetry mode (\texttt{ed\_mode}) requires a distinct
strategy for Green's function construction, implemented in the corresponding
\texttt{ED\_GF\_<ed\_mode>} modules.

\paragraph{{\bf normal}.}
In this mode, all orbital-dependent and spin-diagonal Green's 
functions $G_{\a\b\s\s}$ must be evaluated. The diagonal 
case, where $\a=\b$, is straightforward and involves applying 
the operator $\AA=c_{\a\s}$ to any eigenstate $\ket{\psi_n}$ 
from the global {\tt state\_list} of the low-energy spectrum, 
which is represented by the {\tt sparse\_espace} object. This 
operation relies on the functions {\tt apply\_op\_C/CDG} in 
{\tt ED\_AUX\_FUNX}.

The sector Hamiltonian matrix in this mode is assumed 
to be real symmetric, which simplifies the evaluation 
of the off-diagonal terms by exploiting the following symmetry 
under orbital exchange $G_{\a\b\s\s}(z)=G_{\b\a\s\s}(z)$. 
In this case, the off-diagonal Green's function components can be obtained using just real-valued auxiliary operators such as  
$\OO=c_{\a\s}+c_{\b\s}$, along with the identity 
$G_{\a\b\s\s}=\tfrac{1}{2}(C_\OO - G_{\a\a\s\s} - G_{\b\b\s\s})$,
where $C_\OO$ denotes the DCF associated to $\OO$      . This approach, avoiding the need for complex arithmetic, preserves the efficiency of real-valued computation.  

The diagonal spin, charge and pair susceptibility terms
$\chi^{S^z}_{\a\a}$, $\chi^N_{\a\a}$ and $\chi^\Delta_{\a\a}$
are constructed using the operators $S^z_\a = \sum_{\s\s'} c^\dagger_{\a\s}
\tau^z_{\s\s'} c_{\a\s'}$, 
$N_\a = \sum_{\s\s'} c^\dagger_{\a\s} \tau^0_{\s\s'} c_{\a\s'}$
and $\Delta_\a = c_{\a\dw}c_{\a\up}$,
where $\tau^{a=0,x,y,z}$ are the Pauli matrices.
Off-diagonal terms are  evaluated using the operators
$(S^z_\a + S^z_\b)$, $(N_\a + N_\b)$ or $(\Delta_\a + \Delta_\b)$. 
The excitonic susceptibilities, $\chi^{\vec{E}}_{\a\b}$, are defined with 
respect to the vector operator $E^i_{\a\b} = \sum_{\s\s'}
c^\dagger_{\a\s}\tau^i_{\s\s'}c_{\b\s'}$, where $i=0$ represents the
spin-singlet exciton \cite{Giuli2023PRB}, and $i=x,y,z$  correspond to the spin-triplet exciton \cite{Amaricci2023PRB}.

\paragraph{{\bf superc}.}
In the superconductive case, the orbital-dependent Nambu $s$-wave
Green's function has the form:
\begin{equation}
  \label{GFnambu}
  \hat{G}_{\a\b} =
  \begin{bmatrix}
    G_{\a\b\up\up} & F_{\a\b\up\dw} \\
    \bar{F}_{\a\b\dw\up} & \bar{G}_{\a\b\dw\dw}.
  \end{bmatrix}  
\end{equation}
Exploiting symmetries between matrix components, it is 
sufficient to compute only the top row elements along with 
a few auxiliary functions \cite{Capone2001PRL,Capone2002Science,Capone2004PRLsc,Toschi2005NJP,Toschi2005PRB,Capone2009RMP}.
The diagonal, normal, component $G_{\a\a\up\up}$ can be 
evaluated using the same approach as in the {\tt ed\_mode=}{\bf normal} 
case. However, for the off-diagonal terms $G_{\a\b\up\up}$, 
we forgo symmetry arguments and instead define two 
auxiliary operators, $\OO = c_{\a\up} + c_{\b\up}$ and 
$\PP = c_{\a\up} - i c_{\b\up}$. This allows us to express 
the Green's function as
$G_{\a\b\up\up}=\tfrac{1}{2}[C_\OO + C_\PP -
(1-i)(G_{\a\a\up\up}+G_{\b\b\up\up})]$.

Evaluating the diagonal and off-diagonal anomalous terms 
$F_{\a\b\up\dw}$ requires distinct combinations of 
creation and annihilation operators \cite{Capone2001PRL,Capone2002Science,Capone2004PRLsc,Toschi2005NJP,Toschi2005PRB}. First, the component 
$\bar{G}_{\a\a\dw\dw}$ is evaluated as an auxiliary term 
using $\AA = c^\dagger_{\a\dw}$. Then, we construct the two  
linear combinations
\begin{align*}
\TT &= c_{\a\up} + c^+_{\b\dw}, \\
\RR &= c_{\a\up} - i c^+_{\b\dw},
\end{align*}
which contribute to the auxiliary functions $C_\TT$ and 
$C_\RR$, respectively. The final expression for the 
anomalous function reads
\begin{equation}
F_{\a\b\up\dw} = \tfrac{1}{2} \left[ C_\TT + C_\RR - 
(1-i)(G_{\a\a\up\up} + \bar{G}_{\b\b\dw\dw}) \right].
\end{equation}

\paragraph{{\bf nonsu2}.}
In this case, all Green's function components 
must be explicitly evaluated, as the spin and orbital 
symmetries are in general broken. For the diagonal terms 
$G_{\a\a\s\s}$, the procedure follows the same approach 
outlined in the {\bf normal} case. 

The off-diagonal components $G_{\a\b\s\s'}$, however, 
require a more general treatment. These are computed using 
auxiliary operators defined as:
\begin{align*}
\OO &= c_{\a\s} + c_{\b\s'}, \\
\PP &= c_{\a\s} - i c_{\b\s'},
\end{align*}
which allows the Green's function to be expressed as:
\begin{equation}
G_{\a\b\s\s'} = \tfrac{1}{2} \left[ C_\OO + C_\PP - 
(1-i)(G_{\a\a\s\s} + G_{\b\b\s'\s'}) \right].
\end{equation}
This approach effectively reduces the complexity of 
evaluating the off-diagonal terms by leveraging auxiliary 
functions, despite the absence of full spin symmetry.

\subsection{Observables}\label{sSecObc}
A wide range of predefined impurity observables and local 
static correlations, such as occupation numbers, total 
energy, pair amplitudes, and excitonic order parameters, 
are computed in the \texttt{OBSERVABLES} module. Similar to the
previous cases, this module wraps different implementations 
depending on the operational mode specified by 
{\tt ed\_mode} and is distributed across the corresponding 
files {\tt ED\_OBSERVABLES\_<ed\_mode>}.

Local observables and correlations are generally defined 
through the thermal average 
$\ibra \OO\iket = \tfrac{1}{Z}\Tr\left[e^{-\beta H}\OO\right]$, 
where $\OO$ can be any local operator and  $Z = \sum_n e^{-\beta E_n}$ is the partition function. 
At zero or low temperatures, this can be efficiently 
evaluated using the stored low-energy part of the spectrum, 
taking advantage of the exponential suppression provided by 
the Boltzmann factor:
\begin{equation}
\ibra \OO\iket 
= \frac{1}{Z}\sum_{n=1}^{\infty} e^{-\beta E_n}
\bra{\psi_n} \OO \ket{\psi_n}\simeq \frac{1}{\tilde{Z}}\sum_{n=1}^{N_{\tt state\_list}} e^{-\beta E_n}
\bra{\psi_n} \OO \ket{\psi_n},
\label{eq:thermal_average}
\end{equation}
where $E_n$ and $\ket{\psi_n}$ are the $N_{\tt state\_list}$ low-lying eigenstates 
of the system stored in the {\tt state\_list},  $Z=\sum_n e^{-\beta E_n}$ and  $\tilde{Z}=\sum_{n=1}^{N_{\tt state\_list}}e^{-\beta E_n}$.

\subsection{Impurity reduced density matrix}\label{sSecRDM}
The latest version of \NAME introduces the calculation of the 
impurity Reduced Density Matrix (iRDM, $\rho^\mathrm{imp}$), 
extending the algorithm initially proposed in 
Ref. \cite{BellomiaPhD,Bellomia2024PRB}, to support the analysis of entanglement 
properties of quantum impurities for any value of {\tt ed\_mode}.

For the sake of simplicity, this section focuses on the zero-temperature limit, 
assuming a non-degenerate ground state $\ket{\psi}$ present in the 
{\tt state\_list}. The generalization to the finite-temperature regime 
or degenerate ground states is straightforward, requiring only the  introduction of an
ensemble average $\ibra \rho^\mathrm{imp} \iket$, as defined in 
\equ{eq:thermal_average}.

A pure quantum state $\ket{\psi}$ belonging to a unique symmetry 
sector $\SS_{\vec{Q}}$ can be represented in the Fock basis as:
$
\ket{\psi} = \sum_I a_I \ket{I},
$
where the corresponding pure density matrix $\rho$ is given by:
\begin{equation}\label{DefRho}
\rho = \ket{\psi}\bra{\psi} = \sum_{IJ=1}^{4^{N_\mathrm{s}}} 
a^*_J a_I \ket{I}\bra{J} = \sum_{IJ=1}^{4^{N_\mathrm{s}}} 
\rho_{IJ} \ket{I}\bra{J}.
\end{equation}
The iRDM is obtained by tracing out the bath degrees of freedom:
\begin{equation}
\rho^\mathrm{imp} = \Tr_\mathrm{bath}(\rho).
\end{equation}
However, the memory footprint of $\rho$, as well as the CPU footprint 
of the summations involved in the trace, scale exponentially 
with the system size, i.e.~as $4^{N_\mathrm{s}}$, quickly becoming prohibitive.
To overcome the issue, we implement a fast algorithm that exploits the 
block structure of the Fock space defined by the symmetry sectors and
the subsequent map sparsity, to perform the trace {\it on-the-fly}. 
This approach significantly reduces the computational cost by limiting 
the summation to the size $D_\SS$ of the symmetry sector 
$\SS$, to which the ground state (or any eigenstate) belongs, and avoiding
the storage of the large matrix associated with $\rho$.
We employ a 
{\tt sparse\_map} ${\cal P}$, as detailed in \secu{CodeSparseMap}. 
This structure stores, for each impurity configuration 
(the {\it key}), the 
corresponding bath configurations 
(the {\it values}) that 
satisfy the symmetry constraints over the sector quantum numbers
imposed by the {\tt ed\_mode} variable.
We denote the number of (spin-dependent) keys and values respectively
as ${D}_{i\s}$ and ${D}_{b\s}$.
The fast summation algorithm for evaluating the iRDM differs 
significantly between the {\bf normal} and {\bf superc/nonsu2} modes, 
as outlined in the sections below.

\paragraph{{\bf normal}.}
In this case, we take advantage of spin-resolved sector decomposition, 
as discussed in detail in 
\secu{sSecQNs}. 
The global bitset can be split into spin-resolved 
impurity and bath components as
\begin{equation}
\ket{\vec{n}} =
\ket{{\color{xkcdRed}\vec{n}_\up}}\otimes \ket{{\color{xkcdBlue}\vec{n}_\dw}} =
\ket{{\color{xkcdRaspberry}\vec{i}_\up}}\ket{{\color{xkcdPumpkin}\vec{b}_\up}}\otimes \ket{{\color{xkcdPurple}\vec{i}_\dw}}\ket{{\color{xkcdTeal}\vec{b}_\dw}}.
\label{eq:decomposition_for_rdm_normal}
\end{equation}
Here, creation (destruction) operators  $c^\dagger_{p\sigma}$
($c^\dagger_{p\sigma}$) at position $p$ and for a given spin $\sigma$
act only on the corresponding 
spin subspace, i.e.~they commute with the opposite spin 
subspace $\ket{\vec{n}_{\bar{\s}}}$ without introducing any 
fermionic sign. In this basis \equ{DefRho} takes the form:
\begin{equation}\label{RhoNormal}
\rho = 
    \sum_{{\color{xkcdRaspberry}i_\up}=1}^{D_{i_\up}}\!
    \sum_{{\color{xkcdPumpkin}p_\up}=1}^{D_{p_\up}}\!
    \sum_{{\color{xkcdPurple}i_\dw}=1}^{D_{i_\dw}}\!
    \sum_{{\color{xkcdTeal}p_\dw}=1}^{D_{p_\dw}}\!
    \sum_{{\color{xkcdRaspberry}j_\up}=1}^{D_{j_\up}}\!
    \sum_{{\color{xkcdPumpkin}q_\up}=1}^{D_{q_\up}}\!
    \sum_{{\color{xkcdPurple}j_\dw}=1}^{D_{j_\dw}}\!
    \sum_{{\color{xkcdTeal}q_\dw}=1}^{D_{q_\dw}}\!
    a_{{\color{xkcdRaspberry}i_\up} {\color{xkcdPumpkin}p_\up} {\color{xkcdPurple}i_\dw} {\color{xkcdTeal}p_\dw}}a^*_{{\color{xkcdRaspberry}j_\up} {\color{xkcdPumpkin}q_\up} {\color{xkcdPurple}j_\dw} {\color{xkcdTeal}q_\dw}}
    \ket{{\color{xkcdRaspberry}\vec{i}_\up}}\ket{{\color{xkcdPumpkin}\vec{p}_\up}}\!\otimes\!\ket{{\color{xkcdPurple}\vec{i}_\dw}}\ket{{\color{xkcdTeal}\vec{p}_\dw}}
    \bra{{\color{xkcdTeal}\vec{q}_\dw}}\bra{{\color{xkcdPurple}\vec{j}_\dw}}\!\otimes\!\bra{{\color{xkcdPumpkin}\vec{q}_\up}}\bra{{\color{xkcdRaspberry}\vec{j}_\up}}
\end{equation}
where the coefficients $a_I = a_{{\color{xkcdRaspberry}i_\up} {\color{xkcdPumpkin}b_\up} {\color{xkcdPurple}i_\dw} {\color{xkcdTeal}b_\dw}}$ are the
expansion coefficients of the state on the Fock basis
$\ket{\psi} = \sum_I a_I \ket{{\color{xkcdRaspberry}{i}_\up}}\ket{{\color{xkcdPumpkin}{b}_\up}}\otimes \ket{{\color{xkcdPurple}{i}_\dw}}\ket{{\color{xkcdTeal}{b}_\dw}}$.
Before proceeding further, we show that in the {\bf normal} mode the iRDM is restricted to 
{spin-diagonal} blocks. This follows from the observation 
that the conserved quantum numbers are 
{\it additive} quantities, i.e. 
$N_\s = N_\s^\mathrm{imp} + N_\s^\mathrm{bath}$,
so that the corresponding generator of the symmetry group factorizes as
$U_{N_\s} = U_{N_\s^\mathrm{imp}} \otimes U_{N_\s^\mathrm{bath}}$. 
Since the iRDM acts only on the impurity Fock space, this implies:
\begin{equation}
\begin{aligned}
\rho^\mathrm{imp} &= U_{N_\s}^\dagger \rho^\mathrm{imp} U_{N_\s} = 
U_{N_\s^\mathrm{bath}}^\dagger \otimes U_{N_\s^\mathrm{imp}}^\dagger 
\rho^\mathrm{imp} 
U_{N_\s^\mathrm{imp}} \otimes U_{N_\s^\mathrm{bath}} \\
&= 
U_{N_\s^\mathrm{imp}}^\dagger \rho^\mathrm{imp} U_{N_\s^\mathrm{imp}},
\end{aligned}
\end{equation}
showing that $U_{N_\s^\mathrm{imp}}$ defines a symmetry of 
$\rho^\mathrm{imp}$, which is therefore block diagonal with respect to 
the impurity quantum numbers $N_\s^\mathrm{imp}$.

Using \equ{RhoNormal} the iRDM takes the form:
\begin{equation}
  \label{iRDMnormal}
  \begin{aligned}
  \rho^\mathrm{imp} &= \Tr_{{\color{xkcdCoral}b_\up} {\color{xkcdAzure}b_\dw}}{\rho} = 
  \sum_{{\color{xkcdCoral} b_\up}=1}^{D_{b_\up}}
  \sum_{{\color{xkcdAzure} b_\dw}=1}^{D_{b_\dw}}
  \bra{{\color{xkcdCoral}b_\up}}\otimes\bra{{\color{xkcdAzure}b_\dw}}
    \rho
    \ket{{\color{xkcdAzure}b_\dw}}\otimes\ket{\color{xkcdCoral}{b_\up}}    \cr
    &=
    \sum_{{\color{xkcdCoral} b_\up}=1}^{D_{b_\up}}
    \sum_{{\color{xkcdRaspberry}i_\up}=1}^{D_{i_\up}}\!
    \sum_{{\color{xkcdPumpkin}p_\up}=1}^{D_{p_\up}}\!
    \sum_{{\color{xkcdRaspberry}j_\up}=1}^{D_{j_\up}}\!
    \sum_{{\color{xkcdPumpkin}q_\up}=1}^{D_{q_\up}}\!
    \Bigg(    
    \sum_{{\color{xkcdAzure} b_\dw}=1}^{D_{b_\dw}}
    \sum_{{\color{xkcdPurple}i_\dw}=1}^{D_{i_\dw}}\!
    \sum_{{\color{xkcdTeal}p_\dw}=1}^{D_{p_\dw}}\!
    \sum_{{\color{xkcdPurple}j_\dw}=1}^{D_{j_\dw}}\!
    \sum_{{\color{xkcdTeal}q_\dw}=1}^{D_{q_\dw}}\!
    C_{i,p}\,C_{j,q}\,a_{{\color{xkcdRaspberry}i_\up} {\color{xkcdPumpkin}p_\up} {\color{xkcdPurple}i_\dw} {\color{xkcdTeal}p_\dw}}a^*_{{\color{xkcdRaspberry}j_\up} {\color{xkcdPumpkin}q_\up} {\color{xkcdPurple}j_\dw} {\color{xkcdTeal}q_\dw}} \cr
    &\qquad    
    \langle {\color{xkcdCoral}b_\up}| {\color{xkcdPumpkin}p_\up}\rangle| {\color{xkcdRaspberry}i_\up}\rangle\langle {\color{xkcdRaspberry}j_\up} |\langle {\color{xkcdPumpkin}q_\up}| {\color{xkcdCoral}b_\up}\rangle   \otimes \langle {\color{xkcdAzure}b_\dw}| {\color{xkcdTeal}p_\dw} \rangle|{\color{xkcdPurple}i_\dw}\rangle \langle {\color{xkcdPurple}j_\dw}  |\langle {\color{xkcdTeal}q_\dw}| {\color{xkcdAzure}b_\dw}\rangle \Bigg)
    \cr
&=
    \sum_{{\color{xkcdRaspberry}i_\up}=1}^{D_{i_\up}}
    \sum_{{\color{xkcdRaspberry}j_\up}=1}^{D_{j_\up}}
    \sum_{{\color{xkcdPurple}i_\dw}=1}^{D_{i_\dw}}
    \sum_{{\color{xkcdPurple}j_\dw}=1}^{D_{j_\dw}}
    \underbrace{\left(
    \sum_{{\color{xkcdCoral} b_\up}=1}^{D_{b_\up}}
    \sum_{{\color{xkcdAzure} b_\dw}=1}^{D_{b_\dw}}
    a_{{\color{xkcdRaspberry}i_\up} {\color{xkcdCoral}b_\up} {\color{xkcdPurple}i_\dw} {\color{xkcdAzure}b_\dw}}a^*_{{\color{xkcdRaspberry}j_\up} {\color{xkcdCoral}b_\up} {\color{xkcdPurple}j_\dw} {\color{xkcdAzure}b_\dw}}
    \right)}_{\rho^\mathrm{imp}_{{\color{xkcdRaspberry}i_\up j_\up}} \otimes\, \rho^\mathrm{imp}_{{\color{xkcdPurple}i_\dw j_\dw}} }
    \ket{{\color{xkcdRaspberry}i_\up}}\bra{{\color{xkcdRaspberry}j_\up}}\otimes \ket{{\color{xkcdPurple}i_\dw}}\bra{{\color{xkcdPurple}j_\dw}},
  \end{aligned}
\end{equation}

Notably, the factors $C_{i,b}$, which account for the 
fermionic sign associated to swapping bath and impurity components, are 
trivially 1 due to the absence of cross-spin interference in this 
mode, as enforced by the Kronecker deltas for bath and impurity 
indices.

The numerical implementation relies on the use of the {\tt
  sparse\_map} ($\%sp$) for the ground state sector as reported in the listing below:  
\begin{lstlisting}[style=fstyle,numbers=none,basicstyle={\scriptsize\ttfamily}]
do IimpUp=0,2**Norb-1
  do JimpUp=0,2**Norb-1
    !Finding the unique bath states connecting IimpUp and JimpUp
    call sp_return_intersection(sectorI%H(1)%sp,IimpUp,JimpUp,BATHup,lenBATHup)
    if(lenBATHup==0)cycle
    do IimpDw=0,2**Norb-1
      do JimpDw=0,2**Norb-1
        !Finding the unique bath states connecting IimpDw and JimpDw -> BATHdw(:)
        call sp_return_intersection(sectorI%H(2)%sp,IimpDw,JimpDw,BATHdw,lenBATHdw)
        if(lenBATHdw==0)cycle
        do ibUP=1,lenBATHup
          IbathUp = BATHup(ibUP)
          do ibDW=1,lenBATHdw
            IbathDw = BATHdw(ibDW)
            !Allowed spin Fock space Istates:
            !Iup = IimpUp +  2^Norb * IbathUp
            !Idw = IimpDw +  2^Norb * IbathDw
            iUP= binary_search(sectorI%H(1)%map,IimpUp + 2**Norb*IbathUp)
            iDW= binary_search(sectorI%H(2)%map,IimpDw + 2**Norb*IbathDw)
            i  = iUP + (iDW-1)*sectorI%DimUp
            !Allowed spin Fock space Jstates:
            !Jup = JimpUp +  2^Norb * IbathUp
            !Jdw = JimpDw +  2^Norb * IbathDw
            jUP= binary_search(sectorI%H(1)%map,JimpUp + 2**Norb*IbathUp)
            jDW= binary_search(sectorI%H(2)%map,JimpDw + 2**Norb*IbathDw)
            j  = jUP + (jDW-1)*sectorI%DimUp
            ! 
            io = (IimpUp + 2**Norb*IimpDw) + 1
            jo = (JimpUp + 2**Norb*JimpDw) + 1
            irdm(io,jo) = irdm (io,jo) + psi(i)*psi(j)*weight
          enddo
        enddo
      enddo
    enddo
  enddo
enddo
\end{lstlisting}

\paragraph{{\bf superc/nonsu2}.}
For these lower-symmetry cases, the absence of a clean Fock space 
factorization introduces additional complexities. Specifically, the 
correct evaluation of the off-diagonal iRDM terms, which connect 
different $[N^\mathrm{imp}_\up,\,N^\mathrm{imp}_\dw]$ blocks, 
requires an explicit computation of fermionic permutation signs.

Consider a generic Fock basis state in the form 
$\ket{I} =\ket{{i}_\up\,{b}_\up\,{i}_\dw\,{b}_\dw}$, and express the 
ground state as 
$$
\ket{\psi} = \sum_I a_I \ket{I} = 
    \sum_{i_\up=1}^{D_{i_\up}}
    \sum_{b_\up=1}^{D_{b_\up}}
    \sum_{i_\dw=1}^{D_{i_\dw}}
    \sum_{b_\dw=1}^{D_{b_\dw}}
    a_{i_\up b_\up i_\dw b_\dw}
    \ket{{i}_\up\,{b}_\up\,{i}_\dw\,{b}_\dw}.
$$ 
By tracing over the bath degrees of freedom of the corresponding pure density matrix $\rho$ we obtain:
\begin{equation}
  \begin{aligned}
    \rho^\mathrm{imp} &= \Tr_{b_\up b_\dw}{\rho} =
    \sum_{b_\sigma=1}^{D_{b_\sigma}}
    \bra{b_\up\, b_\dw}
    \,\rho\,
    \ket{b_\dw\, b_\up} =
    \sum_{b_\sigma=1}^{D_{b_\sigma}}
    \ibra{b_\up\, b_\dw}
    \ket{\psi}\ibra{\psi}
    \ket{b_\dw\, b_\up}    \cr
    &=
    \sum_{b_\sigma=1}^{D_{b_\sigma}}
    \sum_{i_\sigma=1}^{D_{i_\sigma}}
    \sum_{p_\sigma=1}^{D_{p_\sigma}}
    \sum_{j_\sigma=1}^{D_{j_\sigma}} 
    \sum_{q_\sigma=1}^{D_{q_\sigma}}
    a_{i_\up p_\up i_\dw p_\dw} a^*_{j_\up q_\up j_\dw q_\dw}
    \langle{b_\up\, b_\dw}|{i_\up\, p_\up\, i_\dw\, p_\dw}\rangle
    \langle{q_\dw \,  j_\dw \, q_\up\, j_\up }|{b_\dw\, b_\up}\rangle\, ,
  \end{aligned}
\end{equation}

To further simplify this expression, it is essential to eliminate the 
sums over the internal bath indices $p_\sigma$ and $q_\sigma$, ideally 
contracting them with the outer bath indices $b_\sigma$. This requires 
reorganizing the bit representation of the basis states to bring the 
bath bitsets before the impurity ones:
$$
\ket{i_\up\, p_\up\, i_\dw\, p_\dw} \rightarrow C_{p_\up,i_\dw} \ket{i_\up\, i_\dw \,p_\up\,p_\dw},
$$
where the fermionic sign factor
$C_{p_\up,i_\dw}  = (-1)^{\#\vec{n}_{p\up} \cdot
  \#\vec{n}_{i\dw}}$
accounts for the exchange of the bath $\ket{p_\up}$  and impurity
$\ket{i_\dw}$ bit configurations.
Here, $\#\vec{n}_{\alpha\sigma}$ denotes the number of ones in the 
occupation vector $\vec{n}_{\alpha\sigma}$. 

Substituting this relation back into the iRDM expression yields:
\begin{equation}
    \rho^\mathrm{imp} =
    \sum_{i_\up=1}^{D_{i_\up}}
    \sum_{j_\up=1}^{D_{j_\up}}
    \sum_{i_\dw=1}^{D_{i_\dw}}
    \sum_{j_\dw=1}^{D_{j_\dw}}
    \underbrace{\left(
    \sum_{b_\up=1}^{D_{b_\up}}
    \sum_{b_\dw=1}^{D_{b_\dw}}
    a_{i_\up b_\up i_\dw b_\dw} a^*_{j_\up b_\up j_\dw b_\dw}
    C_{b_\up,i_\dw}C_{b_\up,j_\dw}\right)}_{
    \rho^\mathrm{imp}_{i_\up i_\dw j_\dw j_\up}}
    \ket{i_\up\, i_\dw}\bra{j_\dw \, j_\up},
  \end{equation}
where the term 
$C_{b_\up,i_\dw}C_{b_\up,j_\dw}$ accounts for the fermionic 
sign associated with the bath-impurity bitset exchange.

In numerical implementations, this approach introduces a few 
key differences with respect to the {\bf normal} mode.
In particular, the 
ordering of the \emph{key-value} pairs in the {\tt sparse\_map} 
must reflect the contiguous bitset structure, which is critical 
for correctly reconstructing the Fock state indices.
See the following listing:
\begin{lstlisting}[style=fstyle,numbers=none,basicstyle={\scriptsize\ttfamily}]
do IimpUp=0,2**Norb-1
  do IimpDw=0,2**Norb-1
    do JimpUp=0,2**Norb-1
      do JimpDw=0,2**Norb-1
        !Build indices of the RDM in 1:4**Norb
        iImp = iImpUp + iImpDw*2**Norb
        jImp = jImpUp + jImpDw*2**Norb
        call sp_return_intersection(sectorI%H(1)%sp,iImp,jImp,Bath,lenBath)
        if(lenBATH==0)cycle
        do ib=1,lenBath
          iBath = Bath(ib)
          !Reconstruct the Fock state Ii map back to sector state i
          ii= iImpUp + iImpDw*2**Ns + iBath*2**Norb
          i = binary_search(sectorI%H(1)%map,ii)
          !Reconstruct the Fock state Jj map back to sector state j
          jj= jImpUp + jImpDw*2**Ns + iBath*2**Norb
          j = binary_search(sectorI%H(1)%map,jj)
          !Build the signs of each component of RDM(io,jo)
          nBup  = popcnt(Ibits(ii,Norb,Norb*Nbath))
          nIdw  = popcnt(Ibits(ii,Ns,Norb))
          nJdw  = popcnt(Ibits(jj,Ns,Norb))
          signI = (-1)**(nIdw*nBup)
          signJ = (-1)**(nJdw*nBup)
          sgn   = signI*signJ
          !  
          io = (iImpUp+1) + 2**Norb*iImpDw
          jo = (jImpUp+1) + 2**Norb*jImpDw
          rdm(io,jo) = rdm(io,jo) + psi(i)*conjg(psi(j))*weight*sgn
        enddo
      enddo
    enddo
  enddo
enddo
\end{lstlisting}

\subsection{Bath Functions}\label{sSecFunc}
The module {\tt ED\_BATH\_FUNCTIONS} implements the on-the-fly 
calculation of the hybridization function 
$\Delta_{\a\b\s\s'}(z)$ and the non-interacting Green's functions 
$G^0_{\a\b\s\s'}(z)$ for arbitrary complex frequencies $z \in \CCC$. 
These are defined as
\begin{equation}
\Delta_{\a\b\s\s'}(z) = \sum_p \left[\hat{V}^p
\left( z\11 - \hat{h}^p \right)^{-1} \hat{V}^p\right]_{\a\b\s\s'},
\end{equation}
\begin{equation}
G^0_{\a\b\s\s'}(z) = 
\left[ (z + \mu)\11 - \hat{h}^0 - 
\Delta(z) \right]_{\a\b\s\s'}^{-1},
\end{equation}
where $V^p_{\a\b\s\s'}$ and $h^p_{\a\b\s\s'}$ are the bath coupling 
and bath Hamiltonian matrices, respectively, and $\mu$ is the chemical 
potential.

The module supports all cases defined by the {\tt ed\_mode} and 
{\tt bath\_type} variables, including both user-supplied baths (provided 
as rank-1 arrays of doubles) and the internally allocated {\tt 
effective\_bath} instance {\tt dmft\_bath}. This flexibility ensures 
compatibility with a wide range of other solvers.

Different routines are available for directly evaluating $\Delta$, 
$G^0$, and its inverse $[G^0]^{-1}$, which are critical for computing 
the self-energy functions on demand (see \secu{sSecIO}). 

In the superconducting case, {\tt ed\_mode}={\bf superc}, 
special care is needed to properly handle the anomalous (off-diagonal) 
components within the Nambu basis. In this case, the bath functions 
include both the normal and anomalous components.

\subsection{Bath Optimization}\label{sSecFit}
In the DMFT-ED framework, the bath parameters must be optimized to 
faithfully reproduce the Weiss field, 
${\GG^{-1}_{0}}_{\a\b\s\s'}(z)$, or its corresponding hybridization 
function 
\begin{equation}
\Theta(z) = (z + \mu)\11 - H^\mathrm{loc} - \GG_0^{-1}(z),
\end{equation}
where $\mu$ is the chemical potential and $H^\mathrm{loc}$ is the 
local non-interacting Hamiltonian. The Weiss field is obtained from the DMFT 
self-consistency equation \cite{Georges1996RMP}, while the bath 
discretization requires a careful fitting of this continuous function 
using a finite set of bath parameters (see \secu{sSecBath}).

Several algorithms have been proposed for this optimization 
step \cite{Garcia2004PRL,Taranto2012PRB,Mejuto-Zaera2020PRB}, each with strengths suited 
to different physical contexts. To maintain flexibility in \NAME, 
we keep the optimization part independent 
from the core impurity solver. However, we also include a fully 
integrated optimization routine based on the conjugate gradient (CG) 
minimization of the cost function
\begin{equation}
\chi = \sum_{n=1}^{L_\mathrm{fit}} 
\frac{1}{W_n} \left\|X(i\omega_n) - 
X^\mathrm{QIM}(i\omega_n; \{V, h\}) \right\|_q,
\label{eq:chiq}
\end{equation}
where the $q$-norm is a suitably chosen distance metric in the 
matrix function space. Here, $X_{\a\b} = {\GG_{0}}_{\a\b\s\s'}$ or 
$\Theta_{\a\b\s\s'}$ are the user-supplied local functions, while 
$X_{\a\b}^\mathrm{QIM} = G^0_{\a\b\s\s'}$ or 
$\Delta_{\a\b\s\s'}$ are the corresponding quantities for the quantum 
impurity model, see \secu{sSecFunc}.

While different optimization methods have been developed for 
truncated algorithms, e.g. sCI \cite{Mejuto-Zaera2020PRB} or DMRG \cite{Bauernfeind2017PRX}, 
which handle systems with a 
larger number of bath levels, the CG minimization approach has proven 
both efficient and flexible for the small to moderate bath sizes 
typical in ED solvers. 

The entire fit procedure is encapsulated in the function 
\texttt{ed\_chi2\_fitgf} provided by the module 
\texttt{ED\_BATH\_FIT}. To exploit the regularity of the bath functions, 
the fit is performed on the Matsubara frequency axis, where the functions 
are smooth and rapidly decaying. The form of $X_{\a\b\s\s'}$ is controlled 
by the input parameter \texttt{cg\_Scheme=Weiss,Delta}.

To provide maximum control over the fitting process, we include several 
tunable parameters:
\begin{itemize}
\item \texttt{cg\_method=0,1} - Chooses the CG algorithm: 
\texttt{0} for a Fletcher-Reeves-Polak-Ribiere variant adapted from 
Numerical Recipes \cite{NumRec77}, and \texttt{1} for the original 
algorithm described in Ref. \cite{Georges1996RMP}, commonly used 
in the DMFT community.
\item \texttt{cg\_grad=0,1} - Sets the gradient calculation method: 
\texttt{0} for analytical gradients (when available), \texttt{1} for 
numerical gradients (required if \texttt{cg\_method=1}).
\item \texttt{cg\_Lfit} - Sets the number of Matsubara frequencies 
$L_\mathrm{fit}$ used in the fit. This can be used to restrict the 
fit to the low-frequency regime, where the function behavior is 
most relevant.
\item \texttt{cg\_Weight=0,1,2} - Determines the frequency weighting 
scheme: \texttt{0} for uniform, \texttt{1} for inverse frequency 
weighting, and \texttt{2} for inverse Matsubara index weighting, which 
can emphasize low-energy contributions.
\item \texttt{cg\_pow} - Sets the power $q$ of the cost function, 
allowing the user to fine-tune the sensitivity of the optimization 
to outliers.
\item \texttt{cg\_norm=elemental,frobenius} - Sets the matrix norm to be used to define the cost function in \equ{eq:chiq}. The default value (\texttt{elemental}) corresponds to a generic element-wise $\chi^q$ norm, defined as 
\begin{equation*}
    ||X-X^\mathrm{QIM}||_q \equiv {\sum_{ij} \left|X_{ij}-X^\mathrm{QIM}_{ij}\right|^q}.
\end{equation*}
The \texttt{frobenius} option implements instead the matrix distance induced by the Frobenius inner product as
\begin{equation*}
    ||X-X^\mathrm{QIM}||_q \equiv 
    \sqrt[q]{\Tr\left[(X-X^\mathrm{QIM})^\dagger(X-X^\mathrm{QIM})\right]} =
    \left(\sum_{ij} \left|X_{ij}-X^\mathrm{QIM}_{ij}\right|^2 \right)^{\!\!\!\frac{1}{q}}.
\end{equation*}
For $q=2$ (see \texttt{cg\_pow}) the Frobenius norm is equivalent
to the Euclidean distance in matrix space. 
We plan to expand the available options for the {tt cg\_norm} parameter in future \NAME updates, since the {\tt nonsu2} and {\tt superc} diagonalization modes involve subtle challenges in optimizing the off-diagonal components of X. These difficulties are related to the coexistence of matrix elements with different physical origin, so that they can live on different orders of magnitude, hence requiring carefully balanced optimization metrics to ensure an accurate bath representation.
\item \texttt{cg\_Ftol} - Controls the fit tolerance, setting the 
convergence threshold for the CG minimization.
\item \texttt{cg\_Niter} - Specifies the maximum number of allowed 
iterations for the CG minimization.
\item \texttt{cg\_stop=0,1,2} - Defines the exit condition of the 
minimization corresponding, respectively, to the options  $C_1\land C_2$, $C_1$ and $C_2$, with
\begin{align*}
C_1 & : |\chi^{n-1} - \chi^n| < \mathtt{cg\_Ftol} (1+\chi^n), \\
C_2 & : \left\|x_{n-1} - x_n\right\| < 
\mathtt{cg\_Ftol} (1+\left\|x_n\right\|),
\end{align*}
where $\chi^n$ is the cost function at the $n^\mathrm{th}$ step, and 
$x_n$ the corresponding parameter vector. The stop condition is a 
logical OR between these criteria, with the parameter value selecting 
which condition to apply.
\end{itemize}

Together, these parameters provide precise control over the fitting 
process, allowing users to balance speed and accuracy according to 
their specific needs.

\subsection{Input/Output}\label{sSecIO}
The \texttt{ED\_IO} module provides comprehensive access to the 
results of the Lanczos diagonalization of the quantum impurity 
problem. Since each instance of the code persists in memory until a 
new calculation is initiated, access to the relevant data is managed 
through a set of dedicated functions. These include routines for 
extracting dynamical response functions, self-energy components, and 
impurity observables, as well as for performing on-the-fly 
recalculation of the impurity Green's functions and self-energies at 
arbitrary points in the complex frequency domain. The latter 
capability is enabled through the efficient use of \texttt{gfmatrix} 
objects for data storage.

A full list of available functions, including parameter descriptions 
and usage examples, can be found in the online documentation (see \href{https://edipack.github.io/EDIpack/}{edipack.github.io/EDIpack}).  
Here, we highlight two representative examples:
The first is the function \texttt{ed\_get\_dens}, which populates the 
provided input array with the orbital impurity occupations 
$\langle n_{\a\s}\rangle$. This function is essential for extracting 
local density information, a key diagnostic in many DMFT studies.

The second example is \texttt{ed\_get\_sigma}, which retrieves the 
normal or anomalous components of the Matsubara or real-frequency 
self-energy function $\Sigma(z)$. This function represents a central 
output of any DMFT calculation, as the self-energy encapsulates the 
full set of local electronic correlations and their frequency 
dependence.

\subsection{EDIpack2ineq: inequivalent impurities}\label{sSecIneq}
In many contexts, it is necessary to solve systems with multiple, 
independent quantum impurities. This situation frequently arises in 
DMFT when modeling lattices with complex unit cells containing 
inequivalent atomic sites, or in supercell calculations where 
translational symmetry is broken, as in heterostructures, disordered 
systems, or multi-orbital setups. 

Given that \NAME allows for only a single instance of the solver at a 
time, a dedicated extension is required to handle these more complex 
cases. To address this, we developed the \NAME{2ineq} 
sub-library, which extends the base functionality of \NAME by managing 
memory and procedures for multiple impurities. This approach ensures 
that all functions remain accessible through standard Fortran 
interfaces, seamlessly integrating both \NAME and \NAME{2ineq} 
without compromising overall software design.

Below, we provide a brief overview of the main features introduced by 
\NAME{2ineq}.

\subsubsection{Structure}\label{ssSecIneqStructure}
The core of the \NAME{2ineq} sub-library consists of several 
Fortran modules, all wrapped under the main interface module 
{\tt EDIPACK2INEQ}. This module provides access to the full range of 
procedures and variables required to solve inequivalent quantum 
impurity problems. To use this extension, the user must include both 
the main \NAME and the \NAME{2ineq} modules, as shown below:
\begin{lstlisting}[style=fstyle,numbers=none,basicstyle={\scriptsize\ttfamily}]
program test
  !Load EDIpack library 
  USE EDIPACK
  !Load the Inequivalent impurities extension
  USE EDIPACK2INEQ
  ...
\end{lstlisting}

A significant part of the \NAME{2ineq} extension is the 
definition of global variables that extend the memory pool of 
\NAME. This includes higher-rank arrays for storing impurity-specific 
data, ensuring that each impurity's state is maintained separately 
during the computation. While dedicated MPI communication could be 
used to manage specific objects in parallel, e.g. {\tt
  effective\_bath} or {\tt gfmatrix}, we opted for a simpler and safer  
file-based approach to maintain compatibility with the standard I/O 
procedures described in \secu{sSecIO}.

\subsubsection{Core routines}\label{ssSecIneqGlobal}
The module {\tt E2I\_MAIN} wraps the key extensions to the main 
algorithms of \NAME, including initialization, diagonalization, and 
finalization, while preserving the original function names for consistency.

\paragraph{{\tt ed\_init}.} {\bf Initialization} of the solver. This function  extends the corresponding one in \NAME  to 
  accept a rank-2 bath array, where the leading dimension specifies 
  the number of inequivalent impurity problems. This ensures each bath 
  is properly initialized for subsequent diagonalization.

\paragraph{{\tt ed\_solve}.} {\bf Diagonalization} of the inequivalent quantum impurity problems. This function is extended to  accept a  
  rank-2 bath array and to manage the diagonalization of each impurity 
  problem. Parallel execution is controlled by the input flag 
  {\tt mpi\_lanc=T/F}, which determines whether the Lanczos 
  diagonalization is performed in parallel or sequentially across 
  different impurities.

\paragraph{{\tt ed\_finalize}.} {\bf Finalization} of the solver. This handles the global memory release extending the corresponding  function in \NAME, which clears the memory pool for all inequivalent impurity instances.

\subsubsection{Inequivalent Baths}\label{ssSecIneqBath}
The bath setup for multiple impurities is managed by the 
{\tt E2I\_BATH} module, which provides functions for defining 
site-specific bath matrices and their corresponding variational 
parameters. This includes support for conventional symmetry operations 
and the handling of replica bath structures. 

In the module {\tt E2I\_BATH\_REPLICA}, we extend the 
matrix basis definition for the variational parameters 
$\vec{\lambda}$, allowing for flexible bath optimization across 
multiple sites. Although the matrix basis is currently shared among 
all impurities, future versions may allow for fully independent 
bath parameterizations.

Additionally, the module {\tt E2I\_BATH\_FIT} extends the generic 
function {\tt ed\_chi2\_fitgf} to support simultaneous, independent 
bath optimization for all impurities using MPI, thereby improving the 
efficiency of large-scale calculations.

\subsubsection{Input/Output}\label{ssSecIneqIO}
A key component of the \NAME{2ineq} extension is the 
enhanced I/O capability for handling impurity-specific observables. 
The {\tt E2I\_IO} module includes a variety of functions for 
retrieving site-resolved quantities, such as local Green's functions, 
self-energies, and density matrices. These functions use the same 
naming conventions as the core \NAME library, ensuring a consistent 
user experience across single and multi-impurity calculations.

\ifSubfilesClassLoaded{
  \bibliography{references}
}{}
\end{document}

%% file: 04_cbinding.tex
\section{Interoperability}\label{SecInterop}
The recent growing availability of state-of-the-art software dedicated
to the solution of quantum impurity problems using different methods \cite{Bulla2008RMP,Parcollet2015CPC,Seth2016CPC,Bauernfeind2017PRX,Ganahl2015PRB,Wallerberger2019CPC,Mejuto-Zaera2020PRB}
poses a challenge to test accuracy and reliability of the
results.
As such, software packages are expected to develop a higher level of interoperability, i.e. the capability to operate with other software, possibly written in different programming languages.
Modern Fortran, which is the language of choice for \NAME, since many
years supports the standardized generation of procedures and
variables that are interoperable with C.
Here we describe the implementation of an interoperability layer aiming at developing APIs for other languages as well as 
integrating \NAME in complex scientific frameworks, e.g. TRIQS \cite{Parcollet2015CPC} or w2dynamics \cite{Wallerberger2019CPC}. 

\subsection{C bindings}\label{sSecInteropCbindings}
The interoperability with C language is provided by the
{\tt ISO\_C\_BINDING} module, which is part of the Fortran
standard since 2003\cite{Reid2003CISE,Reid2007SFF}. The module contains definitions of named
constants, types and procedures for C interoperability.
Alongside, a second key feature essential to expose any Fortran entity to C is the {\tt BIND(C)} intrinsic function.
In \NAME we exploit these features of the language to provide a
complete interface from the Fortran code to C/C++ named {\tt
  EDIPACK\_C}.

\subsubsection{Installation and inclusion}\label{sSecInteropCbindingsInstallation}
The C-binding module is included in the build process of \NAME and
compiled into a dynamical library {\tt
  libedipack\_cbindings.so (.dylib)}. As discussed in the
\secu{sSecInstallBuildInstall}, support for inequivalent
impurities is configured at the build level and propagates to the
C-bindings library as well. An exported variable {\tt has\_ineq} is
defined and exported to C/C++ as a way to query the presence of support for the inequivalent impurities. 
The generated library and header files get installed in the include
directory at the prefix location, specified during configuration
step. The corresponding path is added to the environment variable {\tt
  LD\_LIBRARY\_PATH}, valid of any Unix/Linux system, via any of the
loading methods outlined in \secu{sSecInstallOSloading}. 

The C/C++ compatible functions and variables 
are declared in the header file {\tt e\-di\-pack\_c\-bin\-din\-gs.h}.
Since {\tt ISO\_C\_BINDING} only provides, to date, C compatibility, the 
functions and variables are declared with C linkage, which prevents name mangling.
Function overloading is also not supported, hence all interfaced Fortran functions, for example supporting multiple input variable combinations with different types and ranks, are here represented by multiple alternative functions.

As an example, we consider the {\tt ed\_chi2\_fitgf} Fortran function, which handles the fitting of bath parameters. This is callable in C++ in the following variants:
\begin{itemize}
    \item  {\tt chi2\_fitgf\_single\_normal\_n3}: for {\tt ed\_mode=normal/nonsu2} rank-3 Weiss field/hybridization function arrays
    \item  {\tt chi2\_fitgf\_single\_normal\_n5}: for {\tt ed\_mode=normal/nonsu2} rank-5 Weiss field/hybridization function arrays
    \item  {\tt chi2\_fitgf\_single\_superc\_n3}: for {\tt ed\_mode=superc} rank-3 Weiss field/hybridization function arrays
    \item  {\tt chi2\_fitgf\_single\_superc\_n5}: for {\tt ed\_mode=superc} rank-5 Weiss field/hybridization function arrays
\end{itemize}
Analogous functions for the inequivalent impurities case, i.e. \NAME{2Ineq}, are available. All functions are listed and thoroughly documented in the online manual associated to this release. 
When using \NAME functions in a C/C++ program, care must be taken in the way arrays are passed. Consistently with {\tt ISO\_C\_BINDING}, non-scalar parameters have to be passed as raw pointers. An array of integers containing the dimensions of the former need to be passed as well to allow for proper Fortran input parsing. A working C++ example is provided in the {\tt examples} directory of the \NAME repository and is discussed in Section \ref{SecExamples}.

\subsubsection{Implementation}\label{sSecInteropCbindingsImplementation}
The interface layer is contained in the Fortran module {\tt
  EDIPACK\_C}. It contains a common part and two sets of functions,
one to interface the procedures from \NAME and a second
one to extend the interface to the inequivalent impurities case.
Specifically, the implemented interface functions expose to C through
{\tt bind(C)} statement a number of
procedures composing the Fortran API of \NAME, i.e. contained in {\tt
  ED\_MAIN}. The procedures and shared variables can be divided in four main groups (note that contrary to the Fortran API the functions here lose the prefix {\tt ed\_}):

\paragraph{{\bf Variables}.}
A number of relevant input and shared variables, which are normally
required to setup or to control the calculation, are interfaced to C
directly in the \NAME modules {\tt ED\_INPUT\_VARIABLES} using the {\tt
  bind(C)} constructs. These are implicitly loaded into the C-binding module {\tt
  EDIPACK\_C} through the Fortran {\tt USE EDIPACK} statement and
then further interfaced in the C++ header file.

\paragraph{{\bf Main}.} This group contains interface to the exact
diagonalization method, interfacing the solver  {\tt
  solve\_site/ineq} as well as its initialization
{\tt init\_solver\_site/ineq} and finalization {\tt
  finalize\_solver} procedures. It also includes the functions used to
set the non-interacting part of the impurity Hamiltonian
$h^0_{\a\b\s\s'}$ through the functions {\tt set\_Hloc}, or the
interaction Hamiltonian through {\tt add\_twobody\_operator}.

\paragraph{{\bf Bath}.} In this group we implement a number of
procedures dealing with bath initialization, symmetry operations and
optimization. In particular it contains the function {\tt
  get\_bath\_dimension} returning the
dimension of the bath array on the user side as well as the setup of the matrix basis
$\vec{\Gamma}$ for the replica (general) bath via different instances of
{\tt set\_Hreplica(general)\_\-\{site,lattice\}\_d\{3,5\}}. 
A crucial part of the DMFT self-consistency loop is the optimization of the bath through CG algorithm (see \secu{sSecFit}). To this end we interfaced a
number of functions which
cover all the cases supported in \NAME, conveniently called {\tt
  chi2\_fitgf\_\{single,lattice\}\_\{normal,superc\}\_n\{3,4,5,6\}}. Note that, because the actual
optimization is still performed through the Fortran code, no changes
apply to the outcome of this step.    

\paragraph{{\bf Input/Output}.}
The input and output parts of the software are interfaced in this group
of functions. In particular, {\tt read\_input} exposes to C the input
reading procedure of \NAME, which sets all the internal
variables of \NAME.
Next, in {\tt edipack\{2ineq\}\_c\_binding\_io} we interface all the
functions implementing the communication from the \NAME instance to
the user, namely those to retrieve static observables (e.g. {\tt
  get\_dens}), the impurity Green's functions and self-energies
(e.g. {\tt get\_Sigma\_\{site,lattice\}\_n\{3,5\}}), the
impurity susceptibilities (e.g. {\tt get\_spinChi}), the impurity reduced density matrix as well
as the non-interacting Green's and hybridization functions starting from the user bath array.


\subsection{EDIpack2py, the Python API}\label{sSecInteropEDIpy}
As a first application of the \NAME C-bindings we implemented a
complete Python interface called EDIpack2py. This is a Python module which
enables access to all the library features and unlocks implementation of
further interfaces of \NAME as a plug-in solver for external Python-based software. Detailed documentation can be found online at \href{https://edipack.github.io/EDIpack2py}
{edipack.github.io/EDIpack2py}.

\subsubsection{Installation}\label{sSecInteropEDIpyInstallation}
\paragraph{From source.}
EDIpack2py is available as a platform-agnostic Python module depending on EDIpack.
The code can be obtained and installed from source as:
\begin{lstlisting}[style=mybash]
git clone https://github.com/edipack/EDIpack2py EDIpack2py
cd EDIpack2py
pip install . 
\end{lstlisting}

\paragraph{PyPi.}
The EDIpack2py package is also available in PyPi at
\href{https://pypi.org/project/EDIpack2py/}{pypi.org/EDIpack2py}. The package can be easily installed in any system supporting {\tt pip} as:

\begin{lstlisting}[style=mybash]
pip install edipack2py
\end{lstlisting}


\paragraph{Anaconda.}
Similarly to \NAME, the Python API EDIpack2py is available as an
Anaconda package for GNU/Linux and macOS systems. Packages are available for
Python$\geq 3.10$. The \NAME package contains the {\tt
  EDIpack2py} Python module as well as \NAME and SciFortran
libraries. 
Using Conda or Mamba the installation proceeds as following:
\begin{lstlisting}[style=mybash]
conda create -n myenv  #create a virtual environment called "myenv"
conda activate myenv   #activate it
conda install -c conda-forge -c edipack edipack #install edipack
\end{lstlisting}

\noindent
When the {\tt EDIpack2py}    module is imported, it attempts to load the dynamic library {\tt
  libedipack\-\_cbindings.so (.dylib)} containing the Fortran-C bindings
for \NAME. By default the library search proceeds as follows: 
\begin{enumerate}
\item The user can override the location of the library
  (determined during the \NAME build configuration) by exporting an
  environment variable called {\tt EDIPACK\_PATH}.
\item By default, the Python module detects the location of the
  Fortran libraries via {\tt pkg-config}. Any of the loading methods
  outlined in \secu{sSecInstallOSloading} automatically pushes the
  correct configuration to {\tt PKG\_CONFIG\_PATH}. 
\item As a last resort, the environment variables {\tt
    LD\_LIBRARY\_PATH} and {\tt DYLD\_LIBRARY\_PATH} are analyzed to
  retrieve the correct location. 
\end{enumerate}
If none of the previous attempts succeeds, the module will not load correctly and an error message will be printed.

\subsubsection{Implementation}\label{sSecInteropEDIpyImplementation}
The Python API provided in the EDIpack2py module consists essentially of a
class called for convenience {\tt global\_env}.
This class contains all the global variables inherited from the \NAME
C-bindings library, and implements a number of interface functions
leveraging the Python duck typing to \NAME.  
The variables and the functions of \NAME are exposed to the user and
are accessed as properties and
methods of the {\tt global\_env} class.

The {\tt global\_env} class needs to be imported at the beginning
of the Python script, along with other useful modules. Numpy is
necessary, while mpi4py is strongly recommended.

\begin{lstlisting}[style=mypython]
import numpy as np
import mpi4py
from mpi4py import MPI
from edipack2py import global_env as ed
import os,sys
\end{lstlisting}

The EDIpack2py module supports the solution of problems with independent
impurities, interfacing to the EDIpack2ineq extension of the
library, if present. Should the inequivalent impurities package not be
built, the Python module silently disables the support to it, so that
invoking any related procedure  will result in a {\tt RuntimeError}.
The user can check the availability of the inequivalent impurities
interface by querying the value of  {\tt edipack2py.global\_env.has\_ineq}.

The implementation of the Python API is divided into two main parts. The
first is a set of global variables, the second includes 4 groups of
functions: solver, bath, input/output, auxiliary. 

\paragraph{{\bf Global variables}.}
This includes a subset of the input variables available in \NAME which
are used to control the calculation.
The variables are loaded globally in {\tt EDIpack2py} and can be accessed
or set locally as properties of the class {\tt
  global\_env}. The global variables are initialized, alongside the remaining
default input variables, through a call to procedure {\tt
  edipack2py.global\_env.read\_input} which interface the {\tt
  ed\_read\_input} function in \NAME. 
A given example is reported in the following code extract:

\begin{lstlisting}[style=mypython]
import numpy as np
from edipack2py import global_env as ed
ed.Nspin = 1            # set a global variable
mylocalvar = ed.Nspin   # assign to a local variable
print(ed.Nspin)         # all functions can have global variables as arguments
np.arange(ed.Nspin)     # array of integers from 0 to Nspin-1
\end{lstlisting}

\paragraph{{\bf Solver functions}.}
This group includes a number of functions enabling initialization,
execution and finalization  of the \NAME solver.
\begin{itemize}
  \item {\tt init\_solver} and {\tt set\_Hloc}. The first 
    initializes the \NAME environment for the quantum impurity problem
    solution, sets the effective bath either reading it from a file or
    initializing it from a flat band model. Once this function is
    called, it is not possible to allocate a second instance of the solver.
    {\tt set\_Hloc} sets the
    non-interacting part of the impurity Hamiltonian $h^0_{\a\b\s\s'}$. 
    Either function takes different argument combinations there
    including support for inequivalent impurities.

  \item {\tt solve} This function solves the quantum impurity problem,
    calculates the observables and any dynamical correlation
    function. All results remain stored in the memory and can be accessed
    through input/output functions.

  \item {\tt finalize\_solver} This function cleans up the \NAME
    environment, frees the memory deallocating all relevant arrays and
    data structures. A call to this functions enables a new
    initialization of the solver, i.e. a new call to {\tt
      init\_solver}.  
  \end{itemize}

\paragraph{{\bf Bath functions}.}
This set covers the implementation of utility functions handling the
effective bath on the user side as well as interfaces to specific
\NAME procedures, either setting bath properties or applying
conventional symmetry transformation. Here we discuss 
a pair of crucial functions in this group.
\begin{itemize}
\item {\tt bath\_inspect} This function translates between the
  user-accessible continuous bath array and the bath components
  (energy levels, hybridization and so on). It functions in both ways,
  given the array returns the components and vice-versa. It
  autonomously determines the type of bath and ED mode.

\item {\tt chi2\_fitgf}
  This function fits the Weiss field or hybridization function ($\Delta$)
  with a discrete set of levels. The fit parameters are the bath
  parameters contained in the user-accessible array. Depending on the
  type of system we are considering (normal, superconductive,
  non-SU(2) symmetric) a different set of inputs has to be passed. The specifics
  of the numerical fitting routines are controlled in the input file.
  \end{itemize}

Additionally, the group includes the function {\tt
  get\_bath\_dimension}, returning the correct
dimension for the user bath array to be allocated and {\tt
  set\_H{replica/general}}, which sets the matrix basis
$\vec{\Gamma}$ and initializes the bath variational parameters
$\vec{\lambda}$ for {\tt bath\_type=replica,general}.

\paragraph{{\bf Input/Output functions}.}
This group includes functions that return to the userspace
observables or dynamical correlation functions evaluated in \NAME and
stored in the corresponding memory instance. Each function
provides a general interface, which encompasses all dimension of the
input array there including inequivalent impurities support.
For example, the function {\tt get\_sigma} returns the self-energy
function array (evaluated on-the-fly) for a specified supported shape,
normal or anomalous type and on a specific axis or frequency domain.

\paragraph{{\bf Auxiliary functions}.}
This group includes some auxiliary functions, either interfacing \NAME
procedures or defined locally in Python to provide specific new
functionalities. Among the latter we include {\tt get\_ed\_mode}, which returns an integer index depending on the value of the variable {\tt ed\_mode=normal,superc,nosu2},
and {\tt get\_bath\_type}, which works similarly for {\tt bath\_type}.


\subsection{EDIpack2TRIQS: the TRIQS interface}\label{sSecInteropTRIQS}
A thin compatibility layer between \NAME and TRIQS \cite{Parcollet2015CPC}, i.e. Toolbox for Research on
Interacting Quantum Systems,  called EDIpack2TRIQS is available as a stand-alone project. This is a pure Python package
built upon EDIpack2py (see \secu{sSecInteropEDIpy}) that provides a
limited object-oriented interface to the most important features of \NAME.

EDIpack2TRIQS strives to offer seamless interoperability with program tools
based on TRIQS by adopting data types, conventions and usage patterns
common to other TRIQS-based impurity solvers, such as
TRIQS/CTHYB \cite{Seth2016CPC}. It also enables execution of \NAME calculations
in the MPI parallel mode with no extra effort from the user.
Detailed documentation of the package can be 
found online at \href{https://krivenko.github.io/edipack2triqs/}
{krivenko.github.io/edipack2triqs}.

\subsubsection{Installation}\label{sSecInteropTRIQSInstallation}
The package depends on \NAME, and its dependencies therein, EDIpack2py
and TRIQS version 3.1 or newer. Assuming the three prerequisites are correctly 
installed in the system, the current development version of EDIpack2TRIQS can be 
installed with {\tt pip} from its 
\href{https://github.com/krivenko/edipack2triqs}{GitHub repository} as follows:

\begin{lstlisting}[style=mybash]
git clone https://github.com/krivenko/edipack2triqs
cd edipack2triqs
pip install .
\end{lstlisting}

\paragraph{Anaconda.}
Another option for installing the package is by using the Anaconda package 
manager on Unix/Linux systems.
The following commands will create a new {\tt conda} environment named 
`edipack' and install the most recently released version of EDIpack2TRIQS along 
with its dependencies (\NAME, EDIpack2py and the TRIQS libraries).

\begin{lstlisting}[style=mybash]
conda create -n edipack
conda activate edipack
conda install -c conda-forge -c edipack edipack2triqs
\end{lstlisting}

\subsubsection{Implementation}\label{sSecInteropTRIQSImplementation}
The programming interface of EDIpack2TRIQS is built around the  
singleton Python class {\tt EDIpackSolver}, defined in the module
{\tt edipack2triqs.solver}, whose instance represents the internal state of 
the \NAME library and exposes its functionality through a number of 
attributes and methods.

\paragraph{{\bf Constructor}.}
The constructor of {\tt EDIpackSolver} accepts the Hamiltonian to be diagonalized
in the form of a second-quantized fermionic operator, the TRIQS {\tt Operator}
object described in Sec.~8.7 of Ref.~\onlinecite{Parcollet2015CPC} (impurity 
problems involving bosons are not supported in the current version). In 
addition to the Hamiltonian, four {\it fundamental operator sets}
{\tt fops\_imp\_up}, {\tt fops\_imp\_dn}, {\tt fops\_bath\_up} and
{\tt fops\_bath\_dn} must be provided.
Each of the sets contains pairs of labels {\tt (b,j)} carried
by the operators $c^\dagger_{b,j} / c_{b,j}$ corresponding to either impurity
or bath electronic degrees of freedom with a certain spin projection.
Having these two crucial pieces of information, the constructor initializes
the underlying Fortran library and automatically selects the appropriate
{\tt ed\_mode} (see \secu{sSecQNs}) and
{\tt bath\_type} (see \secu{sSecBath}). In addition, the
constructor accepts a vast array of keyword arguments that allow for fine-tuning of
the diagonalization process. Among others, these include the ED algorithm
selection, the quantum numbers to be used, parameters of the Krylov space,
the spectrum cutoff and various tolerance levels.

\paragraph{{\bf Method {\tt solve()}}.}
The method {\tt solve()} calls
{\tt edipack2py.global\_env.solve()} and therefore performs the bulk of
calculations. It accepts the inverse temperature $\beta$ required to calculate
the expectation value of physical observables, and parameters of energy grids
for Green's function calculations.

\paragraph{{\bf Input parameters}.}
It is possible to read off and change parameters of the Hamiltonian between
successive calls to {\tt solve()} via respective attributes of {\tt 
EDIpackSolver}. Note, however, that the changes that would necessitate a change of {\tt ed\_mode} or {\tt bath\_type} are disallowed. The relevant attributes
are the following.
\begin{itemize}
    \item {\tt nspin} --- the number of non-equivalent spin projections 
          (read-only).
    \item {\tt norb} --- the number of impurity orbitals (read-only).
    \item {\tt hloc} --- matrix $h^0_{\alpha\beta\s\s'}$ of
          the quadratic impurity Hamiltonian \equ{Himp}.
    \item {\tt U} --- 8-dimensional array $U_{\alpha\s_1,\beta\s_2,\gamma\s_3,\delta\s_4}$
          of two-particle interactions as defined in \equ{HintUmat}.
    \item {\tt bath} --- an object representing the bath. Depending on the
          bath type selected upon solver object construction, this object is an
          instance of {\tt BathNormal}, {\tt BathHybrid} or {\tt BathGeneral}
          (all three classes are members of the module
          {\tt EDIpackSolver.bath}). The bath objects support basic arithmetic
          operations so that a mixing scheme within a DMFT calculation can be
          easily implemented. The way to access individual bath parameters is
          specific to each of the three classes, and is described in detail in the       \href{https://krivenko.github.io/edipack2triqs/documentation.html\#module-edipack2triqs.bath}{online API reference}.
\end{itemize}

\paragraph{{\bf Calculation results}.}
After a successful invocation of {\tt solve()}, one can extract results of the 
calculation from the following attributes.
\begin{itemize}
    \item {\tt e\_pot}, {\tt e\_kin} --- thermal average of the potential 
          (interaction) and kinetic energy respectively.
    \item {\tt densities}, {\tt double\_occ} --- lists of average densities and
          double occupancies of impurity orbitals.
    \item {\tt magnetization} --- Cartesian components of the average impurity  
          magnetization vectors, one row per orbital.
    \item {\tt superconductive\_phi} --- Matrix of impurity superconductive
          order parameters
          $\phi_{\alpha\beta} = \langle c_{\alpha\up} c_{\beta\down}\rangle$
          in orbital space.
    \item {\tt g\_iw}, {\tt g\_an\_iw} --- Normal and anomalous components of
          the Matsubara impurity Green's function.
    \item {\tt Sigma\_iw}, {\tt Sigma\_an\_iw} --- Normal and anomalous
          components of the Matsubara impurity self-energy.
    \item {\tt g\_w}, {\tt g\_an\_w} --- Normal and anomalous components of
          the real-frequency impurity Green's function.
    \item {\tt Sigma\_w}, {\tt Sigma\_an\_w} --- Normal and anomalous
          components of the real-frequency impurity self-energy.
\end{itemize}

The Green's functions are returned as TRIQS {\tt BlockGf} containers
with names of the individual blocks determined from the block labels {\tt 'b'}
found in {\tt fops\_imp\_up}, {\tt fops\_imp\_dn} and from solver's
{\tt ed\_mode}. The Matsubara and real-frequency meshes of the TRIQS GF 
containers (see Sec.~8.2 of Ref.~\onlinecite{Parcollet2015CPC}) are constructed
according to the parameters passed to {\tt solve()}.
The anomalous components of the Green's functions and self-energies are only 
available when anomalous bath terms are present in the Hamiltonian.
If this is not the case, an attempt to access these attributes results in a
{\tt RuntimeError}.

\paragraph{{\bf Bath parameter fitting}.}
The method {\tt EDIpackSolver.chi2\_fit\_bath()} is essentially a wrapper around
{\tt edipack2py.global\_env.chi2\_fitgf()}. It accepts the function to fit (either
the hybridization function or the Weiss field) in the {\tt BlockGf} format,
and returns the parameter fit result as a {\tt Bath*} object along with a
{\tt BlockGf} representation of the fitted function. In the superconducting
case ({\tt ed\_mode=superc}), this method accepts and returns pairs of
the {\tt BlockGf} containers corresponding to the normal and anomalous 
components of the quantities in question.

\subsection{w2dynamics interface}\label{sSecInteropw2dyN}
\NAME is supported as an alternative impurity solver in the w2dynamics DMFT package \cite{Wallerberger2019CPC}, available on \href{https://github.com/w2dynamics/w2dynamics}{github.com/w2dynamics/w2dynamics}. 
This integration enables users to 
seamlessly switch from the default hybridization expansion continuous-time QMC  (CT-HYB), included natively
in w2dynamics, to the \NAME ED solver. This requires no changes to the input file and at most minor  adjustments to configuration parameters.

\subsubsection{Installation}\label{sSecInteropw2dyNPrereq}
Use of the interface requires a current version of w2dynamics ($\geq 1.1.6$) and a working installation of EDIpack2py (see 
\secu{sSecInteropEDIpy}). 
Users may build and optionally install w2dynamics using conventional CMake based source installation: 
\begin{lstlisting}[style=mybash]
git clone https://github.com/w2dynamics/w2dynamics.git
cd w2dynamics
mkdir build
cd build
cmake ..
make install
\end{lstlisting}
This enables the default CT-HYB solver. For DMFT calculations 
using the \NAME interface, it is sufficient that the 
\texttt{EDIpack2py} module is available for import at runtime, and no compilation is required on the w2dynamics side.

\subsubsection{Implementation}\label{sSecInteropw2dyNPrereq}
The \NAME interface is implemented as the class \texttt{EDIpackSolver} 
in the Python module \texttt{w2dyn.dmft.edipack\_solver}. It is a subclass of 
\texttt{ImpuritySolver} and can be used as a drop-in replacement for 
\texttt{CtHybSolver}. The w2dynamics DMFT solver provides the hybridization 
function and local Hamiltonian for the auxiliary impurity problem via 
an \texttt{ImpurityProblem} instance passed to \texttt{set\_problem}. It then 
it invokes the \texttt{solve} method to obtain the results, including, 
e.g., the Green's function, as an \texttt{ImpurityResult}.
The \texttt{solve} method sets up the calculation using \NAME by calling 
\texttt{EDIpack2py}, writing input files to a subdirectory, running 
\NAME, and processing both the return values of \texttt{EDIpack2py} 
methods and the output files. It then formats the results following the w2dynamics conventions.

This implementation allows w2dynamics to abstract over the specific 
choice of impurity solver as much as possible. \NAME is only called 
to solve individual impurity problems, while the standard w2dynamics 
code handles higher-level tasks such as the DMFT loop, support for 
multiple inequivalent impurities, and the user-facing interface via 
standard w2dynamics input files and output in its usual HDF5 format.

As a result, features that require explicit support from w2dynamics 
but are not yet implemented cannot be used through the \NAME interface. 
In particular, solving impurity problems in the superconducting phase 
(\texttt{ed\_mode=superc}) is not yet  supported.

\subsubsection{Configuration and Usage}\label{sSecInteropw2dyNConfig}
w2dynamics reads its configuration parameters from a file named \texttt{Parameters.in} by default. To use \NAME methods, the {\tt solver} parameter in the section {\tt General} needs to be set as follows:
\begin{lstlisting}[style=mybash]
[General]
solver = EDIPACK
\end{lstlisting}
Additional parameters specific to \NAME are defined in the accessory section \texttt{[EDIPACK]}. These set many of the \NAME's corresponding input variables (in  uppercase letters). For example,  the number of bath sites {\tt NBATH} can be configured as:
\begin{lstlisting}[style=mybash]
[EDIPACK]
NBATH = 7
\end{lstlisting}

This section includes also the options to control the bath optimization and  diagonalization algorithms. Other input variables that define the model (e.g., {\tt NORB}, {\tt ULOC}, and {\tt BETA}) or relate to functionalities handled by 
w2dynamics itself (e.g., \texttt{NLOOP}) must be provided through 
standard parameters in other sections, such as \texttt{[General]} or 
\texttt{[Atoms]}. An example of a complete properly formatted input file is provided in \secu{SecExamplesBetheDMFTW2D}.
The full set of configuration parameters can be found in the {\tt configspec} file, located in the {\tt w2dyn/auxiliaries} of the w2dynamics repository, see  \cite{Wallerberger2019CPC} for further details.

A calculation can be launched by running the w2dynamics DMFT program 
\texttt{DMFT.py}, which supports parallel execution via MPI to take 
advantage of \NAME's parallelization.

\subsubsection{Output and Results}\label{sSecInteropw2dyNOutput}
Calculation results are stored in an HDF5 file, grouped by DMFT iteration (with group names formatted as {\tt dmft-001}) and inequivalent impurity (for impurity-specific data, using group names like 
\texttt{ineq-001}). 
The content of this file can be accessed using the 
\texttt{hgrep} script provided with w2dynamics or other HDF5 tools. 
Stored quantities include, for example, the impurity self-energy (in 
datasets \texttt{siw-full}) and Green's function (in 
\texttt{giw-full}) on the Matsubara frequency axis, as well as single 
and double occupations (in \texttt{occ}). Additionally, the interface 
provides access to results not available when using the CT-HYB solver, 
such as the impurity self-energy (\texttt{somega}) and Green's function 
(\texttt{gomega}) on the real frequency axis.
Standard \NAME output files for each individual ED impurity solution are also available in subdirectories named like the the main HDF5 output file with iteration and impurity numbers appended.

\subsection{EDIpack2jl, the Julia API}\label{sSecInteropEDIjl}
The C-binding approach is extremely handy, in that it opens the way for interoperability  with a large number of languages and frameworks. 
As a further significant example, an {\bf experimental} Julia API 
is provided for \NAME. Although this is at the present a proof of concept, it is capable of replicating the results obtained with Fortran, C++ and
Python implementations for the Bethe lattice example driver (see \secu{SecExamples}).
Its structure and operation mimic that of the EDIpack2py layer, with minimal language-specific differences.

The \NAME{} Julia API consists of a module called {\tt EDIpack2jl}, which provides access to the global variables and functions contained in {\tt  libedipack\_cbindings.so (.dylib)}. 
The library is searched upon loading the module, with the following order of priority:

\begin{itemize}
\item {\tt EDIPACK\_PATH}: if this environment variable is set, {\tt EDIpack2jl} will look there for the library first
\item {\tt LD\_LIBRARY\_PATH}: for Linux systems
\item {\tt DYLD\_LIBRARY\_PATH}: for macOS systems
\end{itemize}

\subsubsection{Structure}

In partial analogy to the Python API, the global variables, but not the functions, are contained in a  {\tt struct} called {\tt global\_env}.

Similarly to the C++ case, global variables and functions are accessed as raw pointers. As a consequence, it is important that the dimensions of the array-like variables (such as {\tt ULOC}) and, in general, the amount of memory occupied by each variable are correctly accounted for. This is achieved within the Julia module by appropriately casting the variables to compatible types, such as {\tt  Cint, Cbool, Cdouble}.
Functions are called from the dynamic library, making use of the {\tt  ccall} procedure. As previously stated, as a consequence of the C linking conventions multiple alternative version of the interfaced Fortran procedures are present, to account for the different input variable combinations; the EDIpack2jl wrapper functions take care of selecting the appropriate Fortran procedure depending on the set of input parameters provided by the user.

\subsubsection{Installation and usage}

At present, the \NAME Julia interface is not offered as a package. The git repository has to be cloned via

\begin{lstlisting}[style=mybash]
git clone https://github.com/EDIpack/EDIpack2jl.git
\end{lstlisting}
and the location of the source files needs to be included in the user program via
\begin{lstlisting}[style=myjulia]
push!(LOAD_PATH, joinpath(@__DIR__, "PATH/TO/REPO/src"))
\end{lstlisting}

The EDIpack2jl module can then be loaded via
\begin{lstlisting}[style=myjulia]
using EDIpack2jl
\end{lstlisting}
The correct way to access global variables and functions is as in the following example:
\begin{lstlisting}[style=myjulia]
EDIpack2jl.read_input("inputED.conf")
ed = EDIpack2jl.global_env
println("Nspin = ", ed.Nspin)
ed.Nspin = 2
\end{lstlisting}

The names and inputs of the Julia-wrapped functions are entirely analogous to those of the Python API.

We provide an example script for the simple case of the Bethe lattice in the normal phase in the {\tt examples} folder of the EDIpack2jl repository. This script is intended to be run serially. Documentation of the API and further examples are being developed.

\ifSubfilesClassLoaded{
  \bibliography{references}
}{}
\end{document}

%% file: 05_examples.tex
\section{Examples}\label{SecExamples}
In this section we illustrate and benchmark the functionalities and of  \NAME
library and its interfaces as a solver for DMFT through a variety of examples. We also discuss in detail the relevant code parts from different programming languages. 
The codes, the data and the scripts to rework some of the examples presented here, as well as additional upcoming tutorials, can be found in the repository \href{https://github.com/EDIpack/EDIpack2examples}{github.com/EDIpack/EDIpack2examples}.  

Some tasks---particularly those associated with the implementation of the DMFT self-consistency---are inherently independent of \NAME itself, and the impurity solver is agnostic about them. 
Thus, in the implementations using Fortran, Python or other \NAME interfaces to scientific toolboxes, we adopt a reverse communication strategy, which requires the user to independently carry out these parts or, equivalently, to take advantage of existing utilities. 

In our examples based on the Fortran APIs, we rely on external open-source libraries, such as DMFTtools (see \href{https://github.com/aamaricci/DMFTtools}{github.com/aamaricci/DMFTtools}), to perform specific tasks, including evaluating the local Green’s function, implementing the self-consistency, constructing the tight-binding model, or calculating the kinetic energy.
Where appropriate, we include comments that reference these external procedures.

On the other hand, the results obtained with w2dynamics rely on the internal framework provided by the software itself -- see the script \texttt{DMFT.py} \cite{Wallerberger2019CPC}, which seamlessly handles all these tasks.
Similarly, the TRIQS library offers a comprehensive framework for manipulating Green’s functions and related quantities \cite{Parcollet2015CPC}, making these tasks easily accessible while maintaining very high standards of optimization and accuracy.

\subsection{Hubbard model on the Bethe lattice (Fortran/C++ API, {\tt ed\_mode=normal})}\label{SecExamplesBetheDMFT}
The description of the Mott transition using the Hubbard model on the Bethe lattice is conventionally regarded as the standard test bed for any DMFT implementation.  \cite{Georges1996RMP,Rozenberg1999PRL,Kotliar1999EPJB,Kotliar2000PRL,Kotliar2002PRL}.
Here, we present a guided implementation of the DMFT 
solution for the Bethe lattice at $T=0$ using \NAME as impurity
solver.
The model under consideration is the Fermi-Hubbard Hamiltonian:
$$
H = -t \sum_{\langle ij\rangle,\s} c^\dagger_{i\s} c_{j\s} + 
    U \sum_i n_{i\up}n_{i\dw},
$$
where $c^\dagger_{i\s}$ ($c_{i\s}$) are the creation (annihilation) 
operators for an electron at site $i$ with spin $\s$, and 
$n_{i\s} = c^\dagger_{i\s} c_{i\s}$ is the corresponding occupation 
operator. The first sum runs over nearest-neighbor pairs
$\langle ij \rangle$.
We consider the system defined on a Bethe lattice with density of states
$\rho(\e)=\frac{2}{\pi D^2}\sqrt{D^2-\e^2}$,
where $D=2t$ is the half-bandwidth.
Within DMFT framework \cite{Georges1996RMP}, this lattice model is mapped onto a quantum impurity problem with an effective electronic bath that must be determined self-consistently.

Below, we discuss the key components of a basic DMFT implementation 
using \NAME for the Bethe lattice, employing either the Fortran or 
C++ APIs and which can be found 
in the {\tt examples} directory of the \NAME source code. Both 
samples share a similar initial structure, including memory 
allocation, creation of the Bethe DOS and solver initialization:
 \begin{center}
\begin{minipage}[t]{0.49\linewidth}
\textbf{Fortran}
\begin{lstlisting}[style=fstyle,frame=none,numbers=none,basicstyle={\scriptsize\ttfamily}]
program ed_hm_bethe
   USE EDIPACK
   USE SCIFOR
   implicit none
   integer               :: Le=1000
   real(8)               :: wmixing=0.5d0
   real(8)               :: D=1d0
   integer               :: Nb
   real(8),allocatable   :: Bath(:)
   complex(8),allocatable:: Hloc(:,:,:,:)
   real(8),allocatable   :: Ebands(:),Dbands(:)
   complex(8),allocatable:: Smats(:,:,:,:,:)
   complex(8),allocatable:: Delta(:,:,:,:,:)
   !...  
   !EDIpack: Read variables
   call ed_read_input('inputED.conf')
   
   !Solver-specific arrays. Using Rank-5  
   allocate(Smats(Nspin,Nspin,Norb,Norb,Lmats))
   allocate(Delta(Nspin,Nspin,Norb,Norb,Lmats))
   allocate(Hloc(Nspin,Nspin,Norb,Norb))
   Hloc=0d0
   !...
   !Construct Bethe DOS using SciFortran
   !de = 2*D/(Le-1)
   allocate(Ebands(Le),Dbands(Le))
   Ebands = linspace(-D,D,Le,mesh=de)
   Dbands = dens_bethe(Ebands,D)*de
   !
   !EDIpack: Set impurity Hamiltonian
   call ed_set_hloc(Hloc)
\end{lstlisting}
\end{minipage}
\begin{minipage}[t]{0.49\linewidth}
\textbf{C++}
\begin{lstlisting}[style=cstyle,frame=none,numbers=none,basicstyle={\scriptsize\ttfamily}]
#include <edipack_cbindings.h>
using namespace std;
//...
//Main variables
int Le = 1000;
int iloop = 0;
double wmixing = 0.5;
double D = 1.0;

//EDIpack: Read  variables    
char input[] = "inputED.conf"; 
read_input(input);      
//Dimensions
int64_t d[4] = {Nspin,Nspin,Norb,Norb};
int total_size = d[0] * d[1] * d[2] * d[3];
int total_size_n5 = total_size * Lmats;    
//Solver-specific arrays rank5
vector<complex<double>> Hloc(total_size);
vector<complex<double>> Smats(total_size_n5);
vector<complex<double>> Delta(total_size_n5);
//...
//Construct Bethe DOS
vector<double> Ebands, Dbands;

//Locally defined functions: de = 2*D/(Le-1)
Ebands = linspace(-D,D,Le);
Dbands = dens_bethe(Ebands,D,de);

//EDIpack: Set impurity Hamiltonian
ed_set_Hloc_single_N4(Hloc.data(), d);
\end{lstlisting}
\end{minipage}
 \end{center}

The bath is described by the (unknown) function $\GG^{-1}_0$, i.e. 
the Weiss field ({\tt Weiss}). In the ED method implemented in \NAME, 
the bath is approximated using a finite number of discrete energy levels. 
The Weiss field $\GG^{-1}_0$ is then used to determine the bath parameters $\vec{x} = \{\hat{V}, \hat{h}\}$ through the optimization method outlined in \secu{sSecFit}.

The starting point for any calculation is a reasonable initial guess 
for the Weiss field, or equivalently, the bath parameters. In \NAME, 
this is accomplished using the function {\tt
  ed\_init\_solver} (Fortran API) or {\tt init\_solver\_site} (C++ API).

\begin{center} 
\begin{minipage}[t]{0.49\linewidth}
\textbf{Fortran}
\begin{lstlisting}[style=fstyle,frame=none,numbers=none,basicstyle={\scriptsize\ttfamily}]
   !EDIpack: Initialize solver
   Nb=ed_get_bath_dimension()
   allocate(bath(Nb))
   call ed_init_solver(bath)
$\phantom{.}$   
$\phantom{.}$   
\end{lstlisting}
\end{minipage}
\begin{minipage}[t]{0.49\linewidth}
\textbf{C++}
\begin{lstlisting}[style=cstyle,frame=none,numbers=none,basicstyle={\scriptsize\ttfamily}]
//EDIpack: Initialize solver
int Nb;
vector<double> Bath(Nb);
Nb = get_bath_dimension_direct();
int64_t bath_dim[1] = {Nb};
init_solver_site(Bath.data(), bath_dim);
\end{lstlisting}
\end{minipage}
\end{center}

The iterative algorithm to solve the DMFT problem proceeds as follows:
\begin{itemize}  
\item[{\tiny {\bf EDIpack}}] Call the \NAME {\bf impurity solver}
  whose only input is the set of parameters $\vec{x}$ contained in a
  rank-1 array. All  \NAME options are controlled through input file
  specifications.
  
\item[{\tiny {\bf EDIpack}}] Retrieve the self-energy functions $\Sigma_{\a\b\s\s'}(i\omega)$ on the
  Matsubara axis using dedicated function {\tt ed\_get\_sigma} available in the \NAME API.
  
\item[{\tiny {\it User}}] Evaluate the local interacting Green's function
$$
G_\mathrm{loc}(i\omega) = \int_{-D}^D \frac{\rho(\e)}{\zeta -\e}d\e
$$ 
with  $\zeta=i\omega+\mu-h^0-\Sigma(i\omega)$. Note that this step can be performed analytically for the Bethe lattice \cite{Georges1996RMP} or completely substituted by retrieving the impurity Green's function $G_\mathrm{imp}$ via the {\tt ed\_get\_gimp} function.
  
\item[{\tiny {\it User}}] Update the Weiss field via the {\bf self-consistency}
  relation: $\GG^{-1}_0(i\omega) = G^{-1}_\mathrm{loc}(i\omega) +
    \Sigma(i\omega)$. For the Bethe lattice, this simplifies to
    $\Delta = \tfrac{D^2}{4}G_\mathrm{loc}$ or, equivalently, $\Delta = \tfrac{D^2}{4}G_\mathrm{imp}$.
    
  \item[{\tiny {\it User} \textgreater\ {\bf EDIpack}}] Optimize the bath parameters $\vec{x}$ against the updated
    Weiss field using the conjugate gradient procedures supplied by \NAME. Then, restart at step 1.
\end{itemize}

The first two steps are handled directly by \NAME routines, while during the subsequent steps the 
 control returns to the user, who must implement the algebraic updates required to 
close the self-consistency loop and optimize the bath.
Given the critical importance of this step, \NAME provides access to a well-tested implementation of the conjugate gradient method for performing the bath optimization, ensuring stability and reproducibility of the results. This task is conceptually distinct from the  diagonalization of the impurity problem, which remains the core focus of the package.
Alternative optimization methods can also be employed as 
needed \cite{Mejuto-Zaera2020PRB,Huang2023PRB,Nakatsukasa2018SJOSC}. 
An example of implementation is provided in the following listings.

\begin{center} 
\begin{minipage}[t]{0.49\linewidth}
\textbf{Fortran}
\begin{lstlisting}[style=fstyle,frame=none,numbers=none,basicstyle={\scriptsize\ttfamily}]
!DMFT loop
do while(.not.converged.AND.iloop<nloop)
    iloop=iloop+1     
    
    !EDIpack: Call ED solver
    call ed_solve(bath)     
    !EDIpack: Retrieve ${\color{comment-color} \Sigma(i\omega_n)}$
    call ed_get_sigma(Smats,'m')
    
    !Build local Green's function
    wfreq = pi/beta*(2*arange(1,Lmats)-1)
    do i=1,Lmats
       zeta= xi*wfreq(i)+xmu - Smats(1,1,1,1,i)
       Gmats(1,1,1,1,i)=sum(DOS/(zeta-Ene))*de  
    enddo    
    
    !Self-consistency
    Delta = 0.25d0*D*Gmats
    
    !Fitting -> new bath
    call ed_chi2_fitgf(Weiss,bath,ispin=1)    
    
    !Check convergence: from ${\color{comment-color}\mathrm{DMFTtools}}$
    converged=check_convergence(Delta,&
          dmft_error,Nsuccess,Nloop)          
enddo
\end{lstlisting}
\end{minipage}
\begin{minipage}[t]{0.49\linewidth}
\textbf{C++}
\begin{lstlisting}[style=cstyle,frame=none,numbers=none,basicstyle={\scriptsize\ttfamily}]
//DMFT loop
while (iloop < Nloop && !converged) {
  //EDIpack: Call ED solver
  solve_site(Bath.data(),bath_dim,1,1);      
  //EDIpack: Retrieve ${\color{comment-color} \Sigma(i\omega_n)}$
  get_sigma_site_n5(Smats.data(),//
      0,0,wm.data(),Lmats,0);
  //Build local Green's function
  for (int i=0;i<Lmats;++i) {
    zeta= wm[i] + xmu - Smats[i];
    Gmats[i] = complex<double>(0.0,0.0);
    for (int j=0; j< Le; j++) {
      Gmats[i]+=Dbands[j]/(zeta-Ebands[j]);
    }
  }
  //Self-consistency
  for (int i = 0; i < Lmats; ++i) {
    Delta[i] = 0.25 * D * Gmats[i];
  }
  //Fit -> new bath
  chi2_fitgf_single_normal_n5(Delta.data(),//
      delta_dim,Bath.data(),bath_dim,1,0,1);
  //Check convergence: local functions
  converged = check_convergence(Delta,//
      dmft_error, Nsuccess, Nloop, comm);  
}
\end{lstlisting}
\end{minipage}
\end{center}

\paragraph{Results.}
In the following, we present \NAME results for the interaction-driven MIT obtained with previous implementations. The MIT captures the gradual transformation of a partially-filled metallic state into a correlated insulator.

To illustrate this, we report in panel (A) the evolution of the spectral function $-{\rm Im}G_\mathrm{loc}(\omega)/\pi$ as a function of the  interaction strength $U$. 
Despite the inherently {\it spiky} nature of the spectrum, resulting from the finite number of poles in the 
discretized effective bath, the characteristic features of the Mott 
transition are clearly visible. The results in panel (A) have been obtained using a broadening {\tt eps=0.01}. 
At low energies, a renormalized 
quasi-particle peak develops at the Fermi level ($\omega = 0$). 
Simultaneously, the system exhibits the formation of incoherent 
high-energy features, which eventually evolve into well-defined 
Hubbard bands in the Mott insulating phase for $U > U_\mathrm{c}$, 
with $U_\mathrm{c} \simeq 2.8D$.

\begin{figure}[t!]
  \includegraphics[width=\linewidth]{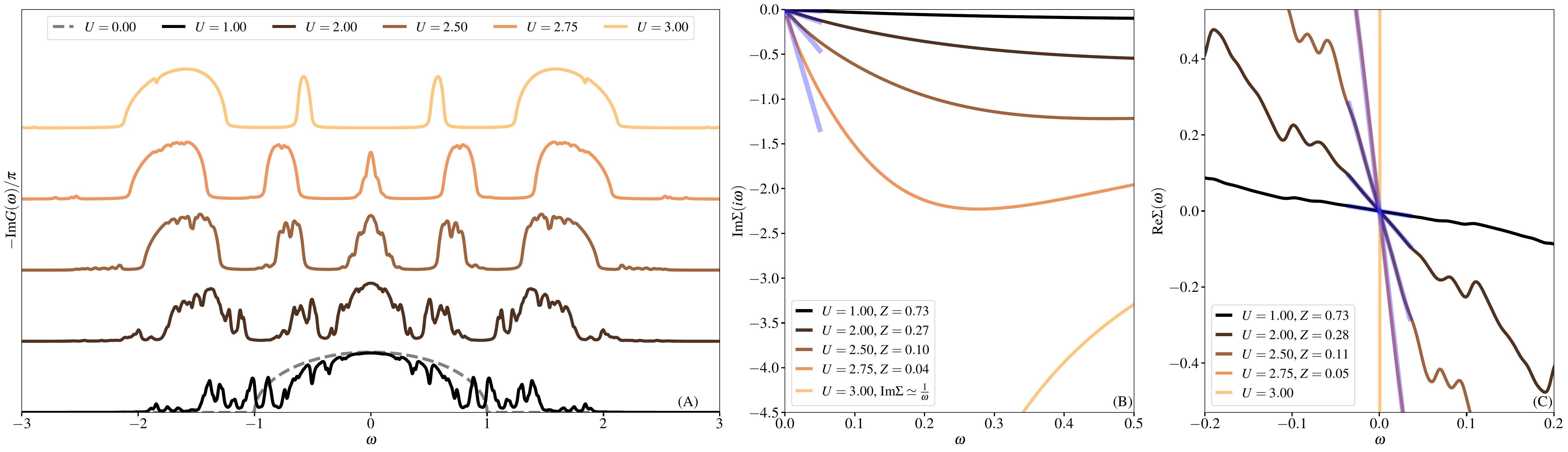}
    \caption{\label{figEx1}%
      \textbf{The metal-insulator Mott transition.}
      (a) Evolution of the spectral function $-\Im{G}(\omega)/\pi$ as
      a function of increasing interaction $U$. The critical
      interaction $U_\mathrm{c}\simeq 2.8D$ separates the correlated metal $U<U_\mathrm{c}$ from
      the Mott insulator $U>U_\mathrm{c}$.
      (b)-(c) The corresponding evolution of the Matsubara self-energy
      $\Im\Sigma(i\omega)$ (b) and
      real-axis one $\Re\Sigma(\omega)$ across the Mott
      transition. For a particle-hole symmetric case, both allow to
      estimate the renormalization constant $Z$ (see main text), using
      linear order expansion in frequency (blue solid lines). The
      values of $Z$ are reported in the legend.
      The Mott insulating solution is associated with a singularity at
      $\omega=0$ of $\Im\Sigma$.
    }
\end{figure}

The formation of a spectral gap separating the Hubbard bands in the 
Mott insulating phase is associated with the divergence of the 
imaginary part of the self-energy at the Fermi level. This divergence 
reflects the complete localization of the electrons, effectively 
suppressing coherent quasi-particle excitations. Causality dictates that the real part of the self-energy must also 
grow significantly near the singularity, making it impossible to 
satisfy the quasi-particle pole equation
$$
\omega+\mu-h^0-\e-\Re{\Sigma}(\omega)=0,
$$
which governs the formation of coherent excitations near the Fermi 
level.
Panels (B) and (C) illustrate this phenomenon by showing the evolution of the self-energy $\Sigma$. In panel (B), we present the Matsubara self-energy ${\rm Im}\Sigma(i\omega)$ in the low-energy regime. As the  interaction strength $U$ increases, this function progressively grows, eventually diverging as the critical interaction threshold $U > U_\mathrm{c}$ is crossed.
This divergence along the Matsubara axis is directly linked to the
particle-hole symmetry of the Bethe lattice, which pins the  ${\rm
  Im}\Sigma$ singularity at $\omega = 0$.
Panel (C) complements this picture by displaying the real part 
$\Re{\Sigma}(\omega)$ on the real-axis near the Fermi level. Here, increasing $U$ leads to a rapid rise of this component, culminating in a discontinuous behavior as the critical point is approached. This 
discontinuity directly reflects the divergence in the imaginary part
on the real-axis, confirming the transition to the Mott insulating state.

A quantitative measure of this transition is provided by the 
quasi-particle renormalization factor $Z$, which can be used to capture the degree of electron delocalization. This parameter ranges from 1 for a non-interacting metal to 0 for a fully localized Mott insulator. It is 
defined through the low-energy expansion of the self-energy as
$$
Z=\left(1-\frac{\partial\Re\Sigma}{\partial\omega}\Biggr|_{\omega\rightarrow
    0}\right)^{-1},
$$
which can also be estimated from the linear behavior of ${\rm Im}\Sigma(i\omega)$ for 
$\omega\to0$ in the metallic regime using the relation:
$$
   \frac{\Im\Sigma(i\omega_n)}{\omega_n}\Biggr|_{\omega_n\rightarrow 0}=
   \frac{1}{\pi}\int_{\mathbb R}d\epsilon \frac{\Re\Sigma(\epsilon)}{\epsilon^2}=
   \frac{\partial\Re\Sigma}{\partial\omega}\Biggr|_{\omega\rightarrow 0}.
$$

The linear fits highlighted in panels (B) and (C), along with the 
corresponding $Z$ values provided in the legends, clearly indicate that 
the slope of the self-energy at low energy increases with $U$ on both 
the Matsubara and real axes. At the transition point, this slope 
diverges, reflecting the onset of complete electron localization as 
$Z \to 0$, consistent with the singular behavior  
$-{\rm Im}\Sigma(\omega\to0) \to\infty$.

\subsubsection{Finite temperature (w2dynamics interface, CT-HYB benchmark)}\label{SecExamplesBetheDMFTW2D}
To demonstrate how the w2dynamics interface integrates with  \NAME, we briefly discuss how to solve the same problem, i.e. the Hubbard model on the Bethe lattice within DMFT using w2dynamics. 
In order to showcase the capabilities of \NAME to address low-temperature problems, we compare the continuous-time Quantum Monte Carlo (CTQMC) solver using the hybridization expansion method (CT-HYB) included in w2dynamics against the \NAME solver at finite temperature.  

Unlike \NAME, which provides only the impurity solver and bath optimization procedures and requires the user to implement the DMFT algorithm themselves (potentially using various methods), w2dynamics adopts a fundamentally different approach: a single Python script, {\tt DMFT.py}, handles the entire DMFT calculation, leveraging dedicated classes that implement the generic self-consistency. The w2dynamics calculation is then entirely controlled by a model-dependent parameters file {\tt Parameters.in}, which contains a number of variable specifications including options to control the ED solver inherited from \NAME. Further information about the functioning of w2dynamics can be found in Ref. \cite{Wallerberger2019CPC}           

Another important difference concerns the initial point of the iterative DMFT solution algorithm: while \NAME starts from a given discrete bath, w2dynamics is initialized with a zero self-energy function or alternatively reads this quantity from a converged solution file. Thus, when using the \NAME interface in w2dynamics, the initial Weiss field is determined using self-consistency and  an initial discrete bath is obtained through the \NAME bath optimization procedure. 

The following is the w2dynamics configuration file used to solve the Bethe lattice problem at finite temperature: 
\begin{lstlisting}[style=mybash,language={},numbers=none,basicstyle={\scriptsize\ttfamily}]
[General]
DOS             = Bethe           # support for the Bethe lattice is built-in
half-bandwidth  = 1               # list of half-bandwidths per orbital
NAt             = 1               # number of impurities
beta            = 100             # inverse temperature
mu              = 1.0             # chemical potential set to achieve half-filling
EPSN            = 0.0             # turns off filling-based chemical potential search
DMFTsteps       = 100             # given no convergence checking, we might want fewer
magnetism       = para            # symmetrize self-energies per spin
FileNamePrefix  = bethe_dmft_U2   # prefix for the output file name
fileold         = bethe_dmft*hdf5 # file to read an initial self-energy from
readold         = 0               # iteration number to read initial self-energy from, 0 turns off
mixing          = 0.5             # mixing, but mixes self-energies and not Weiss fields
mixing_strategy = linear          # linear mixing as in the Fortran example
FTType          = none            # (for CT-HYB): use the NFFT-measured G
solver          = EDIPACK         # use EDIpack as impurity solver, not default CTHYB

[Atoms]
[[1]]                             # one subsection per impurity
Nd              = 1               # number of orbitals
Hamiltonian     = Kanamori        # create a Hubbard-Kanamori interaction Hamiltonian
Udd             = 2.0             # equivalent to ULOC
Vdd             = 1.0             # equivalent to UST, meaningless for 1 orbital
Jdd             = 0.5             # equivalent to JH, JX and JP, also meaningless here

[EDIPACK]                         # further ED parameters, as in the Fortran example
NBATH           = 7               # number of bath sites
ED_TWIN         = True            # use twin symmetry
LFIT            = 2048            # number of Matsubara frequencies used for the bath fit
LANC_NGFITER    = 500             # number of Lanczos iterations for Green's function
CG_FTOL         = 1e-10           # conjugate-gradient tolerance
CG_NITER        = 2048            # maximum number of conjugate-gradient iterations
ED_FINITE_TEMP  = True            # at finite temperature T = 1/beta

[QMC]                             # Parameters for some grid sizes and the CT-HYB calculation
Ntau            = 1024            # imaginary time grid size
Niw             = 4096            # number of positive Matsubara frequencies
# parameters only relevant for CT-HYB follow
MeasGiw         = 1               # enable NFFT measurement of G
NCorr           = 175             # estimate of the autocorrelation length
Nmeas           = 200000          # number of measurements / sample size
Nwarmups        = 1000000         # number of initial warmup steps for Markov chain thermalization
\end{lstlisting}
Assuming the parameters are listed in a plain text file called 
{\tt bethe\_dmft.in}, the DMFT simulation can then by run on {\tt NC} cores via:
\begin{lstlisting}[style=mybash,numbers=none,morekeywords={mpiexec},deletekeywords={in}]
mpiexec -n NC /path/to/w2dynamics/DMFT.py bethe_dmft.in
\end{lstlisting}

The results are collected into an HDF5 \cite{The_HDF_Group_Hierarchical_Data_Format} archive in the usual w2dynamics format, including output quantities inherited from \NAME. Results can be viewed using the script {\tt hgrep} provided with w2dynamics or with other HDF5 tools. In this example we extract the Matsubara self-energy function $\Sigma(i\omega_n)$ for the last DMFT iteration:
\begin{lstlisting}[style=mybash,numbers=none]
/path/to/w2dynamics/hgrep latest siw-full -1
\end{lstlisting}

\begin{figure}
  \includegraphics[width=\linewidth]{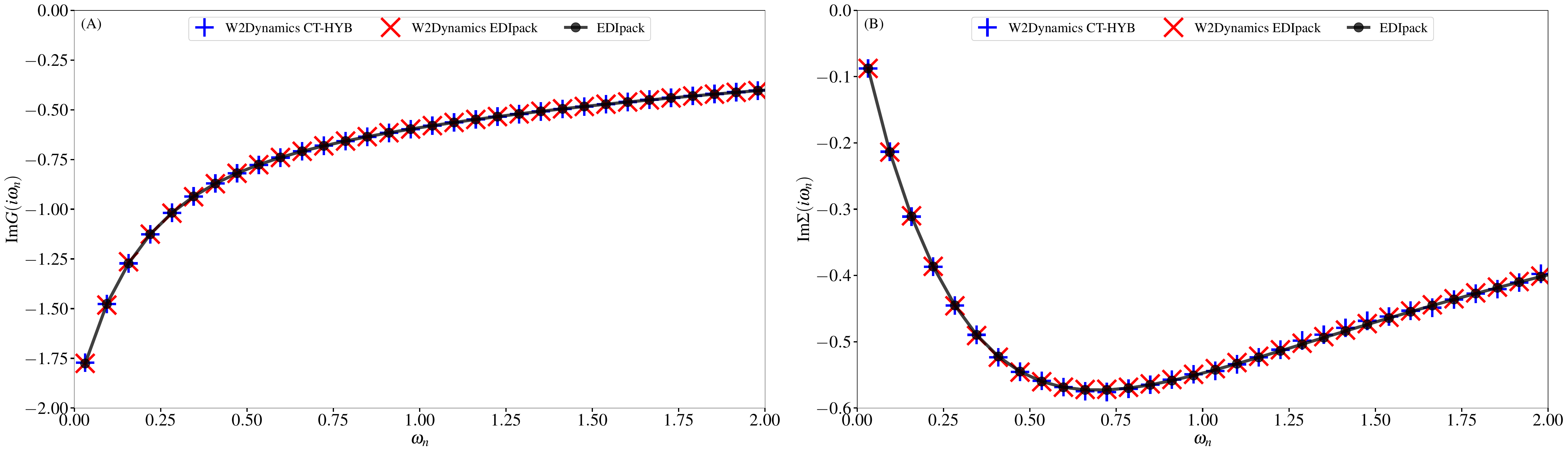}
    \caption{\label{figEx1W}%
    \textbf{Finite temperature DMFT solution.}
    Comparison of the imaginary parts of the Green's function $\Im{G}(i\omega_n)$ and self-energy $\Im{\Sigma}(i\omega_n)$ from different solutions of the Hubbard model on the Bethe lattice using DMFT. Data are for $T/D=0.01$ and $U/D=2.0$. CTQMC results from the solver included with w2dynamics are compared against \NAME ED results, used both through the w2dynamics interface and with a standalone Fortran program. 
        }
\end{figure}

In \figu{figEx1W} we report a comparison of the results obtained using the hybridization expansion CTQMC and the \NAME ED method as solvers in w2dynamics. In addition we compare with results obtained using the Fortran API directly as shown in the previous subsection. Using DMFT, we solve the Hubbard model on a Bethe lattice for $T/D=0.01$ and $U/D=2.00$, corresponding to a correlated metallic state. 

In panel (A) we show the behavior of the imaginary part of the impurity Green's function $\Im{G}(i\omega_n)$ which is the direct output for both methods (note that w2dynamics does not directly have access to the real-axis Green's function when using the CTQMC solver). In panel (B) we show the same comparison for impurity self-energy $\Im{\Sigma}(i\omega_n)$. 
The results obtained from the three calculations are in excellent agreement with each other already for a small number of bath levels ({\tt NBATH=7}) and relatively low QMC statistics ({\tt Nmeas=200000} with {\tt NC=10} parallel processes).

\subsection{Attractive Hubbard model (Python API, {\tt ed\_mode=superc})}\label{SecExamplesAHM}
The second example we present concerns the DMFT description of the 
attractive Hubbard model \cite{Caffarel1994PRL,Toschi2005NJP,Toschi2005PRB} on a two-dimensional square lattice. 
This example has two main goals: (i) to demonstrate \NAME's support for $s$-wave superconductivity, and (ii) to showcase the Python API through a concrete example. 

The 
model Hamiltonian is given by:
$$
H = \sum_{\ka,\s} \epsilon(\ka) c^\dagger_{\ka\s} c_{\ka\s} 
    - U \sum_i n_{i\up} n_{i\dw},
$$
where $U > 0$, $c^\dagger_{i\s}$ 
is the creation operator for an electron at site $i$ with spin $\s$ and $c^\dagger_{\ka\s} = \tfrac{1}{\sqrt{N}} 
\sum_i e^{-i\ka \cdot R_i} c^\dagger_{i\s}$. 
The occupation operator is $n_{i\s} = c^\dagger_{i\s} c_{i\s}$, and 
the energy dispersion relation is $\e(\ka)=-2t[\cos{(k_x a)}+
\cos{(k_y a)}]$, where we set the lattice spacing 
$a=1$  and the choose the energy unit such that $4t=D=1$ for convenience.

The DMFT workflow for this case is largely similar to the previous 
example, but it now operates in the Nambu basis defined by the spinor $\psi_i=[\hat{c}_{i\up}\quad  \hat{c}^\dagger_{i\dw}]^T$ where the symbol $\hat{o}$ indicates the potential multi-orbital nature of the system, which reduces to a scalar in the present single-orbital case.
In this basis, the Green's function takes the matrix form:
\begin{equation}
  {\mathbf G} =
  \begin{pmatrix}
    \hat{G}_{\uparrow\uparrow} & \hat{F}_{\uparrow\downarrow}\\
    \hat{\bar{F}}_{\downarrow\uparrow}  &    \hat{\bar{G}}_{\downarrow\downarrow} \\
  \end{pmatrix}
\end{equation}
The components in the second row, denoted 
by $\hat{\bar{A}}$, are connected to the first row by particle-hole and time-reversal symmetries. The specific relations depend on the 
symmetry of the order parameter (here, $s$-wave) and whether the 
functions are defined on the Matsubara or real-frequency axis:
\begin{equation}
\begin{array}{cc}
  \hat{\bar{G}}(i\omega) = -\hat{G}^*(i\omega)\;; &  \hat{\bar{F}}(i\omega) = \hat{F}(i\omega)\\
  \hat{\bar{G}}(\omega)  = -\hat{G}^*(-\omega) \;; & \hat{\bar{F}}(\omega) = \hat{F}^*(i\omega)\\
\end{array}
\end{equation}  

The code implementation closely follows the structure of the previous 
example, with some notable adjustments related to the Nambu basis. 
These symmetries allow computing only the 
independent components in the first row, reducing 
the computational effort. 
Note that part of the operations required to implement the DMFT cycle are implemented in the Python module {\tt aux\_funx.py}, adapting from {\tt DMFTtools} functions.  
The initial part of the code handles the 
lattice structure and solver initialization, as described below.

\begin{lstlisting}[style=mypython,numbers=none,basicstyle={\scriptsize\ttfamily}]
import numpy as np
#Import EDIpack2py:
from edipack2py import global_env as ed
#Import MPI support 
import mpi4py
from mpi4py import MPI
#Import functions to build ${\color{comment-color}G_\mathrm{loc}}$ and perform DMFT self-consistency in Nambu space
from aux_funx import * 
import os,sys

#Start MPI framework:
comm = MPI.COMM_WORLD
rank = comm.Get_rank()
master = (rank==0)

#Functions: build 2D grid and dispersion ${\color{comment-color}\e(k)}$
def generate_kgrid(Nk):
    b1=2*np.pi*np.array([1.0,0.0])
    b2=2*np.pi*np.array([0.0,1.0])
    n1, n2 = np.meshgrid(np.arange(Nk), np.arange(Nk))
    n1=n1/Nk;n2=n2/Nk
    gridout = np.stack([n1.ravel(), n2.ravel()], axis=-1)
    return np.dot(gridout,[b1,b2])
    
def h_square2d(k,t):
  return -2*t*( np.cos(k[...,0,np.newaxis,np.newaxis])+
                np.cos(k[...,1,np.newaxis,np.newaxis]))*np.eye(ed.Norb)
    
#Read input
ed.read_input("inputAHM.conf")

#Generate ${\color{comment-color}H_k=\e(k)}$ and set Hloc
kgrid   = generate_kgrid(Nk)
Hk      = h_square2d(kgrid,t_hop)
Hloc    = np.sum(Hk,axis=0)/Nk**2
ed.set_hloc(Hloc.astype(complex))

#Build dispersion in Nambu space ${\color{comment-color}\e(k)\tau^z}$
HkNambu = np.array([h_square2d(kgrid,t_hop),-np.conj(h_square2d(-kgrid,t_hop))])

#Setup ED Solver
Nb=ed.get_bath_dimension()
bath = ed.init_solver()
\end{lstlisting}

The iterative scheme for the solution of DMFT closely follows the
sequence already discussed in \secu{SecExamplesBetheDMFT}:   
\begin{itemize}
\item[{\tiny {\bf EDIpack}}] Call the exact diagonalization {\bf impurity solver} {\tt
    ed.solve} providing the set of bath parameters $\vec{x}=\{V,h\}$  as input. 

\item[{\tiny {\bf EDIpack}}]  Use the dedicated
  input/output \NAME procedures to retrieve the self-energy functions  
  $\hat{\Sigma}(i\omega)$ and $\hat{S}(i\omega)$ on the 
  Matsubara axis.
  
\item[{\tiny {\bf EDIpack}}]
  Evaluate the interacting local Green's functions $\hat{G}_\mathrm{loc}$ and
  $\hat{F}_\mathrm{loc}$:
  \begin{equation}
  {\mathbf G}_\mathrm{loc}(i\omega) =
  \int_\RRR d\e \rho(\e)
  \begin{pmatrix}
    (i\omega +\mu)\hat{\11} - \hat{h}^0 - \hat{\Sigma}(i\omega) -\e & -\hat{S}(i\omega) \\
    -\hat{S}(i\omega) & (i\omega +\mu)\hat{\11} + \hat{h}^0 +
    \hat{\Sigma}^*(i\omega) +\e\\
  \end{pmatrix}^{-1}
\end{equation}

\item[{\tiny {\it User}}] Update the Weiss field's components, respectively 
  $\GG_0^{-1}$ and $\FF_0^{-1}$, through the {\bf self-con\-sis\-ten\-cy}
  relation: $\mathbfcal{G}^{-1}_0(i\omega) = {\mathbf G}^{-1}_\mathrm{loc}(i\omega) +
  {\mathbf \Sigma}(i\omega)$ in Nambu space.
  
\item[{\tiny {\it User}}\textgreater\ {\tiny {\bf EDIpack}}] Optimize the bath parameters $\vec{x}$ to best describe the updated
    Weiss fields, potentially using the \NAME provided conjugate gradient  fit
    procedures.
  \end{itemize}
The corresponding implementation in Python reads:
\begin{lstlisting}[style=mypython,numbers=none,basicstyle={\scriptsize\ttfamily}]
#DMFT CYCLE
converged=False;iloop=0
while (not converged and iloop<ed.Nloop):
    iloop=iloop+1
    #Solve quantum impurity problem for the current bath
    ed.solve(bath)    

    #Retrieve the Matsubara self-energy components ${\color{comment-color}\Sigma(i\omega_n) }$ and ${\color{comment-color}S(i\omega_n) }$
    Smats = np.array([ed.get_sigma(axis="m",typ="n"),ed.get_sigma(axis="m",typ="a")])   
    
    #Perform self-consistency levaraging {\tt aux\_funx.py} procedures:
    Gmats = get_gloc(wm*1j,ed.xmu,HkNambu,Smats,axis="m")
    Weiss = dmft_weiss_field(Gmats,Smats)    
    
    #Fit Weiss field and update the bath
    bath = ed.chi2_fitgf(Weiss[0],Weiss[1],bath)

    #Error check
    err,converged=ed.check_convergence(Weiss,ed.dmft_error)
ed.finalize_solver()
\end{lstlisting}

\begin{figure}[t!]
  \includegraphics[width=\linewidth]{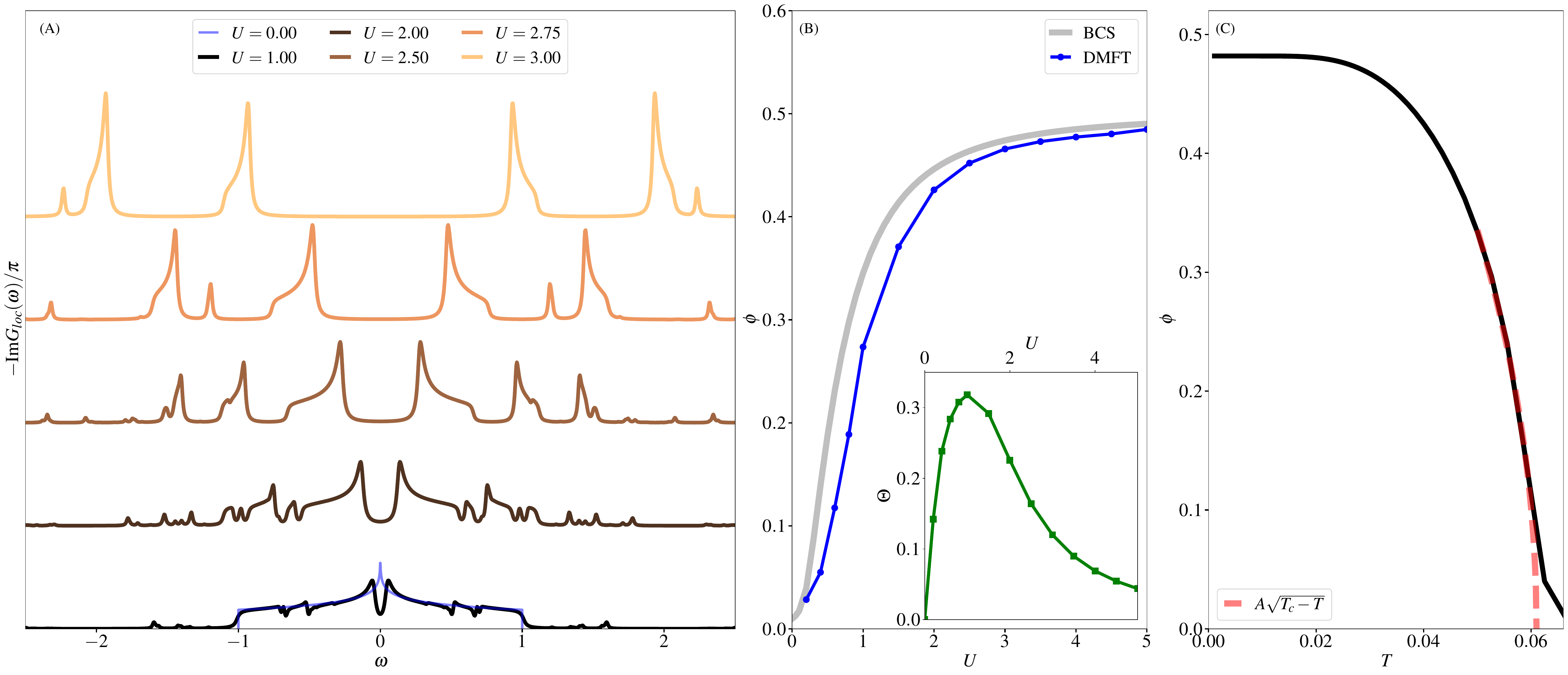}
    \caption{\label{figEx2}%
      \textbf{The BCS to BEC crossover.}
      (A) Evolution of the spectral functions
      $-\Im{G_\mathrm{loc}(\omega)}/\pi$ as a function of increasing
      attraction $U$. 
      (B) The order parameter $\phi=\langle c_\up c_\dw\rangle$ as a
      function of the attraction $U$. Data for BCS (gray) is compared
      to DMFT results (blue line and symbols). Inset: the correlation strength
      $\Theta$ (see main text) as a function of the
      attraction $U$ across the BCS-BEC crossover. 
      (C) Superconducting order
      parameter $\phi$ as a function of temperature across the
      superconductor-to-normal phase transition. Data for $U=4$. The
      fit highlights the critical behavior with a mean-field exponent
      $\beta=1/2$ (red dashed line) and parameters $A\simeq 3.7$, $T_\mathrm{c}=0.61$.       
        }
\end{figure}

\paragraph{Results.}
Here we showcase some results for the DMFT solution of the 
attractive Hubbard model across the BCS-to-BEC crossover regime \cite{Toschi2005PRB,Toschi2005NJP,Amaricci2014PRA} 
illustrating the capability of \NAME to handle $s$-wave 
superconductivity at both zero and finite temperatures.

To begin, panel (A) of \figu{figEx2} shows the evolution of the 
spectral density, obtained from the local normal Green's function as 
$-\tfrac{1}{\pi}\Im G_\mathrm{loc}(\omega)$, as a function of the 
attraction $U$. For any finite $U$, the Van Hove peak near the Fermi level characteristic of the 2D square lattice (visible at 
$U = 0$) is split by the formation of a superconducting gap. The latter reflects the emergence of a finite 
order parameter $\phi = \langle c_\up c_\dw \rangle$ and onset of superconducting coherence.

The evolution of the order parameter with attraction $U$ is presented in panel (B). The figure  
highlights the crossover from the weak-coupling BCS regime to the 
strong-coupling BEC regime. In the BCS limit, $\phi$ displays the 
characteristic exponential growth with $U$, known to be 
computationally challenging. In the opposite, strong-coupling, limit 
$\phi$ saturates at its theoretical maximum of $\phi \rightarrow 1/2$. The comparison with the BCS mean-field result 
(gray line) reveals the effect of local dynamical fluctuations, which  slightly suppress the order parameter, particularly in the intermediate regime. 
To further quantify these dynamical effects, we plot in the inset of the same panel the 
{\it correlation strength} \cite{Amaricci2015PRL,Amaricci2016PRB}
$\Theta=|S(i\omega\to 0)-S(i\omega\to\infty)|/S(i\omega\to\infty)$. 
Large values of $\Theta$ indicate an anomalous self-energy with 
significant dynamical effects, hence a more correlated superconducting state.
Our results show that the DMFT solution significantly departs from the BCS picture in the strongly correlated intermediate-coupling, a regime known to host the highest critical temperature \cite{Toschi2005NJP,Toschi2005PRB}.

Finally, we demonstrate the capability of \NAME to describe finite 
temperature effects. Panel (E) shows the temperature dependence of 
the order parameter $\phi(T)$ across the superconducting-to-normal 
transition. The mean-field nature of this transition is evident from 
the scaling near the critical temperature 
$\phi \sim (T_\mathrm{c} - T)^\beta$ with $\beta = 1/2$, consistent 
with  Ginzburg-Landau theory.

These results collectively illustrate the versatility of \NAME in 
handling both ground state and finite temperature superconducting 
phases, for instance capturing the complex physics of the BCS-BEC crossover with high accuracy.

\begin{figure}[ht!]
    \includegraphics[width=\linewidth]{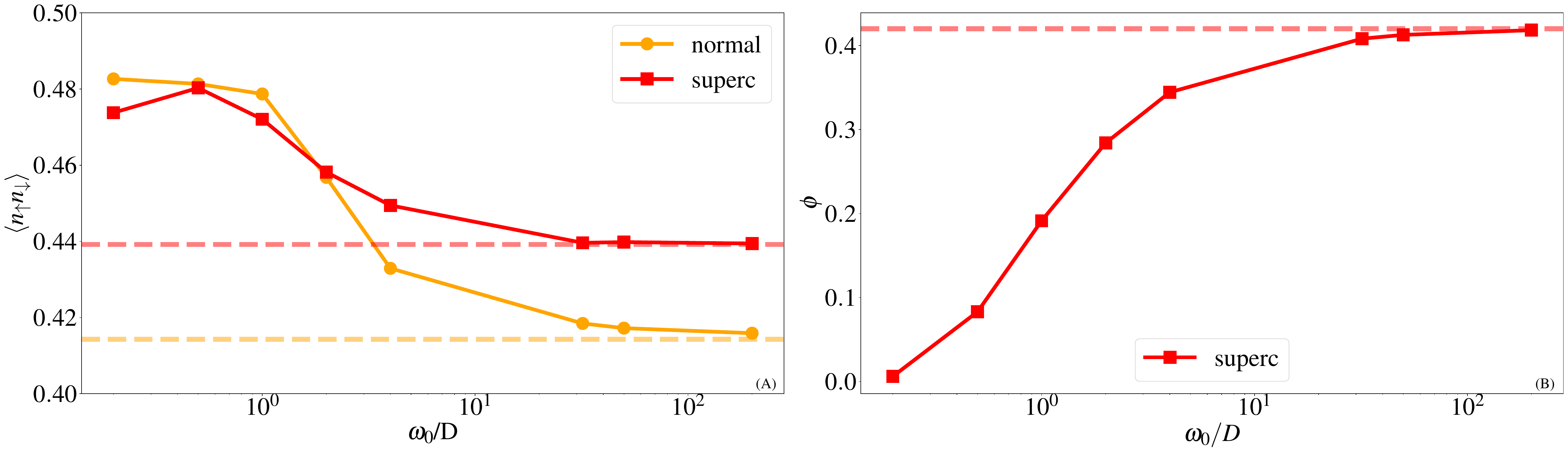}
    \caption{\label{figEx5}
      Evolution of the double occupancy (A) and superconductive order parameter (B) as a function of $\omega_0/D$ for $\lambda=1$ in the normal (orange) and superconductive (red) phase. The horizontal broken lines are the values for the corresponding Hubbard model in the anti-adiabatic limit.}
\end{figure}

\subsection{Holstein model on the Bethe lattice (electron-phonon coupling)}

One of the new features introduced in \NAME is the support for local phonons in combination with superconductivity,
i.e. {\tt ed\_mode=superc}. In order to illustrate this property using a simple application, in this example we discuss the normal and
superconductive solution of the pure Holstein model on the Bethe lattice within DMFT.
Note that the code
implementation for this case is essentially identical to the listings in the previous sections
\ref{SecExamplesBetheDMFT} ({\tt ed\_mode=normal}) and
\ref{SecExamplesAHM} ({\tt ed\_mode=superc}), provided electron-electron interaction is set to zero and phonon parameters are properly configured.  

We consider the model introduced in \secu{SecExamplesBetheDMFT}
with $U=0$ and the additional phononic and electron-phonon terms:
\begin{equation} \label{eqex:H_Holstein}
    H_\mathrm{int} = \sum_i \Big[\omega_0 b^\dagger_i b_i + g(b^\dagger_i +
    b_i)\sum_{\sigma}\left(c^\dagger_{i\sigma}c_{i\sigma}
    -\frac{1}{2}\right)\Big]. 
\end{equation}
We focus on the half-filling regime of the particle-hole symmetric Bethe lattice DOS. We set the half-bandwidth as our energy unit $D=1$ and introduced the electron-phonon coupling $\lambda = \tfrac{2g^2}{\omega_0}$.  
The iterative DMFT solution algorithm follows the same principles illustrated in the previous sections. 

\paragraph{Results.}
In the following we discuss the adiabatic ($\omega_0\to0$) to
anti-adiabatic ($\omega_0\to\infty$) crossover for
the uniform solution of the Holstein model at constant coupling $\lambda=1.0$.
In the anti-adiabatic limit, the Holstein interaction takes a
particularly simple form:
\begin{equation}\label{HlikeAttraction}
    H_\mathrm{int} \overset{ \omega_0 \rightarrow \infty}{ \longrightarrow } -\frac{\lambda}{2} \sum_i \Big[\sum_\sigma\left(c^\dagger_{i\sigma}c_{i\sigma} -\frac{1}{2}\right) \Big]^2,
\end{equation}
which describes a local Hubbard-like attraction among electrons
(mediated by local phonons).
In the  adiabatic limit, $\omega_0 \rightarrow 0$ the system
enters a Bipolaronic Insulating phase for our
choice of the coupling \cite{Capone2006PRB}.

We characterize the model solution by showing the evolution of the
double occupation $\langle n_{\up}n_{\dw}\rangle$ as a function of the
phonon frequency $\omega_0$, see panel (A) of \figu{figEx5}. In this panel we compare
the behavior for the normal phase ({\tt ed\_mode=normal}) and the superconductive
phase ({\tt ed\_mode=superc}). Our results capture the whole
crossover from adiabatic to anti-adiabatic regime. In the former
regime, the double occupation takes a similar value for the two
phases. However, approaching the anti-adiabatic regime, the two
solutions reach the limiting values corresponding to the residual
attraction \equ{HlikeAttraction}.   
To better characterize the nature of the superconducting
phase in the Holstein model, we report the evolution of the anomalous order
parameter $\phi$ in the adiabatic to anti-adiabatic crossover, see panel (B) of \figu{figEx5}.
The DMFT results obtained with \NAME show a rapid increase in the
superconducting order parameter as phonon frequency grows
large. Finally, in the anti-adiabatic regime, $\phi$ saturates to a
finite value corresponding to the attractive Hubbard-like interaction
of strength $\lambda/2$.

\begin{figure}[t!]
    \includegraphics[width=0.5\linewidth]
        {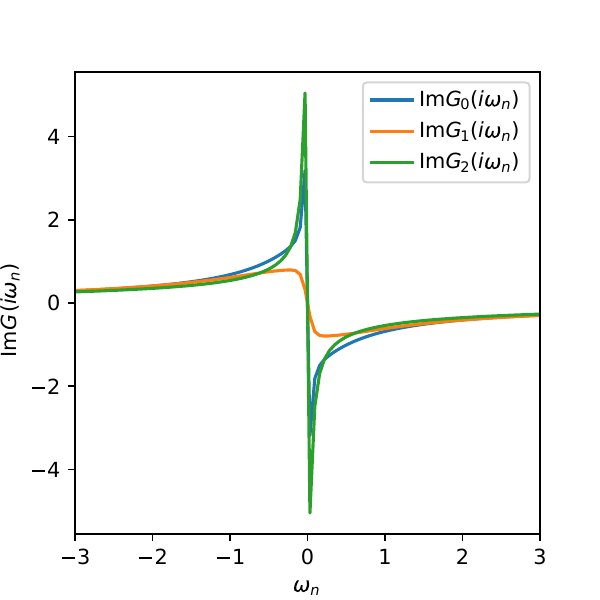}
    \includegraphics[width=0.5\linewidth]
        {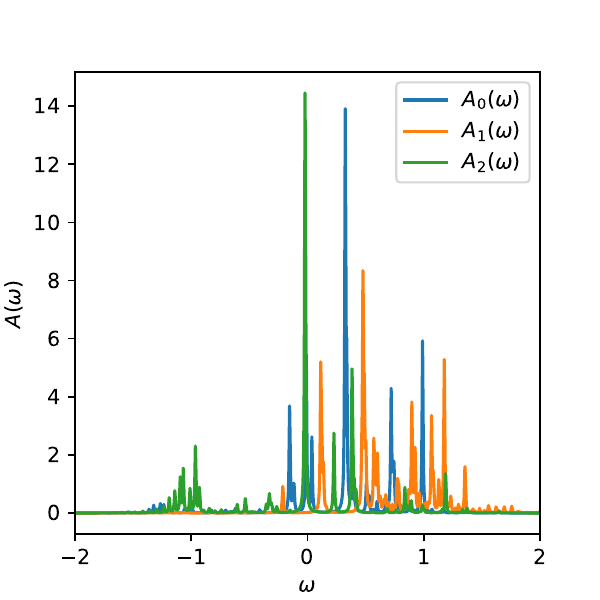}
    \caption{\label{figEx3}%
        Imaginary part of the Matsubara Green's function $G_\alpha(i\omega_n)$ (left) and the
        corresponding orbital-resolved spectral function $A_\alpha(\omega) = -\Im{G_\alpha(\omega)} / \pi$ (right) computed for a
        three-orbital impurity model with an interaction of the Hubbard-Kanamori
        type (\ref{Hint}). This illustration is produced by the EDIpack2TRIQS
        example script presented in \secu{SecExamplesTRIQS}.
    }
\end{figure}

\subsection{Multi-orbital impurity with Kanamori
  interaction (TRIQS interface)}
\label{SecExamplesTRIQS}

We proceed to demonstrate how to use the EDIpack2TRIQS compatibility
layer (\secu{sSecInteropTRIQS}) to solve a quantum system comprised by a
3-orbital correlated impurity coupled to a few non-interacting bath sites.
The interaction term of the impurity Hamiltonian is in the Hubbard-Kanamori
form of \equ{Hint}.

A Python script implementing a calculation that makes use of EDIpack2TRIQS
generally begins with a few module imports.
\begin{lstlisting}[style=mypython,
                 numbers=none,
                 basicstyle={\scriptsize\ttfamily},
                 firstline=1, lastline=13]
# Import common Python modules
from itertools import product
import numpy as np

# Initialize the MPI library before creating the solver object
from mpi4py import MPI

# Import TRIQS many-body operator objects: annihilation, creation and
# occupation number operators
from triqs.operators import c, c_dag, n

# Import the solver object
from edipack2triqs.solver import EDIpackSolver
\end{lstlisting}

One then proceeds to defining the system under consideration. In this case we
consider an impurity atom with three correlated orbitals and two bath states
per each impurity orbital, which corresponds to {\tt bath\_type=normal} in \NAME. This information must be encoded in the fundamental
set objects that are later used to construct the solver.
\begin{lstlisting}[style=mypython,
                 numbers=none,
                 basicstyle={\scriptsize\ttfamily},
                 firstline=15, lastline=26]
spins = ['up', 'dn']    # Names of spin projections
orbs = [0, 1, 2]        # List of impurity orbitals
bath_sites = [0, 1]     # List of bath sites, each carrying 3 orbitals

# Fundamental sets for impurity spin up/down degrees of freedom
fops_imp_up = [('up', orb) for orb in orbs]
fops_imp_dn = [('dn', orb) for orb in orbs]

# Fundamental sets for bath spin up/down degrees of freedom
# There are a total of 6 bath states for each spin projection
fops_bath_up = [('B_up', i) for i in range(3 * 2)]
fops_bath_dn = [('B_dn', i) for i in range(3 * 2)]
\end{lstlisting}
The next step is to define a TRIQS many-body operator expression that represents
the Hamiltonian to be diagonalized.
\begin{lstlisting}[style=mypython,
                 numbers=none,
                 basicstyle={\scriptsize\ttfamily},
                 firstline=28, lastline=67]
# Non-interacting part of the impurity Hamiltonian
h_loc = np.diag([-0.7, -0.5, -0.7])
H = sum(h_loc[o1, o2] * c_dag(spin, o1) * c(spin, o2)
        for spin, o1, o2 in product(spins, orbs, orbs))

# Interaction part
U = 3.0     # Local intra-orbital interactions U
Ust = 1.2   # Local inter-orbital interaction U'
Jh = 0.2    # Hund's coupling
Jx = 0.15   # Spin-exchange coupling constant
Jp = 0.1    # Pair-hopping coupling constant

H += U * sum(n('up', o) * n('dn', o) for o in orbs)
H += Ust * sum(int(o1 != o2) * n('up', o1) * n('dn', o2)
               for o1, o2 in product(orbs, orbs))

H += (Ust - Jh) * sum(int(o1 < o2) * n(s, o1) * n(s, o2)
                      for s, o1, o2 in product(spins, orbs, orbs))
H -= Jx * sum(int(o1 != o2) * c_dag('up', o1) * c('dn', o1) * c_dag('dn', o2) * c('up', o2)
              for o1, o2 in product(orbs, orbs))
H += Jp * sum(int(o1 != o2) * c_dag('up', o1) * c_dag('dn', o1) * c('dn', o2) * c('up', o2)
              for o1, o2 in product(orbs, orbs))

# Bath part

# Matrix dimensions of eps and V: 3 orbitals x 2 bath sites
eps = np.array([[-0.1, 0.1],
                [-0.2, 0.2],
                [-0.3, 0.3]])
V = np.array([[0.1, 0.2],
              [0.1, 0.2],
              [0.1, 0.2]])

# Dispersion of the bath states
H += sum(eps[o, nu] * c_dag("B_" + s, nu * 3 + o) * c("B_" + s, nu * 3 + o)
         for s, o, nu in product(spins, orbs, bath_sites))
# Coupling between the impurity and the bath
H += sum(V[o, nu] * (c_dag(s, o) * c("B_" + s, nu * 3 + o)
                     + c_dag("B_" + s, nu * 3 + o) * c(s, o))
         for s, o, nu in product(spins, orbs, bath_sites))
\end{lstlisting}
Finally, one creates a solver object and performs the actual calculation by
calling its method {\tt solve()}. This is normally the most time- and
memory-consuming step.
\begin{lstlisting}[style=mypython,
                 numbers=none,
                 basicstyle={\scriptsize\ttfamily},
                 firstline=69, lastline=83]
# Create a solver object
solver = EDIpackSolver(H, fops_imp_up, fops_imp_dn, fops_bath_up, fops_bath_dn)

# Solve the impurity model
beta = 100.0                 # Inverse temperature
n_iw = 1024                  # Number of Matsubara frequencies for GF calculations
energy_window = (-2.0, 2.0)  # Energy window for real-frequency GF calculations
n_w = 4000                   # Number of real-frequency points for GF calculations
broadening = 0.005           # Broadening on the real axis for GF calculations

solver.solve(beta=beta,
             n_iw=n_iw,
             energy_window=energy_window,
             n_w=n_w,
             broadening=broadening)
\end{lstlisting}
The computation results are readily available as attributes of the solver
object. The following code snippet shows how to access measured expectation
values of the static observables, and how to employ the plotting framework of
TRIQS to visualize the obtained Matsubara and real-frequency impurity Green's
functions (\figu{figEx3}).
\begin{lstlisting}[style=mypython,
                 mathescape=false,
                 numbers=none,
                 basicstyle={\scriptsize\ttfamily},
                 firstline=85, lastline=113]
# On master MPI node, output some calculation results
if MPI.COMM_WORLD.Get_rank() == 0:
    # Print values of the static observables
    print("Potential energy:", solver.e_pot)
    print("Kinetic energy:", solver.e_kin)
    print("Densities (per orbital):", solver.densities)
    print("Double occupancy (per orbital):", solver.double_occ)

    # Use TRIQS' extensions to matplotlib to plot computed Green's functions
    from triqs.plot.mpl_interface import plt, oplot

    # Plot the Matsubara Green's functions (imaginary part)
    plt.figure(figsize=(4, 4))
    for orb in orbs:
        oplot(solver.g_iw['up'][orb, orb], mode='I',
              label=r"${\rm Im} G_" + str(orb) + r"(i\omega_n)$")
    plt.xlim((-3, 3))
    plt.ylabel(r"${\rm Im} G(i\omega_n)$")
    plt.legend()
    plt.savefig("G_iw.pdf")

    # Plot the orbital-resolved spectral functions
    plt.figure(figsize=(4, 4))
    for orb in orbs:
        oplot(solver.g_w['up'][orb, orb], mode='S', label=f"$A_%i(\\omega)$" % orb)
    plt.xlim(energy_window)
    plt.ylabel(r"$A(\omega)$")
    plt.legend()
    plt.savefig("A_w.pdf")
\end{lstlisting}

\subsection{Interacting Bernevig-Hughes-Zhang model (Fortran API, {\tt ed\_mode=nonsu2})}
In this section, we focus on the DMFT solution of the interacting
Bernevig-Hughes-Zhang (BHZ) model.
As shown in Ref. \cite{Amaricci2023PRB}, this model exhibits an excitonic phase at moderate
interactions in which opposite orbital electrons and holes across the band gap bind and form a coherent phase \cite{Knolle2017PRL,Jia2022NP,Varsano2020NN,Blason2020PRB,Amaricci2023PRB,Giuli2023PRB}

We consider a system of two-orbital electrons on a square
lattice, interacting via a Hubbard-Kanamori term. This system realizes a quantum spin Hall insulator \cite{Kane2005PRL,Bernevig2006S,Hasan2010RMP,Hohenadler2011PRL,Amaricci2015PRL,Tang2017NP,Amaricci2023PRB,Paoletti2024PRB}.
We consider a suitable matrix basis in terms of the Dirac
matrices $\Gamma_{a\a}=\sigma_a\otimes \tau_\a$, where $\sigma_a$ and
$\tau_\a$ are Pauli matrices, respectively, in the spin and orbital
pseudo-spin space. The  model Hamiltonian reads
$$
H = \sum_{k}\psi_{k}^\dagger H(k)\psi_{k} + H_{\rm int},
$$
where $\psi_{k}=[c_{1\uparrow k}, c_{2\uparrow k},
c_{1\downarrow k}, c_{2\downarrow k} ]^T$ is the spinor collecting
annihilation operator $c_{a\sigma k}$ destroying an electron at
orbital $a=1,2$ with spin  $\sigma=\up,\dw$ and lattice momentum
$k$. The non-interacting part of the Hamiltonian is:
$$
H(k) = \left[M-2t(\cos{k_x}+\cos{k_y}) \right]\Gamma_{03} +
   \lambda\sin{k_x}\Gamma_{31} -   \lambda\sin{k_y}\Gamma_{02},
$$
where $M$ is the mass term, which plays the role of a crystal
field splitting among the orbitals. The presence of this term breaks
the symmetry in the orbital pseudo-spin channel.
The  interaction reads: 
$$
   H_{\rm int} = (U-J)\frac{\hat{N}(\hat{N}-1)}{2} - J\left( \frac{1}{4}\hat{N}^2 +
   \hat{S_z}^2 - 2 \hat{T_z}^2\right),
 $$
 where $\hat{N}=\tfrac{1}{2}\psi_i^\dagger \Gamma_{00}\psi_i$ is the
total density operator,
$\hat{S_z}=\tfrac{1}{2}\psi_i^\dagger \Gamma_{30}\psi_i$ is the total
spin polarization operator and $\hat{T_z}=\tfrac{1}{2}\psi_i^\dagger
\Gamma_{03}\psi_i$ is the orbital pseudo-spin polarization operator.
This form corresponds to the density-density part of the
Kanamori interaction. We neglect the pair-hopping and spin-flip purely
for numerical reasons \cite{Amaricci2022CPC}. 
In the non-interacting regime this model describes a
quantum spin Hall insulator (QSHI) for $M<4t$ and a trivial Band
Insulator (BI) for $M>4t$.
The transition point at $M=4t$ describes the formation of a gapless Dirac state at  $k=[0,0]$.

The code implementation follows the same guidelines discussed above for the other examples. The user is required to generate the Hamiltonian $H(k)$ on a discretized Brillouin zone. 
This is employed to evaluate the local interacting Green's function:
$$
G_\mathrm{loc}(z) = \tfrac{1}{N_k}\sum_k \left[z+\mu-H(k)-\Sigma(z)
\right]^{-1}, 
$$
entering in self-consistency conditions $\GG^{-1}=G_\mathrm{loc}^{-1}+\Sigma$. 
Here  $\z\in\CCC$, $N_k$ is the number of $k$-points   and $\Sigma(z)$ is the self-energy matrix  obtained from the DMFT solution of the problem.  
Moreover, the $H(k)$ is used to  construct the {\it renormalized} topological Hamiltonian \cite{Wang2010PRL,Wang2012PRX,Blason2023PRB}
$$
H_\mathrm{top} = \sqrt{Z}[H(k) + \Re\Sigma(\omega\to0)]\sqrt{Z}, 
$$
where $Z$ is the renormalization constant matrix. The matrix $H_\mathrm{top}$ describes the
low-energy properties of the many-body solution and can be used to characterize the topological nature of the interacting system \cite{Gurarie2011PRB,Wang2012PRX,Wagner2023NC,Blason2023PRB,Bau2024PRB}.

\begin{figure}[t!]
  \includegraphics[width=\linewidth]{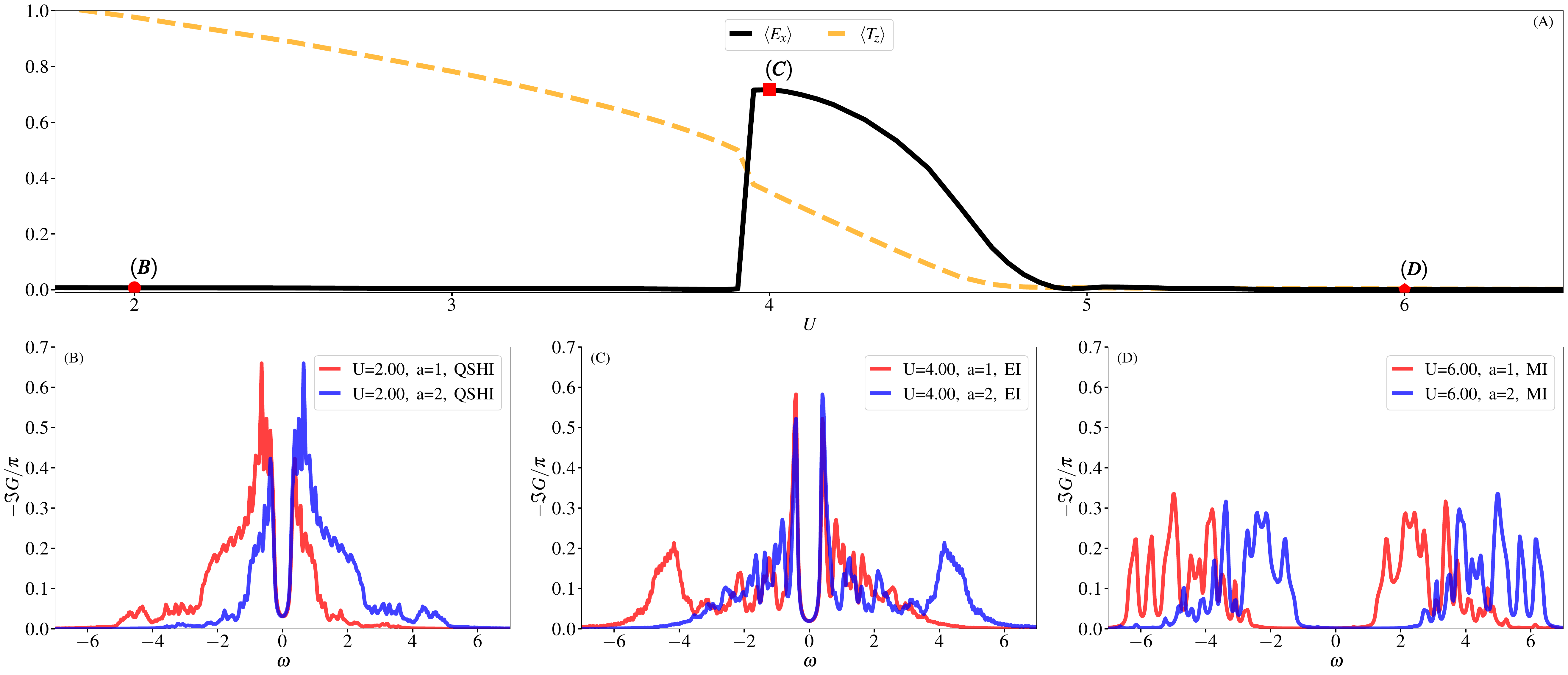}
    \caption{\label{figEx4}%
      \textbf{Topological and Exciton Transition.}
      (A) Evolution of the spin-triplet,
      in-plane excitonic order parameter $\langle E_x\rangle$ (black solid line) and
      orbital polarization $\langle T_z\rangle$ (orange dashed line) as a function of the
      interaction $U$.
      (B-D) Spectral functions of the two orbital electrons for three
      values of the interaction $U=2.00$ (red circle), $4.00$ (red
      square) and $6.00$ (red diamond) capturing, respectively, the
      QSHI \textbf{\textit{(B)}}, the Excitonic Insulator \textbf{\textit{(C)}} and the Mott Insulator
      \textbf{\textit{(D)}}. 
    }
\end{figure}

\paragraph{Results}
To capture the possible symmetry breaking in any excitonic channel
driven by the local interaction, we consider the vector order
parameter
$\vec{E}=[E_0,E_x,E_y,E_z]$, where $E_a=\langle\psi^\dagger
\Gamma_{a1}\psi \rangle$ \cite{Budich2014PRL,Kunes2014PRB,Kaneko2015JOPCS,Kunes2015JOPCM,Knolle2017PRL,Guerci2019PRM,Geffroy2019PRL,Mazza2020PRL,De-Palo2023PRB}.

The first component $E_0$ describes the singlet excitonic state,
whereas the remaining ones correspond to the triplet states with
different spin orientation \cite{Blason2020PRB,Amaricci2023PRB}.   
The analysis of the strong-coupling regime as well as impurity
excitonic susceptibility available in \NAME suggest the possible
instability to an in-plane spin-triplet exciton phase, i.e. with $E_x$
and/or $E_y$ different from zero, see \cite{Amaricci2023PRB}.
Interestingly, this state breaks several symmetries including
time-reversal and spin SU(2), which protect the topological state.

Here we showcase the DMFT description of the excitonic phase in the interacting BHZ model as a way to
illustrate the \NAME ability to capture matter phases with lowered spin-symmetry ({\tt ed\_mode=nonsu2}) and the use of {\tt
  bath\_type=replica}.

The initial part of the code implementation is a simple generalization of the previously discussed cases. 
The next non-trivial steps are: i) construct the lattice Hamiltonian $H(k)$ and the corresponding local part $H_\mathrm{loc}$, which determines the impurity properties, and ii) construct a matrix basis representation for the replica bath. These two steps can be implemented as follows: 

\begin{lstlisting}[style=fstyle,numbers=none,basicstyle={\scriptsize\ttfamily}]
   !> Set $\smash{{\color{comment-color}H_\mathrm{loc}}}$ 
   allocate(Hloc(Nso,Nso))
   Hloc = sum(Hk,dim=3)/Lk
   where(abs(dreal(Hloc))<1d-6)Hloc=zero

   !EDIpack: set the impurity Hamiltonian: $\smash{{\color{comment-color}H_\mathrm{loc}\to h^0}}$
   call ed_set_hloc(Hloc)

   !EDIpack: set the replica bath matrix basis $\smash{{\color{comment-color}  \vec{\Gamma}}}$ and $\smash{{\color{comment-color}  \vec{\lambda}}}$
   !Here Nsym=4.
   !Build the basis and init the variational parameters:
   allocate(lambdasym_vector(Nbath,4))
   allocate(Hsym_basis(Nso,Nso,4))
   Hsym_basis(:,:,1)=Gamma03 ;lambdasym_vector(:,1)= Mh
   Hsym_basis(:,:,2)=Gamma01 ;lambdasym_vector(:,2)= sb_field
   Hsym_basis(:,:,3)=Gamma31 ;lambdasym_vector(:,3)= sb_field
   Hsym_basis(:,:,4)=Gamma11 ;lambdasym_vector(:,4)=-sb_field
   
   !EDIpack: set the basis and initial values of $\smash{{\color{comment-color}  \vec{\lambda}}}$
   call ed_set_Hreplica(Hsym_basis,lambdasym_vector)
   
   !EDIpack: get bath dimension and allocate user bath to this size
   Nb=ed_get_bath_dimension(4)
   allocate(Bath(Nb))
   
   !EDIpack: Initialize the ED solver
   call ed_init_solver(bath)
\end{lstlisting}
Here we use the local non-interacting Hamiltonian 
$H_\mathrm{loc}=\tfrac{1}{N_k}\sum_k H(k)$ to set 
$h^0_{\a\b\s\s'}$, i.e. the impurity Hamiltonian. To anticipate the
possibility of forming an excitonic ordered phase, the replica bath
is constructed out of a matrix basis with 4 distinct elements
$\Gamma_{03}$, $\Gamma_{01}$, $\Gamma_{31}$, $\Gamma_{11}$ which are proportional, respectively, to the mass term, the exciton singlet
$E_0$, the exciton triplet along easy-axis $E_z$ and in-plane
$E_x$ (we assume $E_y=0$ leveraging on residual $U(1)$ in-plane symmetry). 
In the following we consider the case of $M=1$, which corresponds to a QSHI in the non-interacting limit and $J/U=0.25$.
The implementation is nearly identical to the cases discussed above and
we report it here for completeness:
\begin{lstlisting}[style=fstyle,numbers=none,basicstyle={\scriptsize\ttfamily}]
   iloop=0;converged=.false.
   do while(.not.converged.AND.iloop<nloop)
     iloop=iloop+1
     
     !EDIpack: Solve the impurity problem
     call ed_solve(bath)

     !EDIpack: Retrieve ${\color{comment-color} \Sigma(i\omega)}$
     call ed_get_sigma(Smats,axis="mats")
     
     !Get $\color{comment-color}G_\mathrm{loc}$ using $\color{comment-color}\mathrm{DMFTtools}$
     call get_gloc(Hk,Gmats,Smats,axis="m")
     
     !Update the Weiss field (self-consistency) using $\color{comment-color}\mathrm{DMFTtools}$
     call dmft_self_consistency(Gmats,Smats,Weiss)

     !Linear mixing the Weiss fields
     if(iloop>1)Weiss = wmixing*Weiss + (1.d0-wmixing)*Weiss_;Weiss_=Weiss

     !EDIpack: Fit to update the bath
     call ed_chi2_fitgf(Weiss,bath,ispin=1)
     
     !Check convergence: using $\color{comment-color}\mathrm{DMFTtools}$
     converged = check_convergence(Weiss(1,1,:),dmft_error,nsuccess,nloop)
   enddo  
 \end{lstlisting}

The main effect of the interaction on the topological properties is contained in the mass term renormalization.  
The real part of the self-energy being proportional to $\Gamma_{03}$
corrects the mass term with respect to its bare value: $M_\mathrm{eff}=M+\tfrac{1}{4}\Tr{\Re\Sigma(i\omega\to0)}$. 
To leading order (mean-field) this correction is proportional to $\tfrac{U-5J}{4}\langle
\hat{T}_z\rangle$. So, for our choice of parameters, the effect of
interaction would be to effectively reduce the mass term. Thus, in the strong-coupling limit the two orbitals get populated by one electron per site, reaching the
conditions for the formation of a Mott insulating state.
This effect is highlighted in panel (A) of \figu{figEx4}, where we report the progressive reduction of the
orbital polarization  $\langle T_z\rangle$ and, thus, of the effective mass.

In the same panel (A), we show that at intermediate coupling 
the DMFT solution features the formation of a
region of exciton condensation with $\langle E_x\rangle>0$, $\langle
E_0\rangle=\langle E_z\rangle=0$. 

Unlike the static mean-field description \cite{Blason2020PRB}, the
transition from the QSHI topological state to the Excitonic Insulator 
(EI) becomes of first-order upon including the local dynamical fluctuations \cite{Paoletti2024PRB,BellomiaKMH} contained in DMFT. 
The EI continuously evolves into a Mott Insulator (MI)
for larger interaction (neglecting for simplicity any 
antiferromagnetic ordering that would naturally occur in this
regime).
The \NAME solution of this model nicely captures the two transitions which involve breaking of the spin symmetry group SU(2).

To further illustrate the nature of the three distinct phases of this
system, we rely on the direct access to the real-axis spectral
function provided by \NAME solver. In panels (B)-(D) of \figu{figEx4}
we report the spectral functions $-\Im{G}_{a\up,\mathrm{loc}}(\omega)/\pi$ for
$a=1,2$, and three distinct values of the interaction
$U$ placing the solution, respectively, in the QSHI, the EI and the MI.

The correlated QSHI is characterized by the presence of an 
inverted band gap featuring a substantial orbital spectral mixing. The
EI spectrum is instead characterized by a narrow gap, related to the
finite order parameter, flanked by two sharp resonances and featuring
larger high-energy weight.
Finally, in the MI the two-orbital spectral function describes the two 
characteristic Hubbard bands separated by a large Mott gap.

\subsection{Interacting Kane-Mele model (Fortran API, EDIpack2ineq, iRDM)}

To complete the overview of \NAME's features, we consider an additional interacting quantum spin-Hall insulator model, defined on the honeycomb lattice, with two inequivalent sites in the unit cell. Neglecting the Rashba spin-orbit interaction and discarding any ionic character of the unit cell, the noninteracting Kane-Mele Hamiltonian \cite{Kane2005PRLa,Kane2005PRL} can be written in momentum space as
\begin{align*}
    H_\mathrm{KM} = \sum_{{k}} \psi^\dagger_{{k}} H({k}) \psi_{{k}},
\end{align*}
with $\psi_{{k}} = [c_{{k},\mathrm{A},\up},\,c_{{k},\mathrm{B},\up},\,c_{{k},\mathrm{A},\dw},\,c_{{k},\mathrm{B},\dw}]$ and
    \begin{align}
    H({k}) 
    &= x_{{k}}\Gamma_{01} + y_{{k}}\Gamma_{02} + \delta_{{k}}\Gamma_{33}, \label{eq:Hk_kanemele}\\[1mm]
   x_{{k}}&=t\bigl[\cos({k}\cdot{u})+\cos({k}\cdot{v})+1\bigr],\nonumber\\
   y_{{k}}&=t\bigl[\sin({k}\cdot{u})+\sin({k}\cdot{v})\bigr],\nonumber \\
   \delta_{{k}}&=2\lambda_\mathrm{so}\bigl[\sin({k}\cdot{u})-\sin({k}\cdot{v}) - \sin({k}\cdot{u} -{k}\cdot{v})\bigr], \nonumber
\end{align}
where ${u} = (3a/2, \sqrt{3}a/2)$, and ${v} = (3a/2, -\sqrt{3}a/2)$ 
are suitable basis vectors for the honeycomb lattice, $t$ is the
nearest-neighbor hopping amplitude and $\lambda_\mathrm{so}$ is 
the amplitude of a complex next-nearest-neighbor hopping,
with a spin-dependent $\pm\tfrac{\pi}{2}$ chiral phase arising from spin-orbit coupling, which is well-known to open a topological gap at the Fermi level \cite{Kane2005PRLa,Kane2005PRL}.
The spinorial $4\times4$ matrices $\Gamma_{a\ell} = \sigma_a\otimes\tau_\ell$ are 
defined in terms of Pauli matrices $\sigma_a$, $\tau_\ell$ referred, respectively 
to spin and sublattice degrees of freedom.
To investigate the effects of electron-electron repulsion we consider a local Hubbard term \cite{Hohenadler2013ROPIP,Rachel2018ROPIP}:
\begin{equation}
    H_\mathrm{KMH}  = H_\mathrm{KM} 
    + U\sum_{i}\Biggl[
                        \left(n_{i,\up} - \frac{1}{2}\right)
                        \left(n_{i,\dw} - \frac{1}{2}\right)
                      \Biggr],
    \label{eq:KMHmodel}
\end{equation}
where $n_{i,\sigma}=c^\dagger_{i,\sigma}c_{i,\sigma}$ are the 
local spin-density operators. The interaction is written in an explicitly 
particle-hole symmetric form so that the model is at half-filling for zero
chemical potential.

The essential features of the ground state phase diagram have been established by intensive multi-method analyses \cite{Rachel2018ROPIP}.
For $\lambda_\mathrm{so}=0$ the model describes a Dirac semimetal, up to large repulsion, and magnetizes to an isotropic Néel state above a critical $U/t$ ratio \cite{Sorella1992ELE,Castro-Neto2009RMP}.
For a finite spin-orbit coupling $\lambda_\mathrm{so}\neq0$, the quantum spin-Hall state results increasingly stable against magnetic ordering, due to its symmetry-protected nontrivial topology. Furthermore, the long-range ordering
eventually stabilized at large $U/t$, is characterized by a lowered
spin symmetry, due to the inherent coupling of spin and lattice
degrees of freedom that favors in-plane easy-axis magnetization \cite{Griset2012PRB}, as predicted by Hartree-Fock and $t$-$J$ perturbative expansions \cite{Rachel2010PRB} and
later confirmed by Auxiliary-Field Quantum Monte Carlo (AFQMC) 
simulations \cite{Hohenadler2013ROPIP}.

To treat the in-plane Néel ordering with \NAME, we can once again
exploit the \texttt{ed\_mode=nonsu2} option, allowing off-diagonal 
spin amplitudes in all dynamical matrices and so eventually in the
resulting impurity self-energy.
Referring to the two inequivalent sublattices as
A and B, whereas a standard off-plane ($S_z$)
calculation would enforce Néel symmetry (AFM$_\perp$)  as
$\Sigma^\mathrm{A}_{\up\up} = \Sigma^\mathrm{B}_{\dw\dw}$, 
$\Sigma^\mathrm{A}_{\dw\dw} = \Sigma^\mathrm{B}_{\up\up}$,
with
$\Sigma^\mathrm{A}_{\up\dw} = \Sigma^\mathrm{A}_{\dw\up} = \Sigma^\mathrm{B}_{\up\dw} = \Sigma^\mathrm{B}_{\dw\up} = 0$,
an in-plane magnetization (AFM$_\parallel$) corresponds to
$\Sigma^\mathrm{A}_{\up\up} = \Sigma^\mathrm{B}_{\up\up}$, 
$\Sigma^\mathrm{A}_{\dw\dw} = \Sigma^\mathrm{B}_{\dw\dw}$,
$\Sigma^\mathrm{A}_{\up\dw} = -\Sigma^\mathrm{B}_{\dw\up}$, 
$\Sigma^\mathrm{A}_{\dw\up} = -\Sigma^\mathrm{B}_{\up\dw}$.

The same off-diagonal spin symmetry must be realized also in
the bath Hamiltonian, which can be conveniently achieved with
either the \texttt{replica} and \texttt{general} choices for 
the \texttt{bath\_type} option. The most robust strategy to
verify that the ground state of the model is magnetized in the
plane also in DMFT is to allow either an $S_z$ or an $S_x$ (equivalently $S_y$) magnetization in the bath, and then
compare the energies of the respective solutions at $T=0$. 
The corresponding Hamiltonian forms are:
\begin{align}
    H^\mathrm{bath}_\perp &= 
        \sum_{p=1}^{N_\mathrm{b}}(\epsilon_p \sigma_0 + m_p \sigma_z), 
        \label{eq:afmz_dmft_ansatz} \\[1mm]
    H^\mathrm{bath}_\parallel &= 
        \sum_{p=1}^{N_\mathrm{b}}(\epsilon_p \sigma_0 + m_p \sigma_x).
        \label{eq:afmx_dmft_ansatz}
\end{align}
which corresponds to $N_\mathrm{sym}=2$, with
$\lambda_{1,p} \equiv \epsilon_p$ as the degenerate
bath levels that are Zeeman split by $\lambda_{2,p} \equiv m_p$, respectively in the $S_z$ and $S_x$ bases 
(see \secu{sSecBath}).

\begin{lstlisting}[style=fstyle,numbers=none,basicstyle={\scriptsize\ttfamily}]
program dmft_kane_mele_hubbard
  !Load EDIpack library 
  USE EDIPACK
  !Load the inequivalent impurities extension
  USE EDIPACK2INEQ
  ...
  !Setup bath and solver for anisotropic AFM phases
  !> define the symmetry-basis:
  Nineq = 2
  Nsym = 2
  allocate(Hsym_basis(Nspin*Norb,Nspin*Norb,Nsym))
  Hsym_basis(:,:,1) = pauli_0
  select case(ansatz)
    case('off-plane');Hsym_basis(:,:,2) = pauli_z
    case('in-plane') ;Hsym_basis(:,:,2) = pauli_x
  end select
  !> initialize the symmetry parameters
  allocate(lambdasym_vectors(Nineq,Nbath,Nsym))
  call build_replica_band(onsite_band,ed_hw_bath,Nbath)
  ! (user-built guess of the starting bath levels)
  lambdasym_vectors(:,:,1) = onsite_band 
  lambdasym_vectors(:,:,2) = 0d0 ! paramagnetic
  !> define a symmetry-breaking AFM kick, to help numerics
  if(afmkick)then
     lambdasym_vectors(1,:,2) = +sb_field
     lambdasym_vectors(2,:,2) = -sb_field
  endif
  !> setup H_bath inside the solver
  select case(bath_type)
    case('replica')
        call ed_set_Hreplica(Hsym_basis,lambdasym_vectors)
    case('general')
        call ed_set_Hgeneral(Hsym_basis,lambdasym_vectors)
  end select
  !> finally initialize the two solver instances, one per sublattice
  !  (the EDIpack2ineq layer automatically dispatches it for the two
  !   impurity models, if we feed a Bath with the [Nineq,Nb] shape!)
  Nb = ed_get_bath_dimension(Hsym_basis)
  allocate(Bath(Nineq,Nb))
  call ed_init_solver(Bath)
  ... 
\end{lstlisting}

The DMFT loop can then be implemented, with the explicit solution of two
independent impurity problems (prone to numerical noise) or by solving
only one impurity problem and building the full solution by exploiting 
the appropriate Néel symmetry:

\begin{lstlisting}[style=fstyle,numbers=none,basicstyle={\scriptsize\ttfamily}]
   iloop=0;converged=.false.
   do while(.not.converged.AND.iloop<nloop)
      iloop=iloop+1
      !
      !> Solve the two inequivalent impurity problems
      if(neelsym)then
        !> solve just one sublattice and get the other by Neel symmetry
        call ed_solve(Bath(1,:))
        call ed_get_sigma(Smats(1,:,:,:,:,:),axis='m')
        select case(ansatz)
            case('off-plane')
             Smats(2,2,2,:,:,:) = Smats(1,1,1,:,:,:) 
             Smats(2,1,1,:,:,:) = Smats(1,2,2,:,:,:) 
             if(master)write(*,*) ">>> Enforcing AFMz symmetry"
            case('in-plane')
             Smats(2,1,1,:,:,:) = Smats(1,1,1,:,:,:)   
             Smats(2,2,2,:,:,:) = Smats(1,2,2,:,:,:)   
             Smats(2,1,2,:,:,:) = -Smats(1,1,2,:,:,:)  
             Smats(2,2,1,:,:,:) = -Smats(1,2,1,:,:,:) 
             if(master)write(*,*) ">>> Enforcing AFMx symmetry"
        end select
      else
         !> solve both sublattices independently using EDIpack2ineq:
         !  mpi_lanc=T => MPI lanczos, mpi_lanc=F => MPI for ineq sites
         call ed_solve(Bath,mpi_lanc=.true.)
         !> retrieve all self-energies:
         call ed_get_sigma(Smats,Nineq,axis='m')
         !
      endif
      !
      !> Get ${\color{comment-color}G_{loc}(i\omega_n) }$ : using ${\color{comment-color}\mathrm{DMFTtools}}$
      call dmft_gloc_matsubara(Hk,Gmats,Smats)
      !
      !> Update local Weiss fields $\smash{{\color{comment-color}\GG^{-1}_0 = G^{-1}_{loc} + \Sigma}}$: using ${\color{comment-color}\mathrm{DMFTtools}}$
      call dmft_self_consistency(Gmats,Smats,Weiss)
      !
      !> Fit the new bath:
      !  - normal mode: normal/AFMz and we fit spin-components independently
      !  - nonsu2 mode: broken Sz-conservation and we fit
      !                 both spin components together
      select case(ed_mode)
       case("normal")
         call ed_chi2_fitgf(Bath,Weiss,ispin=1)
         call ed_chi2_fitgf(Bath,Weiss,ispin=2)
       case("nonsu2")
         call ed_chi2_fitgf(Bath,Weiss,Hloc)
      end select
      !
      !> linear mixing and convergence check: using ${\color{comment-color}\mathrm{DMFTtools}}$
      if(iloop>1) Bath = wmixing*Bath + (1.d0-wmixing)*Bath_prev
      Bath_prev = Bath
      converged = check_convergence(Weiss(:,1,1,1,1,:),dmft_error,nsuccess,nloop)
      !
   enddo
   !
   !> Compute Kinetic Energy: using ${\color{comment-color}\mathrm{DMFTtools}}$
   call dmft_kinetic_energy(Hk,Smats)
\end{lstlisting}

After convergence is reached, we evaluate some quantities of interest, e.g. computing the 
lattice kinetic energy (as implemented in the {DMFTtools} 
library). The potential energy, as well as the AFM order parameters
and impurity observables are automatically computed by \NAME, 
and saved to files with the appropriate inequivalent-impurity label.

\begin{figure}
\hspace{1cm} (a) \hspace{6.5cm} (b)\\
    \includegraphics[width=0.47\linewidth,trim={0 0 0 5mm},clip]{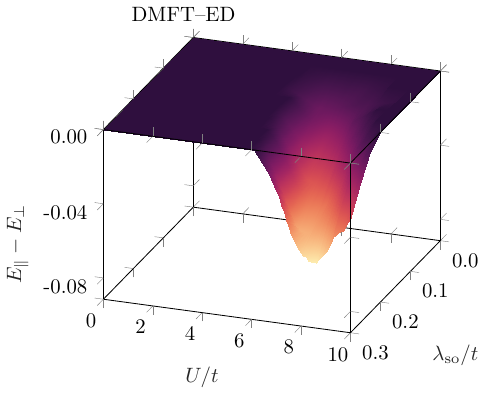}\hfill
    \includegraphics[width=0.45\linewidth]{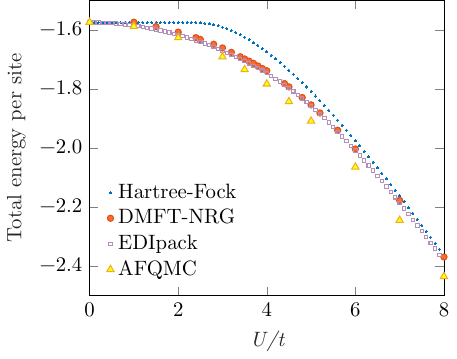}\\
    \caption{In panel (a) we report the total energy difference between 
    the in-plane and the
    off-plane AFM solutions found with \NAME at $T=0$. Negative values
    signal a ground state with in-plane magnetization. The plateau at zero
    energy difference marks the topological phase of the model (a 
    paramagnetic quantum spin-Hall insulator), at all points except on the $\lambda_\mathrm{so}=0$ line, where the ground state is an interacting
    Dirac liquid, at weak coupling, or an isotropic antiferromagnet at strong coupling. In panel (b) we compare the total energies, across the 
    $\lambda_\mathrm{so}=0$ line, to published data for Hartree-Fock,
    DMFT-NRG and AFQMC solutions \cite{Raczkowski2020PRB}.}
    \label{fig:KMenergy}
\end{figure}

We evaluate the total energy of both the AFM$_\perp$
and AFM$_\parallel$ calculations and compare them to establish which  is favored as the ground state of the system. In \figu{fig:KMenergy}(a)
we show results for the energy difference $E_\parallel-E_\perp$ across a wide range of $U/t$ and $\lambda_\mathrm{so}$ values. To assess the
degree of accuracy of our total energy calculation, we report in
\figu{fig:KMenergy}(b) a $\lambda_\mathrm{so}=0$ benchmark against
Hartree-Fock (simple static mean-field, but inherently variational), 
DMFT with NRG solver and lattice AFQMC data, as taken from 
Ref. \cite{Raczkowski2020PRB}. 

\begin{figure}
    \centering
    \includegraphics[width=0.8\linewidth]{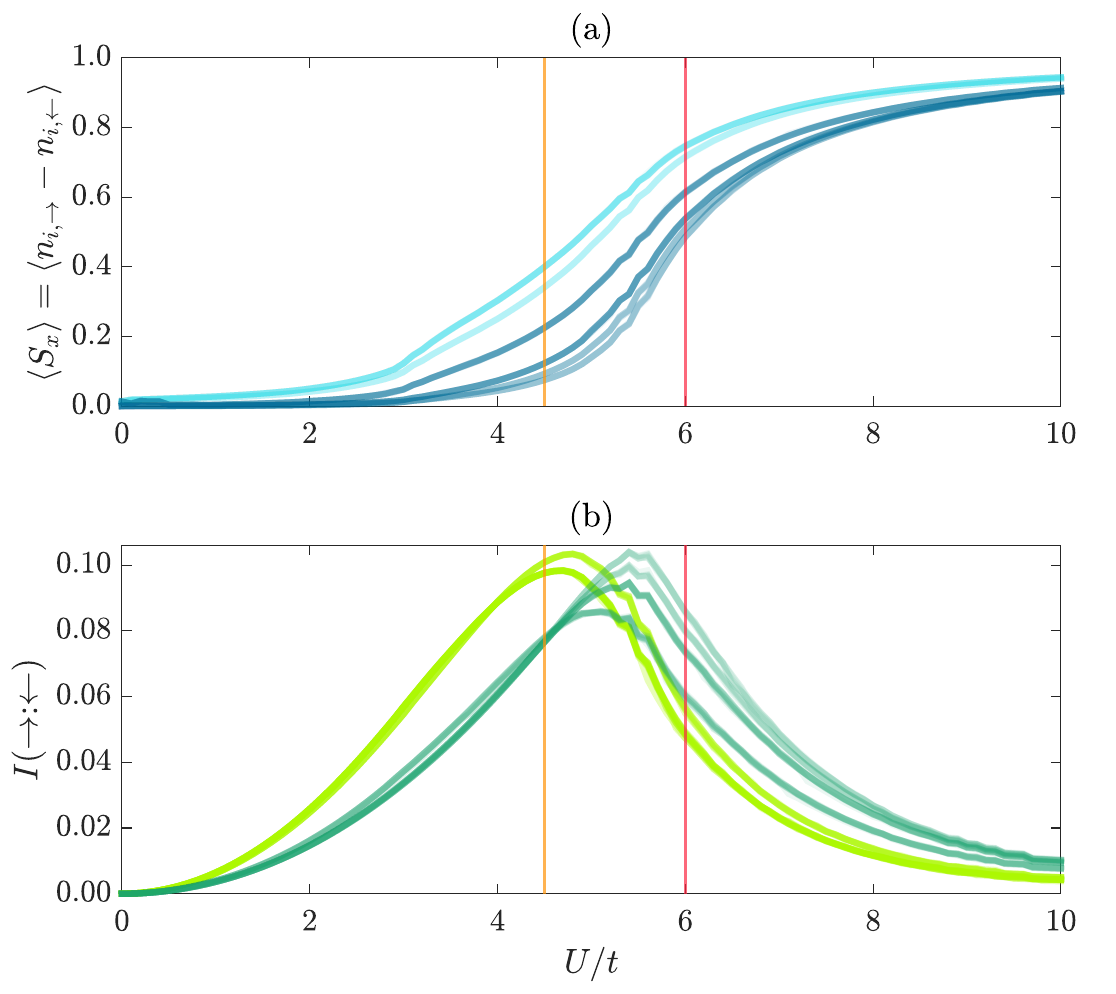}\\[1mm]
    (c) \hspace{10cm} (d) \\
    \hspace{1.2cm}
    \includegraphics[width=0.33\linewidth]{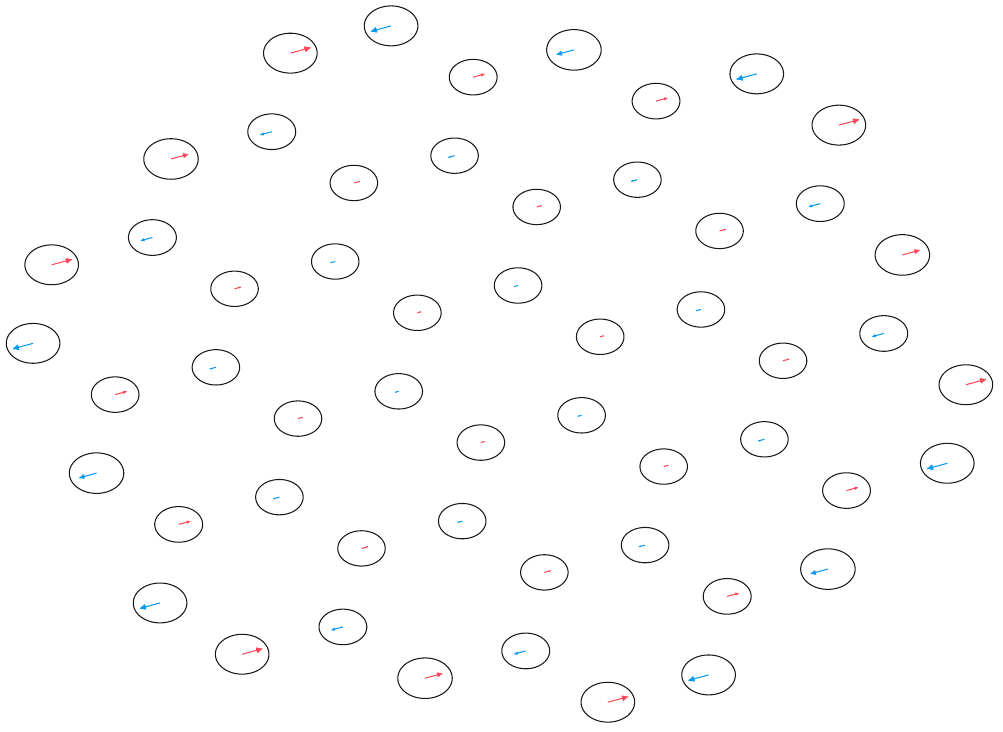} \hspace{1.2cm}
    \includegraphics[width=0.33\linewidth]{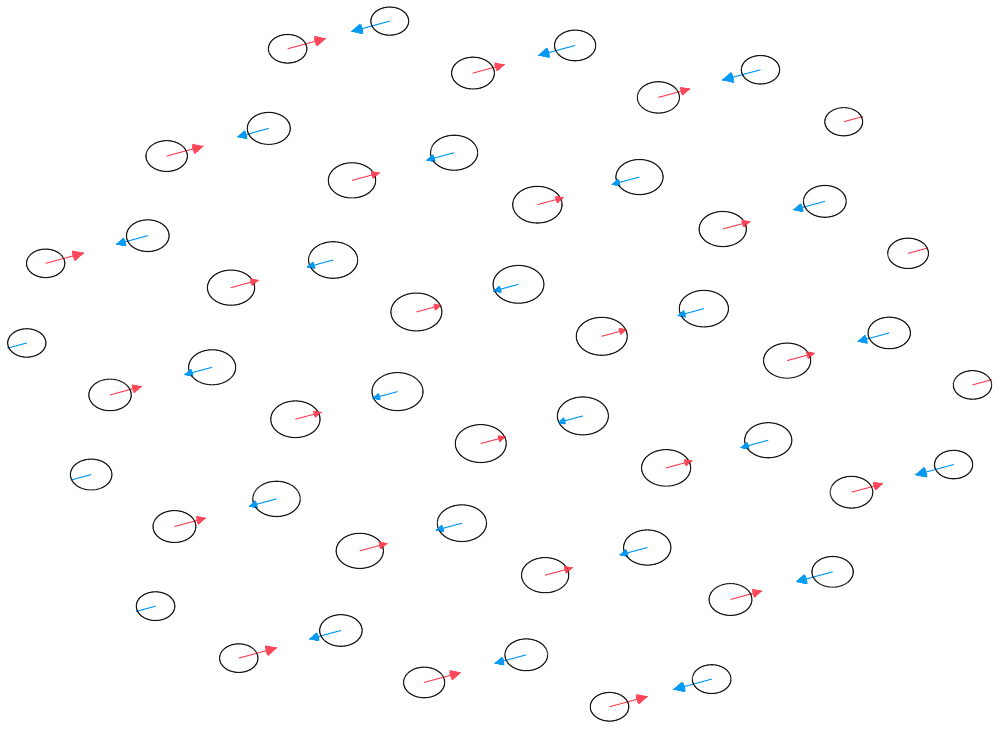} \vspace{3mm}
    \caption{Kane-Mele-Hubbard model at the nanoscale.
    In (a) the local in-plane
    magnetization $\ibra S_x \iket = \ibra n_{i,\rightarrow}-n_{i,\leftarrow}\iket$. In
    (b) the local correlation as measured by the mutual 
    information between local in-plane spin eigenstates
    $I(\rightarrow:\leftarrow)$ \cite{BellomiaPhD,BellomiaKMH,Bellomia_intracorr}. 
    Both the (a) and (b) panels mark the edge and ``bulk'' 
    sites in lighter and darker colors, respectively.
    In (c) and (d) a visual representation of the given 
    nano-system at different values of $U/t$, as marked 
    respectively by the orange and red vertical lines 
    in panels (a) and (b). The area of the 
    circles is proportional to $I(\rightarrow:\leftarrow)$, 
    the length of the arrows to $\ibra S_x \iket$.
    All data are taken at $\lambda_\mathrm{so}=0.3t$ and $T=0$.}
    \label{fig:KMflake}
\end{figure}

\subsubsection{Kane-Mele-Hubbard model at the nanoscale}
The capabilities of the EDIpack2ineq layer go well beyond the
simple case of a two-site unit cell. Building on the previous example, we can readily adapt the calculation to deal with a nanoscopic \cite{Amaricci2014PRA} system with $N_\mathrm{ineq}$ inequivalent atoms, such as hexagonal flakes
\cite{Valli2016PRB,Valli2018NL}.
We consider a 
$N_\mathrm{ineq}\times N_\mathrm{ineq}$ local Hamiltonian 
$H_\mathrm{loc}$ with open boundary conditions, and implementing
the self-consistent equations with a trivial single momentum 
$k=0$. In this configuration the system develops an inhomogeneous magnetization, so that in general no particular lattice symmetry group, e.g. N\'eel, can be exploited. 

In \figu{fig:KMflake} we report results for a nanosystem of 
$N_\mathrm{ineq}=54$ sites, solved with the bath parametrization 
described in \equ{eq:afmx_dmft_ansatz}, for the in-plane 
antiferromagnetic state.
Panel (a) shows the interaction dependence of the local magnetization,
computed from the impurity ground state as  
$ \ibra S_x \iket = \ibra c^\dagger_{i,\up} c_{i,\dw} + c^\dagger_{i,\dw} c_{i,\up} \iket \equiv \ibra n_{i,\rightarrow} \iket - \ibra n_{i,\leftarrow} \iket$,
so that we can define the $S_x$ spin occupation numbers as
\begin{equation*}
    \ibra n_{i,\rightarrow} \iket \equiv \frac{\ibra n_{i,\up} + n_{i,\dw} \iket +
    \ibra c^\dagger_{i,\up} c_{i,\dw} + c^\dagger_{i,\dw} c_{i,\up} \iket}{2},
    \qquad
    \ibra n_{i,\leftarrow} \iket \equiv \frac{\ibra n_{i,\up} + n_{i,\dw} \iket -
    \ibra c^\dagger_{i,\up} c_{i,\dw} + c^\dagger_{i,\dw} c_{i,\up} \iket}{2},
\end{equation*}
which trivially satisfy the conservation of the local average charge $\ibra n_{i,\rightarrow} + n_{i,\leftarrow} \iket = \ibra n_{i,\up} + n_{i,\dw} \iket$.
Having a direct expression for $\ibra n_{i,\rightarrow} \iket$ and $\ibra n_{i,\leftarrow} \iket$
we can verify that the impurity reduced density matrix, computed by 
tracing the bath states in the \texttt{ed\_mode=nonsu2} representation 
(see \ref{sSecRDM}), once diagonalized, takes the familiar form
\cite{Zanardi2002PRA,Su2013MPLB,Walsh2019PRL} 
\begin{equation}
    \rho^\mathrm{imp} = 
     \begin{pmatrix}
            \left\ibra\left(1-n_{i,\rightarrow}\right)\left(1-n_{i,\leftarrow}\right)\right\iket   &    0          &       0       &   0 \\
                    0       &    \left\ibra n_{i,\rightarrow}\left(1-n_{i,\leftarrow}\right)\right\iket  &       0       &   0 \\
                   0       &       0   &    \left\ibra\left(1-n_{i,\rightarrow}\right)n_{i,\leftarrow}\right\iket      &   0 \\
                    0       &       0   &       0          &    \left\ibra n_{i,\rightarrow} n_{i,\leftarrow}\right\iket
        \end{pmatrix} 
\end{equation}
in the basis of the {\it natural} spin-orbitals
\cite{BellomiaPhD,BellomiaKMH,Bellomia_intracorr}, 
whose occupation numbers
are given by $n_{i,\rightarrow}$ and $n_{i,\leftarrow}$.
As  discussed in \cite{BellomiaPhD,BellomiaKMH,Bellomia_intracorr},
by defining the spin traces of $\rho^\mathrm{imp}$, as
\begin{equation}
    \rho^\mathrm{imp}_\sigma = 
    \Tr_{\bar{\sigma}} \left[\rho^\mathrm{imp}\right] =
    \begin{pmatrix}
        \ibra n_{i,\sigma} \iket & 0 \\
        0 & \ibra n_{i,\bar{\sigma}} \iket
    \end{pmatrix}
\end{equation}
we can directly measure local correlations by defining the mutual information
between the impurity spin-orbitals
$I(\rightarrow:\leftarrow) = 
    S(\rho^\mathrm{imp}_{\rightarrow}) + 
    S(\rho^\mathrm{imp}_{\leftarrow}) -
    S(\rho^\mathrm{imp})$,
where $S(\cdot)$ denotes the von Neumann entropy of its argument.

In panel (b) of \figu{fig:KMflake} we report the spatially resolved
interaction dependency of $I(\rightarrow:\leftarrow)$, as a direct 
observation of the enhanced local correlations at the boundary of the
finite system. While it is usually assumed that edge sites must be
more correlated than internal (``bulk'') sites, based on the reduced
lattice coordination, the inter-spin mutual information quantifies directly the phenomenon: the edge correlations,
marked in lighter green, are decidedly higher as long as the system 
remains in a weakly magnetized regime. When the local magnetization
exceeds half of its saturation value, all local correlations start to decrease,
as spin fluctuations are damped by ordering. Significantly, the edge 
correlations get lower than the bulk correlations as the maximum is 
crossed, consistently with the observation that their magnetization
is always higher and so their fluctuations are frozen faster.
Finally, we observe that the nanosystem of \figu{fig:KMflake} has 
finite local magnetization for arbitrary low interaction strengths,
in striking contrast with the infinite lattice. This is a well-known
effect of quantum confinement, as proposed in early works on graphene
flakes \cite{Valli2016PRB,Valli2018NL}.

\ifSubfilesClassLoaded{
  \bibliography{references}
}{}
\end{document}

%% file: 06_conclusions_acknowledgement_appendix.tex
\section{Conclusions}\label{SecConclusions}
We have presented \NAME{}, a versatile and high-performance ED solver for generic quantum impurity problems. Building on the massively
parallel algorithms introduced by its predecessor \cite{Amaricci2022CPC}, this version of
the library features new capabilities allowing for different broken-symmetry solutions within a unified
framework that provides, for instance, reliable evaluations of a variety of local dynamical correlation
functions of an arbitrary complex frequency ranging from the Green's functions to superconducting, excitonic and magnetic response functions.

This new version allows for an efficient evaluation of 
reduced density matrices of the impurity directly from Fock space quantities and thus enables  
quantum information analysis of correlated systems, which is becoming an essential tool to address emergent phases of quantum materials.

A central feature of \NAME is its strong focus on interoperability,
achieved through modern Fortran constructs, C/C++-bindings, and
comprehensive APIs for Python (EDIpack2py). 
These interfaces enable seamless integration with broader
computational frameworks, like TRIQS \cite{Parcollet2015CPC} and
w2dynamics \cite{Wallerberger2019CPC}, expanding the
functionality of these platforms and providing a robust foundation for
reproducible research.

We thoroughly discussed the implementation of the \NAME library and its
most important algorithms and classes. We presented in
detail the third-party interfaces which extend ED
capabilities beyond the domain of the library itself. 
Finally, we showcased the use of the \NAME software in different
contexts through elaborated examples, which can serve as a reference for potential users. 

The modular and extensible design of \NAME provides a
natural foundation for future extensions.
For instance it provides a suitable basis to implement cluster-DMFT \cite{Capone2004PRB,Kotliar2006RMP,Park2008PRL} methods, where the
impurity problem is generalized to include clusters of interacting
sites. This approach includes spatial correlations beyond the single-site approximation, capturing effects
such as $d$-wave pairing, charge ordering, and complex spin textures in strongly correlated systems.
A different outlook is the development of an integration layer with other quantum embedding methods, such as the ghost Gutzwiller scheme \cite{Lanata2015PRX},
which heavily rely on solving effective impurity problems.    
We anticipate that \NAME will become a valuable tool for the computational condensed matter community, supporting a wide range of
studies on strongly correlated materials and providing a reliable
reference platform for quantum impurity solvers.


\section*{Acknowledgements}
We acknowledge helpful discussions occurring at various stages of development of \NAME with: C.~Mejuto-Zaera, P.~Villar~Arribi, M.~Chatzieleftheriou, S.~Adler, F.~Paoletti, M.~Collura, C.~Weber, A.~Sartori, H.~Choi, A.~Valli, M.~Rozenberg.    

\paragraph{Author contributions}
A.A. designed the project, developed the main structure of the software and conceived the computational algorithms in the core library. A.A., L.C. and M.C. conceived the diagonalization, correlation function  calculation and bath optimization strategies implemented in the library. A.A., G.M., A.S., S.G. and M.C. devised the inclusion of the electron-phonon coupling and of the superconducting state. 
L.C., G.M., F.P., A.S., S.G. and G.B. equally contributed to the development and implementation of many different aspects of the software, including code writing, configuration, installation, testing and continuous integration procedures. 
L.C. devised and developed the C-binding layer, the Python, Julia and Monicelli interface together with Python specific deployment and installation tools. 
I.K. developed the TRIQS interface and contributed to the testing part. 
A.K., M.W. and G.S. developed the integration with w2dynamics.  
A.A., M.C., L.d.M. and G.S designed additional algorithms for extended functionalities of the core software.   
A.A., L.C., I.K., A.K., S.G. and G.B. performed the calculations, analyzed the results and prepared the data presentation in the examples.
A.A. and L.C. wrote the backbone of the manuscript.
All authors participated in writing this manuscript. 

\paragraph{Funding information}
L.C. acknowledges support from Deutsche Forschungsgemeinschaft (DFG, German Research Foundation) thr\-ough the cluster of excellence ``CUI: Advanced Imaging of Matter" of the Deutsche Forschungsgemeinschaft (DFG EXC 2056, Project ID 390715994) and gratefully acknowledges the scientific support and HPC resources provided by the Erlangen National High Performance Computing Center (NHR@FAU) of the Friedrich-Alexander-Universität Erlangen-Nürnberg (FAU) under the NHR project b158cb. NHR funding is provided by federal and Bavarian state authorities. NHR@FAU hardware is partially funded by the German Research Foundation (DFG)– 440719683.
I.K. acknowledges support from the European Research Council (ERC) under the European Union's Horizon 2020 research and innovation programme (Grant agreement No.~854843-FASTCORR).
G.S. and A.K. acknowledge financial support by the Deutsche Forschungsgemeinschaft (DFG, German Research Foundation) through FOR 5249 (project-id 449872909, Project P05), and the W\"urzburg-Dresden Cluster of Excellence on Complexity and Topology in Quantum Matter-ct.qmat (EXC 2147, project-id 390858490).
F.P. acknowledges support from the Swiss National Science Foundation via NCCR Marvel and SNSF Grant No. 200021-196966.
M.W. acknowledges support from the Austrian Science Fund (FWF) through grant DOI 10.55776/P36332.
G.M. acknowledges support from the MUR - Italian Ministry of University and Research - through the Rita-Levi Montalcini program.
L.dM. acknowledges support by the European Commission through the ERC-CoG2016, StrongCoPhy4Energy, GA No. 724177.
M.C., S.G. and G.B. acknowledge financial support from the National Recovery and Resilience Plan PNRR
MUR Project No.~CN00000013-ICSC and by MUR via PRIN 2020 (Prot.~2020JLZ52N-002) and PRIN 2022 (Prot.~20228YCYY7) programmes.
A.A. and M.C. acknowledge financial support from the National Recovery and Resilience Plan PNRR MUR Project No.~PE0000023-NQSTI.

\section{Appendix A: Monicelli interface}\label{appendixA}
In order to further showcase the interoperability capabilities of
\NAME provided by the C-bindings module, we describe a simple interface to {\tt Monicelli}.

{\tt Monicelli} is an esoteric programming language based on the LLVM
toolchain.
It is written in C++ and offers wrappers for the basic
types and operations thereof. The syntax replicates a cultural Italian
phenomenon known as ``supercazzola", a rambling nonsensical discourse
which gives the false impression of carrying an actual meaning,
introduced in the movies trilogy ``Amici Miei'', directed by Mario
Monicelli (see \href{https://it.wikipedia.org/wiki/Amici_miei}{it.wikipedia.org/wiki/Amici\-\_\-miei}). 

The {\tt Monicelli} language can be found at
\href{https://github.com/esseks/monicelli.git}{github.com/esseks/monicelli.git}
and can be installed using CMake. The software depends on C++
compiler with {\tt stdlib}, LLVM and in some cases {\tt libz} library. The installation gives access to the {\tt
  Monicelli} compiler {\tt mcc}, which statically links LLVM. 

We believe the discourse flow of this language is best captured by 
illustrating its features and \NAME interface via a paraphrase of the original supercazzola, an epic non-sensical dialogue between a traffic policeman (V), count Mascetti (M) and G. Perozzi (P) (see
\href{https://it.wikipedia.org/wiki/Supercazzola#Origine}{it.wikipedia.org/wiki/Supercazzola}
and \href{https://www.youtube.com/watch?v=SF8YUFdP6eU}{Movie Scene}). 

\begin{dialogue}
\speak{\MakeUppercase{V}} {\it Lei ha clacsonato!}

\speak{\MakeUppercase{P}} {\it Tu ha clacsonato?}
\par\lips\par

\speak{\MakeUppercase{M}} {\it Tarapia tapioco, prematurata l'interfaccia, o scherziamo?}

\speak{\MakeUppercase{V}} {\it Prego}?

\speak{\MakeUppercase{M}} {\it Scusi noi siamo in Monicelli, come fosse un linguaggio esoterico basato su C++ e utilizzante la toolchain di LLVM anche per Linux e Unix soltanto in due, oppure in quattro anche scribai con il file
sorgente {\tt hm\_bethe.mc}? \\ Come {\tt
  github.com/lcrippa/prematurata\_la\_dmft}, per esempio?}

\speak{\MakeUppercase{V}} {\it Ma che DMFT, mi faccia il piacere! Questi signori stavano programmando loro, non si intrometta!}

\speak{\MakeUppercase{M}} {\it Ma no, dico, mi porga il file {\tt bagaglio.cpp}. Le vede le funzioni? Lo vede che interfacciano gli array, non supportati da {\tt Monicelli}, e prematurano anche! 
Ora io le direi, anche con il rispetto per l'autorità, anche solo le
due parole come install {\tt Monicelli} from
\href{https://github.com/esseks/monicelli.git}{github.com/esseks/monicelli.git}
e compila il file {\tt hm\_bethe.mc}, per esempio.}

\speak{\MakeUppercase{V}} {\it Basta così! Mi seguano nel programma di test!}

\speak{\MakeUppercase{P}} {\it No, no, no, attenzione! Il loop DMFT completo \`e supportato secondo la Ref. \cite{Georges1996RMP}, abbia pazienza,
senn\`o.... Plotta i dati, anche un pochino di Green's function e
Self-energy in prefettura.}

\speak{\MakeUppercase{M}} {\it Senza contare che {\tt prematurata\_la\_dmft} ha perso i contatti con il tarapia tapioco, dopo.}

\par\lips\par

\speak{\MakeUppercase{V}} {\it Ho bello che capito. Si farà finta di passar da bischeri!}
\end{dialogue}

We shall now present a fully functioning DMFT code for
the solution of the Bethe lattice written in Monicelli. It requires a minimal C++
interface to handle arrays, complex numbers and properly interfacing
the EDIpack functions.
The first part of the DMFT code loads all the required functions from
{\tt bagaglio.cpp}, in particular it loads the interface to the \NAME
solver functions {\tt ed\_init\_solver} and {\tt ed\_solve}:

\lstset{breaklines=true, breakindent=1em}
\begin{lstlisting}[style=MonicelliStyle,numbers=none]
bituma le funzioni ausiliarie che vengono dal bagaglio
...
blinda la supercazzola leggi o scherziamo?
blinda la supercazzola iniziailsolver o scherziamo?
blinda la supercazzola risolvi o scherziamo?
blinda la supercazzola prendilasigma o scherziamo?
blinda la supercazzola prendilbagno con l`elemento Necchi, il valore Sassaroli o scherziamo?
...
\end{lstlisting}

After a long initialization, we follow the structure already presented
in \secu{SecExamplesBetheDMFT} for the Fortran implementation.
The DMFT code starts with the opening {\tt Lei ha clacsonato} (see
above). Next, we read the input file and set the dimensions of some
array describing the bath, the self-energy and Green's
functions. We start a DMFT iteration loop using the command {\tt
  stuzzica}, which includes an internal frequency loop as {\tt
  Monicelli} does not support array algebra.
The loop contains a call to the \NAME solver function, then proceeds
by retrieving the self-energy function which is used to obtain the
local Green's function. These enter the self-consistency equation
which updates the Weiss field. Finally bath optimization is performed
and if error condition is met the loop exit.
The main step of the implementation reads:

\begin{lstlisting}[style=MonicelliStyle,numbers=none]
Lei ha clacsonato
#Read the input file
    prematurata la supercazzola dimmilfile o scherziamo?
    prematurata la supercazzola leggi o scherziamo? 
...                                                  
#Init the ED solver:
    voglio il sapone, Necchi come se fosse
    prematurata la supercazzola ilbagnoepronto o scherziamo?
    prematurata la supercazzola lavati con il sapone o scherziamo?
        
    prematurata la supercazzola prendilah o scherziamo?
    prematurata la supercazzola iniziailsolver o scherziamo?
...
#DMFT loop:
stuzzica    
      prematurata la supercazzola risolvi o scherziamo?
      
      prematurata la supercazzola prendilag o scherziamo?
      prematurata la supercazzola prendilasigma o scherziamo?
#Get $G_{loc}$ and update $\GG^{-1}$: frequency loop
      stuzzica                   
          il contatoredue come fosse 0
          prematurata la supercazzola prendi con 108, contatore, 0.0, 0.0, 0 o scherziamo?
...
      e brematura anche, se il contatore minore di frequenze
...      
      prematurata la supercazzola lavatiancora o scherziamo?
#Fit the bath:       
      prematurata la supercazzola spiaccica o scherziamo?
#Mix the bath           
      prematurata la supercazzola failmischiotto con il sapone, il frullatore o scherziamo?
#Check error      
      la fine come se fosse prematurata la supercazzola cisiamo o scherziamo?         
 e brematura anche, se la fine minore di 1
\end{lstlisting}

\paragraph{Usage.}
The source code can be retrieved from
\href{https://github.com/lcrippa/prematurata_la_dmft}{github.com/lcrippa/prematurata\_la\_dmft}. 
The {\tt src} directory contains several files, including the source
code {\tt hm\_bethe.mc} implementing the DMFT algorithm, an auxiliary file {\tt bagaglio.cpp} which
contains a number of functions implementing complex algebra, array
construction and interfacing the \NAME procedures. The directory also
contains an example of input file and a converged bath parametrization
for the Bethe lattice solution at $U=2$.

The code is compiled using standard Make invocation in the source
directory:
\begin{lstlisting}[style=mybash,numbers=none]
git clone https://github.com/lcrippa/prematurata_la_dmft
cd prematurata_la_dmft/src
make
\end{lstlisting}
A simple run using the provided input file {\tt inputED.conf} will
re-converge the solution within a few loops. 

\ifSubfilesClassLoaded{
  \bibliography{references}
}{}
\end{document}